%% file: paper.tex
\documentclass[fleqn,usenatbib]{mnras}

\usepackage{newtxtext,newtxmath}
\usepackage[T1]{fontenc}
\usepackage{ae,aecompl}

\usepackage[referable]{threeparttablex}
\usepackage{mathtools}
\usepackage{threeparttable}
\usepackage{standalone}
\usepackage{booktabs, dcolumn}
\usepackage{afterpage}
\usepackage{xspace}
\usepackage{color}
\usepackage{graphicx}	
\usepackage{amsmath}	
\usepackage{amssymb}
\usepackage{longtable}

\newcommand{\s}{\textit{S}}
\newcommand{\n}{\textit{N}}
\newcommand{\mr}{\textit{MR}}
\newcommand{\bc}{\textit{BC}}

\newcommand\T{\rule{0pt}{2.6ex}}      
\newcommand\B{\rule[-1.2ex]{0pt}{0pt}} 
\newcommand{\oii}{\ion{O}{II}}
\newcommand{\cii}{\ion{C}{II}}
\newcommand{\nii}{\ion{N}{II}}
\newcommand{\oiii}{\ion{O}{III}}
\newcommand{\om}{$\Omega$/4$\pi$}
\newcommand{\hp}{H$^+$}
\newcommand{\op}{O$^+$}
\newcommand{\opp}{O$^{++}$}
\newcommand{\oppp}{O$^{+3}$}
\newcommand{\hb}{H$\beta$}

\newcommand{\Draft}[1]{#1}
\newcommand{\DraftTwo}[1]{#1}
\newcommand{\DraftThree}[1]{{#1}}

\title[Bi-abundance photoionization models of PNe]{Bi-abundance photoionization models of planetary nebulae: determining the amount of Oxygen in the metal rich component}

\author[V. G\'omez-Llanos \& C. Morisset]{
V. G\'omez-Llanos\thanks{E-mail: \url{mailto:vgomez@astro.unam.mx}, ORCID: \url{https://orcid.org/0000-0002-1825-8267} }
and C. Morisset\thanks{E-mail: \url{mailto:chris.morisset@gmail.com}, ORCID: \url{https://orcid.org/0000-0001-5801-6724}}
\\
Instituto de Astronomia (IA), Universidad Nacional Aut\'onoma de M\'exico,  
Apartado postal 106, C.P. 22800 Ensenada, 
\\ Baja California, M\'exico.}

\date{Accepted XXX. Received YYY; in original form ZZZ}

\pubyear{2019}

\begin{document}
\label{firstpage}
\pagerange{\pageref{firstpage}--\pageref{lastpage}}
\maketitle

\begin{abstract}
We study the hypothesis of high metallicity clumps being responsible for the abundance discrepancy found in planetary nebulae between the values obtained from recombination and collisionaly excited lines. We generate grids of photoionization models combining cold metal-rich clumps emitting the heavy element recombination lines, embedded in a normal metallicity region responsible for the forbidden lines. 
The two running parameters of the grid are the metallicity of the clumps and its volume fraction relative to the whole nebula. We determine the density and temperatures (from the Balmer jump and the [\ion{O}{III}]~5007/4363~\AA\ line ratio), and the ionic abundances from the collisional and recombination lines, as an observer would do. The \Draft{metallicity} of the near-to-solar region is recovered, while the metallicity of the clumps is systematically underestimated, by up to 2 orders of magnitude. This is mainly because most of the H$\beta$ emission is coming from the "normal" region, and only the small contribution emitted by the metal-rich clumps should be used.
We find that a given ADF(O$^{++}$) can be reproduced by a small amount of rich clumps, or a bigger amount of less rich clumps. 
Finally, comparing with the observations of NGC~6153 we find 2 models that reproduce its ADF(O$^{++}$) and the observed electron temperatures. We determine the fraction of oxygen embedded in the metal-rich region (with a fraction of volume less than 1\%) to be roughly between 25\% and 60\% of the total amount of oxygen in the nebula (a few 10$^{-3} M_\odot$).
\end{abstract}


\section{Introduction}

Determining the physical properties and chemical abundances of the interstellar medium can help us to get a better understanding about the formation and evolution of galaxies. Photoionized regions such as planetary nebulae (PNe) and \ion{H}{ii} regions, give us information about the past and the present composition of the interstellar medium in which they were formed. The method for estimating chemical abundances in these regions is well detailed in \citet{2017Peimbert_pasp129}. However, there are two big problems in the determination of physical parameters and chemical abundances in photoionized regions. The first problem is that in most cases the temperature T(BJ) estimated using the Balmer jump of H is lower than the temperature T([\ion{O}{iii}]) determined with the emission line ratio [\ion{O}{iii}]~4363/(4959+5007) \citep{1971Peimbert_Bole6}. The second problem is that the ionic abundances estimated with optical recombination lines (ORLs) are systematically higher than the ones estimated with collisionally excited lines (CELs) for the same ion. The ratio of these two ionic abundances is defined as the abundance discrepancy factor (ADF). \cite{2001Liu_mnras327} shows that there is a positive correlation between the ADF(O$^{++}$) and the difference in temperatures T([\ion{O}{iii}])-T(BJ) for a sample of 11 PNe. \Draft{\citet{2004Tsamis_mnra353} show that there is some evidences for a cold plasma using the ORLs observations.} Some mechanisms have been proposed to explain the abundance discrepancy, such as temperature fluctuations in the gas \citep{1967Peimbert_apj150}, non-thermal electron energy distribution \citep{2012Nicholls_apj752}, density condensations (Viegas \& Clegg 1994)\nocite{1994Viegas_mnra271}, and abundance inhomogeneities \citep{1990Torres-Peimbert_aap233, 2000Liu_mnra312, 2002Pequignot_12}.

Chemically homogeneous photoionization models of PNe can not reproduce both CELs and ORLs in the cases of high ADFs \Draft{\citep{2009Bohigas_rmxa45}}. Models combining chemically inhomogeneous regions have been able to reproduce the CEL and ORL simultaneously \citep[][for the cases of NGC 6153, M 2-36, M 1-42, Abell 30, 30 Doradus, and PN SMC N87]{2002Pequignot_12, 2003IAUS..209..389T, 2003Ercolano_mnra344, 2005Tsamis_mnra364, 2006pnbm.conf..192T}. 
Afterward, a 3D chemically inhomogeneous model developed by \citet{2011Yuan_mnra411} for the PN NGC~6153, with an ADF$\sim$10, successfully reproduces the observed intensities of the CELs and ORLs. In this chemically inhomogeneous model, the H-poor (or metal-rich) region that only has less than 2\% of the total mass contributes to the majority of the emission of heavy elements ORLs, while the emission of UV and optical CELs of heavy elements comes mainly from the gas with "normal" \Draft{metallicity}. If the amount of metals in the H-poor region increases, the emissivity of the ORLs will be higher. The same line emission could then be obtained with a more massive and not so metal-rich region or a less massive and more metal-rich region, leading to a degeneracy between the metallicity of the metal-rich region and its mass contribution. 

Observational evidence of chemical inhomogeneities has been shown by \citet{2016Garcia-Rojas_apjl824}, where narrow-band images of PN NGC 6778 are obtained. The images show that the regions emitting the ORLs \ion{O}{ii} 4649+51 is located in the central parts of the nebula, while the CEL [\ion{O}{iii}]~5007 comes from the outer parts of the nebula that coincides with the H$\alpha$ emission. \citet{2017Pena_mnras472} measured the expansion velocity (V$_{exp}$) for different ions using ORLs and CELs, for 14 PNe, finding that in most cases the kinematics is different for ORLs and CELs (V$_{exp}^{ORLs}$ < V$_{exp}^{ CELs}$). They suggest that the ORLs are emitted by a plasma that was ejected in a different event that the plasma emitting the CELs, causing the difference in expansion velocities.

An important parameter when chemical inhomogeneities are present, is the electronic temperature of the H-poor region, since it can give us an idea of how H-poor is the gas. However, determining this temperature it is not an easy task, given that ORLs have a low dependence on temperature, and are very faint. Small uncertainties in the recombination line ratios will give a large variations in the temperature estimation.

In this work we want to investigate the ADF values (determined as an observer would do) for models composed of different metallicity regions: a gas with "normal" \Draft{metallicity} and a metal-rich gas. We study the degeneracy between the metallicity and the mass of the metal-rich gas, and the consequences of the overestimation of H$\beta$ when determining abundances with ORLs. We show that the total mass of oxygen involved in the metal-rich region is not strongly affected by the degeneracy and we determine a range of values for this parameter in the case of PN NGC 6153.

The structure of the paper is outlined here. In sect. \ref{sec:models} we describe how we make the models, its components and the hypothesis that are assumed. In sect. \ref{sec:results} we show the estimations of plasma physical conditions and chemical abundances for the outputs of the bi-abundance models. In sect.~\ref{sec:compare_mod_object} we compare the results of our models to an observed object.
In sect. \ref{sec:discussion} we discuss the obtained results and detail the limitations of the work. Finally in sect. \ref{sec:conclusions} we present the conclusions of the work.

\section{Modeling strategy}
\label{sec:models}
We use the code Cloudy \DraftThree{v.17.02} \citep{2017Ferland_rmxaa53}, via the python library pyCloudy \citep{2013Morisset_}, to compute the photoionization models for this work. This code solves the thermal equilibrium, the photoionization equilibrium and the radiative transfer for each concentric cell in a spherical symmetric simulation. \Draft{The code outputs the electron temperature and density, the ionic fractions of all the elements, as well as the line emission for each cell of the model.} \DraftTwo{The line emissions are obtained with PyNeb \citep{2012Luridiana_283} in the case of: \oii\ from \citet{2017Storey_mnra470}, and \cii\ and \nii\ from \citet{1991Pequignot_aap251}. We use the add\_emis\_from\_pyneb method from the CloudyModel class of the pyCloudy library, which use the electron temperature and density, and the ionic abundance of the corresponding element, to compute the emissivities of lines which are not computed by Cloudy. It calls the PyNeb corresponding RecAtom class and computes the emission of the lines for each zone of the Cloudy model. The total intensity is obtained by integrating over the volume, in the same way than it is done for the lines computed by Cloudy itself.}

\begin{figure}
\centering
\includegraphics[scale = 0.55]{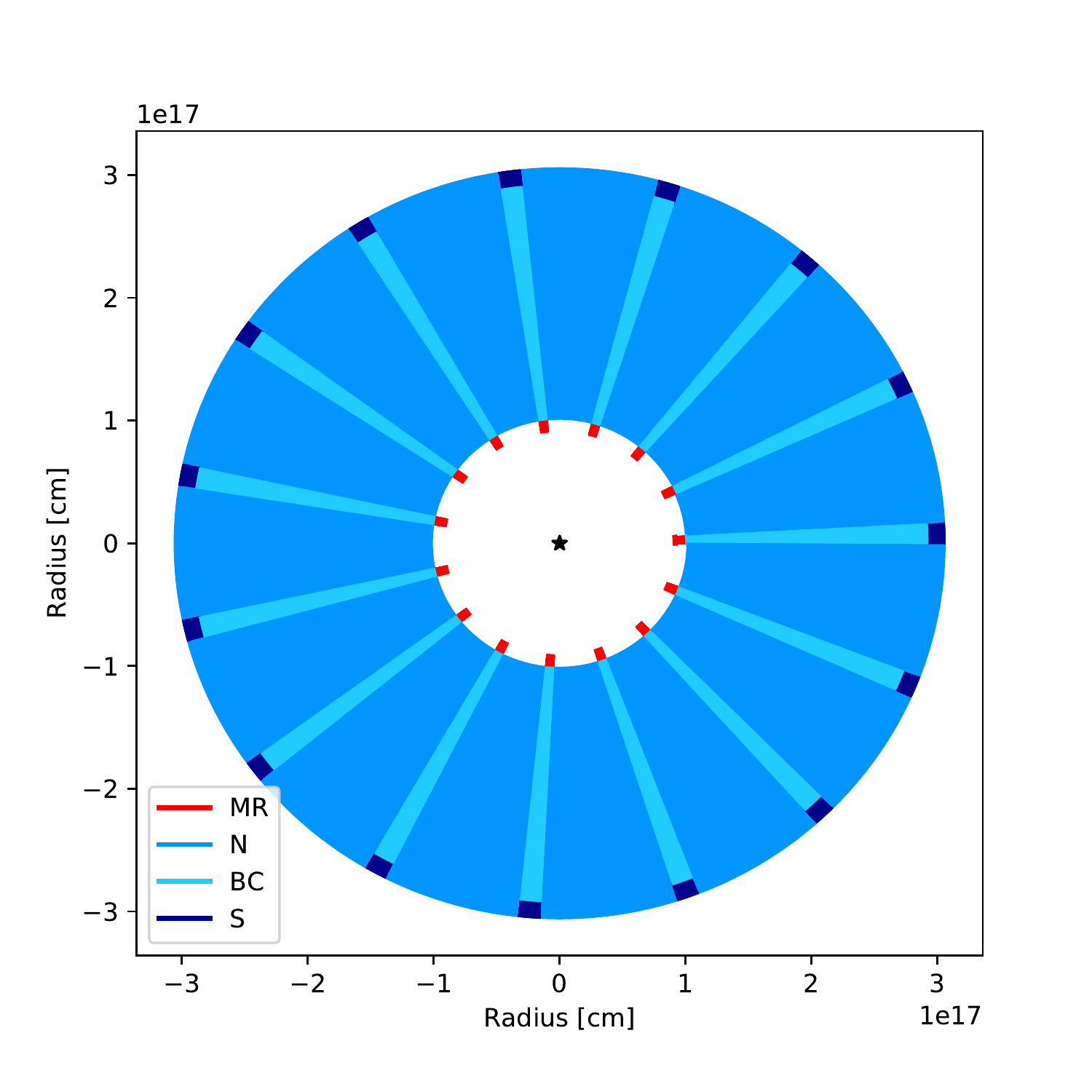}
\caption{Schematic representation of the morphology of the 4 components of our chemically inhomogeneous PN. All blue regions are gas with close to solar \Draft{metallicity}, and the red regions represents metal-rich (\mr) clumps that in this case are about 630 times more metallic. The region of middle darkness blue is the "normal" gas (\n), the light blue region is the gas behind the clumps (\bc), and the dark blue region is a shadow (\s) ionized by the diffuse of the normal component. The fraction of solid angle for: \mr, \bc\ and \s\ is $\Omega$/4$\pi$~=~0.25.} 
\label{fig:squeme}
\end{figure}

To simulate a PN one needs to describe the properties of the ionizing star and its surrounding gas. We simplify the emission of the central star to a black body at a given effective temperature (Teff) and luminosity (L). The gas is described with the following properties: its morphology, its \Draft{metallicity} and its density distribution. We assume a spherical morphology with constant hydrogen density in the whole nebula. In the figure \ref{fig:squeme} the different regions of the modeled PN are schematically represented in different colors. The nebula is made of a close to solar \Draft{metallicity} gas represented in blue with different darkness (see below) and some metal-rich clumps shown in red. To model such a chemically inhomogeneous nebula with a 1D photoionization code, we actually run different models that will be combined afterward. One model (M$_1$) corresponds to the "normal" gas (\n, middle darkness blue in Fig.~\ref{fig:squeme}). Another model (M$_2$) deals with the metal-rich clumps (\mr, in red) and the gas behind these clumps (\bc, light blue). This model M$_2$ is in fact computing 2 regions at once. 
It appears that in some conditions this \bc\ gas recombines at a smaller radius than the \n\ gas outer radius: this implies the existence of a fourth region: a shadowed component (\s, in dark blue of fig.~\ref{fig:squeme}) of same \Draft{metallicity} than the \n\ component, that requires its own model (M$_3$).

As a summary we present in table \ref{tab:summary_names} the names of the different components, their corresponding abbreviations, the reference of the photoionization model, and the regime of metallicity for each one of the four components. These four gas components are detailed in the following sections.

\begin{table}
    \centering
    \caption{Some properties of the four components used to build the PN model in this work.}
    \begin{tabular}{c c c c}
    \hline \hline
    Name & Abbreviation & Model & Metallicity \\ \hline \hline
    "normal" & \n & M$_1$ & close to solar \\
    meta-rich & \mr & M$_2$ & rich \\
    behind clump & \bc & M$_2$& close to solar\\
    shadow & \s & M$_3$&  close to solar \\
     \hline \B
    \end{tabular}
    \label{tab:summary_names}
\end{table}

\subsection{"Normal" (\n) component}
\label{sec:normal-model}

The "normal" (\n) region is the base component (with close to solar metallicity) of our nebular models. We choose to compare our models with the PN NGC~6153 (see section \ref{sec:obs}), because it has UV, optical and IR observations \citep{1984Pottasch_apjl278, 2000Liu_mnra312} and a previous 3D bi-abundance model has been published by \citet[][here after Y11]{2011Yuan_mnra411}. The parameters for the \n\ component are based on the Bn component of the 3D bi-abundance model by Y11. From the column of the Bn component of their table 2, we take the effective temperature (92~kK), luminosity (1.3$\times 10^{37}$~erg/s) and chemical abundances \Draft{(shown in table \ref{tab:N_abunds}).} 
The inner and outer radius of our \n\ component are approximated to 5 and 15~arcsec based on the observations and models presented in Y11. The distance to NGC~6153 used by Y11 is of 1.5~kpc \citep{2003Pequignot_209}, but more recent data has been published by \citet{2018Gaia_vizier1345} resulting in distance of 1.36~kpc. With such a distance the inner and outer radius of the nebula are 1.02$\times 10^{17}$~cm and 3.05$\times 10^{17}$~cm, respectively. The hydrogen density is simplified to a constant value of 3,000~cm$^{-3}$, taken as an approximation of the the density distribution of Y11. The \hb\ flux from the model with the described parameters is compared to the observed value of NGC~6153, finding that the \hb\ flux in the model is about twice the observed one. To lower the \hb\ flux to about half in the same physical size, we add a filling factor (ff) and diminish the hydrogen density (nH). The adopted ff is 0.6 and the nH explored are: 1000, 1500, 2000, and 2500~cm$^{-3}$). We find the best fit of \hb\ with an hydrogen density of 2500~cm$^{-3}$ with filling factor of 0.6. In this model the ionization stage (measured with line rations such as [\ion{S}{iii}]/[\ion{S}{ii}] y [\oiii]/[\oii]) is too high, to fix this we lower the luminosity (to 1.07$\times 10^{37}$~erg/s) by a factor equal to the square of the ratio of the distances 1.36 \citep{2018Gaia_vizier1345} and 1.5~kpc \citep{2003Pequignot_209}. Since the ionization stage was still slightly higher, we refine the fit by decreasing the luminosity to 9$\times 10^{36}$~erg/s. With this parameters the total \hb\ flux and ionization stage are simultaneously fitted. We include graphite and silicate dust grains, each following a -3.5 slope size distribution with ten sizes from 0.005$\mu$m to 0.25$\mu$m; the total dust to gas ratio is 4.2$\times$10$^{-3}$ by mass (about 2/3 of the canonical dust for ISM in Cloudy), see Sec.~\ref{sec:compare_mod_object} for detailed justification.

\begin{table}
    \centering
    \caption{Chemical abundances of the \n\ region in units of 12+log(X/H).}
    \begin{tabular}{c c c}
    \hline \hline
    \input{abunds.tex}
    \end{tabular}
    \label{tab:N_abunds}
\end{table}

\begin{figure*}
\centering
\includegraphics[scale = 0.65]{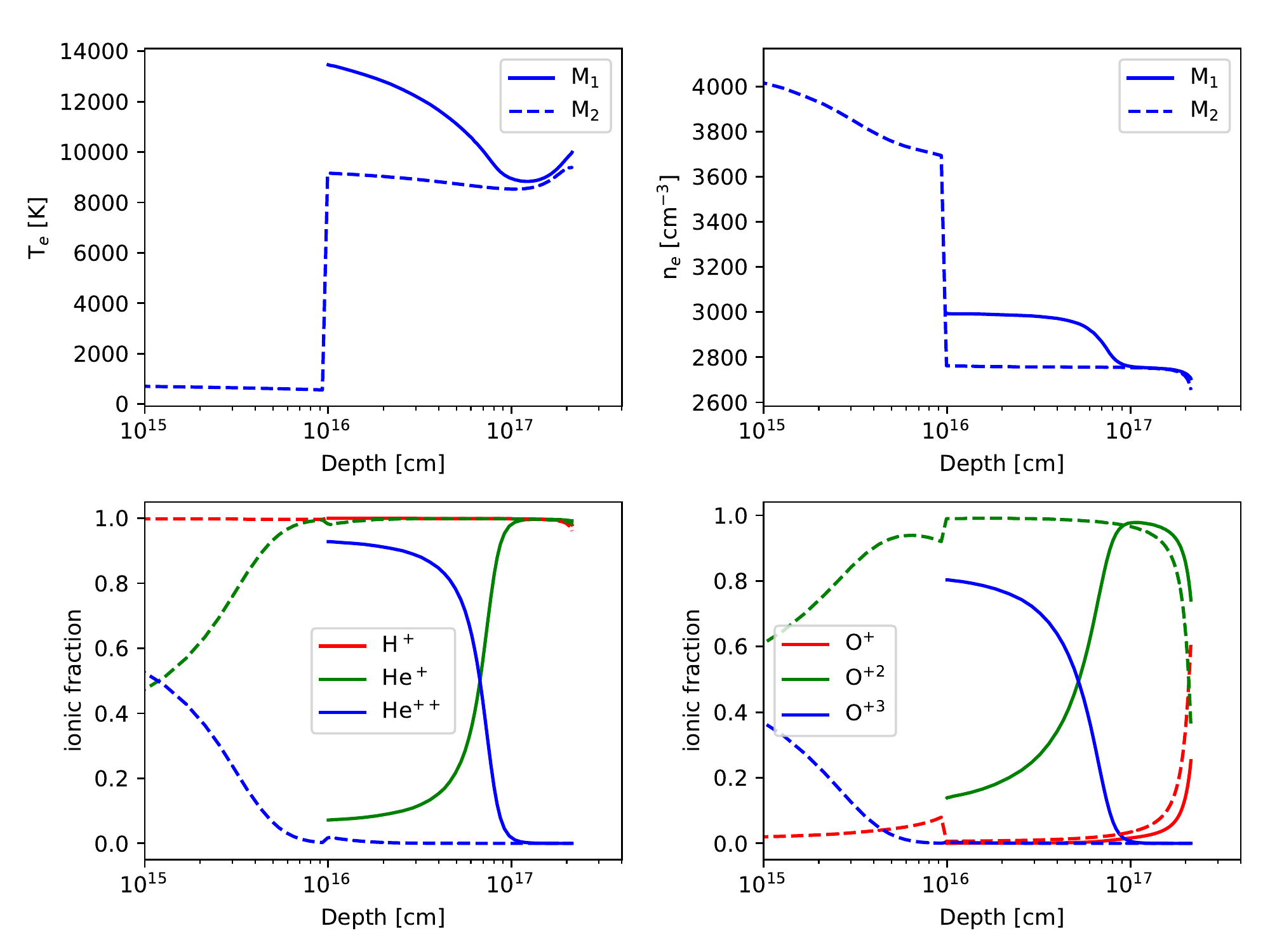}
\caption{Solid line is for model M$_1$ that represents the \n\ component (sec. \ref{sec:normal-model}) and dashed line is for model M$_2$ that represents: the \mr\ component with log ACF = 2.1 (sec. \ref{sec:rich-model}) and the \bc\ component (sec. \ref{sec:behind-clumps-region}). The parameter depth is the radius of the model minus the inner radius of the \mr\ component. Top left panel: electron temperature as a function of depth. Top right panel: electron density as a function of depth. Bottom left panel: ionic fraction of H$^+$, He$^+$ and He$^{++}$ in red, green and blue, respectively. Bottom right panel: ionic fraction of O$^+$, O$^{+2}$ and O$^{+3}$ in red, green and blue, respectively. }
\label{fig:normal_model}
\end{figure*}

The gas modeled here will be referred to as the \n\ component, it has the main contribution in terms of mass in the bi-abundance model (see section \ref{sec:bi-model}). The electron temperature, density and the ionic fraction of H$^+$, He$^+$, He$^{++}$, O$^+$, O$^{+2}$ and O$^{+3}$ of the different components are shown as a function of \DraftThree{depth} in figure \ref{fig:normal_model}. In particular, solid lines are drawing the behaviour of the \n\ component generated by the model M$_1$. At \DraftThree{depth $\sim$ 6$\times$10$^{16}$~cm} we can see the small drop in electronic density (upper right panel) corresponding to the recombination of He$^{++}$.

\subsection{Metal-rich (MR) component}
\label{sec:rich-model}
The metal-rich (\mr) component can be seen as a collection of clumps in the inner parts of the nebula. Given the large amount of free parameters, we decided to fix the inner and outer radius of the clumps at 4.5\arcsec\ and 5\arcsec\ (or 9.2$\times$10$^{16}$~cm and 1$\times$10$^{17}$~cm at a distance of 1.36~kpc), respectively. We only vary the fraction of volume that is occupied by the clumps (but see the Sec.~\ref{sec:discussion} where we explore different values for the density and the inner radius of this \mr\ region). The effective temperature, the hydrogen density and the helium abundance are the same than in the \n\ component. \citet{2003IAUS..209..389T} did a 2-component model of NGC~6153 and found that a better fit is obtained when both components have similar densities. The \mr\ and \n\ regions are not supposed to be in contact and then not pressure equilibrium hypothesis is used.
The mean ionization parameter\footnote{$U(r) = Q_0 / 4\pi r^2 N_e c $, where $Q_0$ is the ionizing photon rate of the central star per second, $r$ is the distance to the central star, $N_e$ is the electron density and $c$ the light velocity. \DraftTwo{The mean value is obtained over the volume of the nebula: $<U> = \int U dV / V.$}} in this region is \DraftTwo{log~<U>} $\sim$ -1.6. The abundances we adopt for the metals are detailed bellow.  

For a given element X, the abundance in the \mr\ region is defined by the following equation:
\begin{equation}
    X/H_{MR} = X/H_{N} \cdot ACF(X)
\label{eq:ACF}
\end{equation}
where ACF(X) is the abundance contrast factor between the \n\ and the \mr\ components. In the following we will mainly use the 10-logarithmic value of the ACF (expressed in "dex"). The enhancement of the abundance in the \mr\ clumps is certainly not uniform (it may differ from one element to another one), but it is out of the scope of this paper to take this effect into account. Thus for a given ACF, the same factor is applied to all the metals. \DraftThree{The dust content included in the close-to-solar components (\n, \bc\ and \s) is less than the canonical D/G for the ISM. The IR emission from this regions reproduces the IRAS observations (see sec. \ref{sec:em_spec}), so we don't include dust in the \mr\ region, assuming all the dust is in the close-to-solar gas. The opposite extreme assumption that all the dust is in the \mr\ component is discussed in sec. \ref{sec:dust_mr}.}

In the figure \ref{fig:normal_model}, the \mr\ component is drawn as dashed lines and corresponds to \Draft{depth $\leq$ 9$\times$10$^{15}$~cm}. Its temperature (upper-left panel) is close to 650~K in this case (ACF = 2.1 dex). The electron density (upper-right panel) is higher than in the \n\ component, due to the contribution of the metallic ions to the free electrons soup.  

\begin{figure*}
    \centering
    \includegraphics[scale = 0.68]{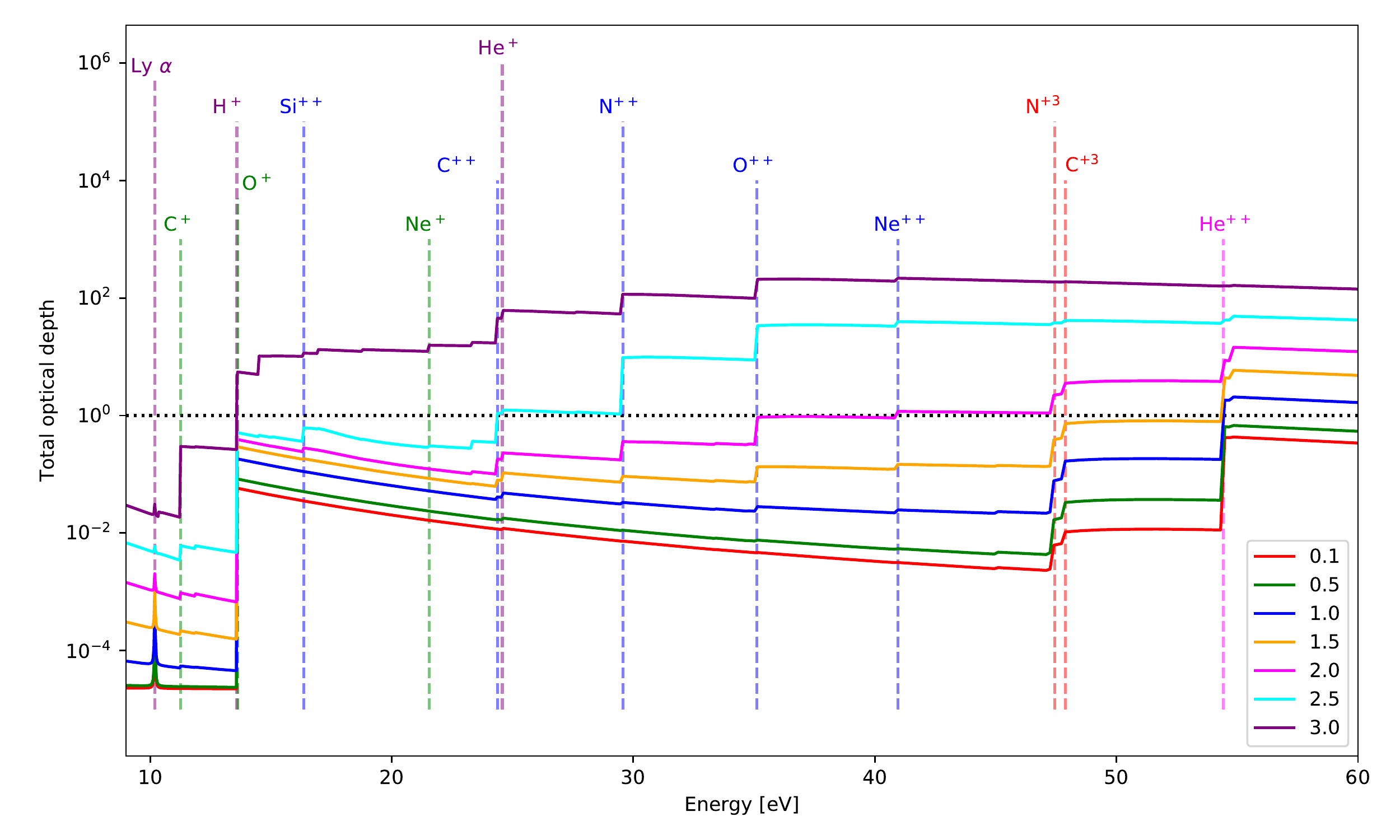}
    \caption{Total optical depth as a function of energy. This value is measured at the outer radius of the \mr\ component for ACF(X): 0.1, 0.5, 1.0, 1.5, 2.0, 2.5 and 3.0~dex. The horizontal black dotted line indicates the optical depth equal to 1. The vertical dashed lines denote ionization potentials, for H and He: purple (magenta) represents once (twice) ionized ions, for metals: the colors green, blue and red are for once, twice and tree times ionized ions respectively.}
    \label{fig:opd_IP}
\end{figure*}

\Draft{The ionization structure of a given element is determined by the balance between recombination and ionization, with the more ionized ions being closer to the ionization source. In the lower-right panel of figure \ref{fig:normal_model} we can see that at the inner part of model M$_2$, where the \mr\ component is present, the dominant ion is O$^{+2}$, while for the model M$_1$ of the \n\ component, the dominant ion at the inner part is O$^{+3}$. Although the ionization source is the same for both models, the O/H abundance is not the same, being 2.1 dex larger for the \mr\ component (in this example) making the amount of photons with energies larger than 54.9~eV not enough for the O$^{+3}$ to be the dominant ion in the \mr\ component. We can see that just before the region of the M$_2$ model where the abundance goes back to the same value than in the \n\ component (at depth $9\times 10^{15}$cm), the O$^+$ abundance begins to increase while the O$^{+2}$ and O$^{+3}$ are decreasing, because of the large absorption of photons with energies larger than 35~eV. The decrease in the metal abundances at this transition zone has an effect on the structure of the oxygen ionization, with an abrupt increase of the O$^{+2}$ ions, and a decrease of the O$^+$ ions.}

In order to confirm that the decrease of O$^{++}$ is due to the lack of 35~eV we show in fig.~\ref{fig:opd_IP} the total optical depth at the outer radius of the \mr\ component as a function of energy. We clearly see the effect of the metals dominating the changes in the optical depth at high metallicities. We can see that for ACF $\gtrsim$ 2.0 dex the optical depth starts to be larger than 1 at 35~eV.

\begin{figure}
    \includegraphics[scale = 0.43]{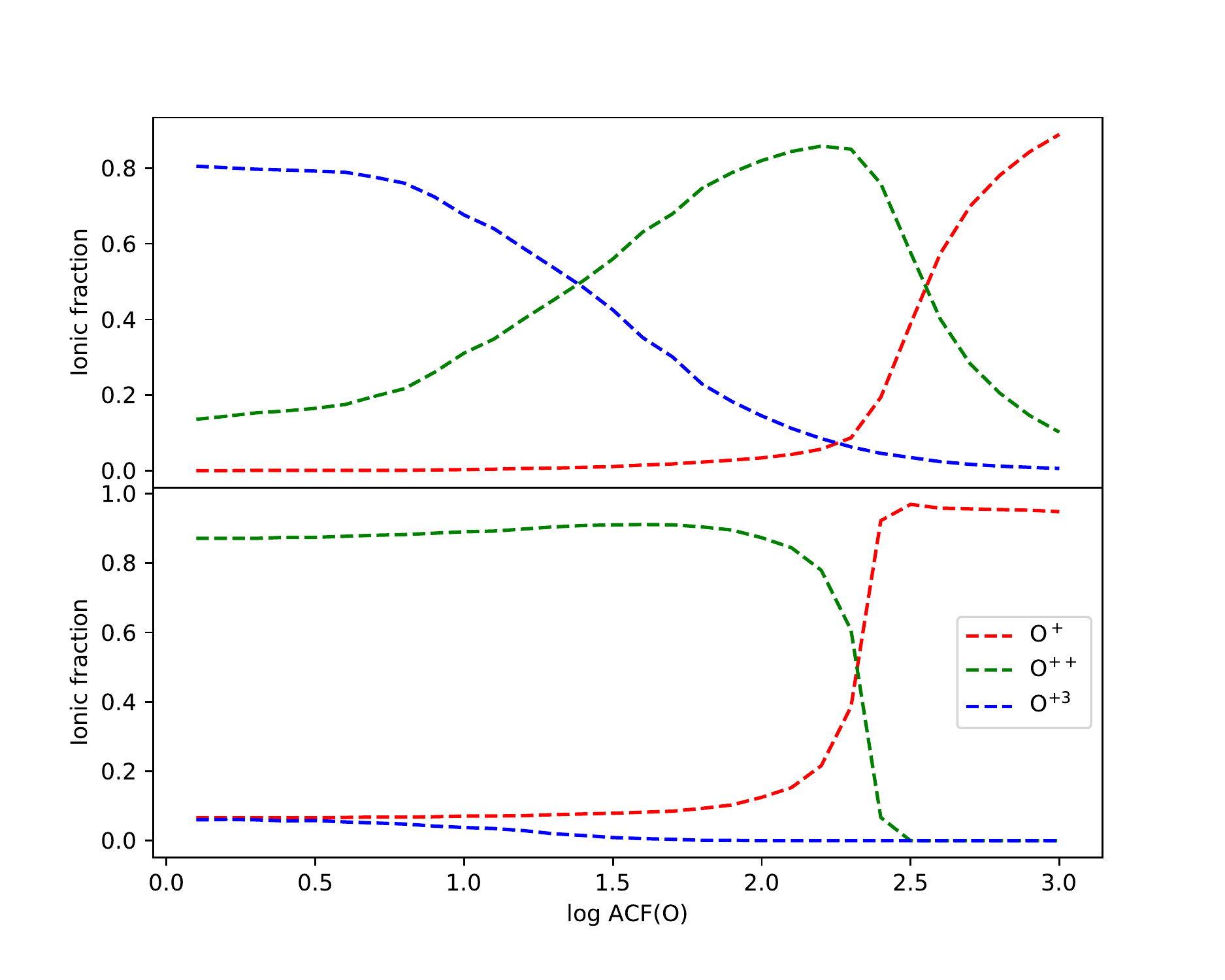}
    \caption{Distribution of the ionic fraction integrated over the volume of O$^+$, O$^{++}$, and O$^{+3}$ in red, green and blue dashed lines, respectively, as a function of the log~ACF(O) for the \mr\ component in the top panel, and the \bc\ component in the bottom panel.}
    \label{fig:acf_ion_frac}
\end{figure}

To illustrate this behaviour in more details, we show the ionic fractions (integrated over the volume) of oxygen as a function of the ACF(O) in the top panel of fig. \ref{fig:acf_ion_frac}. When the ACF increases, we observe a decrease of the mean ionization of the \mr\ region. We fixed the size of this region, which is then matter-bounded (we consider here only the \mr\ region. There is actually close to solar metallicity gas behind this region, it will be discussed in sec.~\ref{sec:behind-clumps-region}). Increasing the metallicity increases the opacity of the gas, reducing its Str\"omgren size. In the case of the smallest metallicity, the relatively small \mr\ region only includes the inner part of what would be the \mr\ region if not matter-bounded, corresponding to the highest ionization. On the other side, when the metallicity is very high, the \mr\ region is almost radiation-bounded and shows a global lower ionization.

\subsection{Behind clumps (\bc) component}
\label{sec:behind-clumps-region}
Behind the \mr\ component we found gas with the same chemical abundances \Draft{and dust} than the \n\ component but with a different ionization stage, we will call this gas "behind clump" (\bc) gas. Cloudy is able to model the \bc\ region in the same run than the \mr\ region. So for each \mr\ gas a \bc\ gas will be modeled, in the same run. \DraftThree{We use the \textit{function} command of Cloudy, to include dust only in the \bc\ component of the model.}
The model including the \mr\ and \bc\ components will be referred to as the M$_2$ model. The \bc\ component has the same hydrogen density, chemical abundances and inner radius than the normal component but a different outer radius and ionization parameter since the \mr\ gas that is at a smaller radius than the \bc\ gas is absorbing some of the ionizing radiation. 

\Draft{In the bottom panel of fig. \ref{fig:acf_ion_frac} we show the \bc\ component ionic fractions of oxygen integrated over the volume as a function of the ACF(O). We can see that at log~ACF $\sim$ 2.3 the dominant ion changes from O$^{++}$ to O$^+$.} 
For values of ACF smaller than 2.3~dex, there is still some photons more energetic than 35~eV entering the \bc\ region and ionizing all the O$^+$. Above 2.3~dex, no more photons ionizing O$^+$ escape the \mr\ region, and the \bc\ region is purely O$^+$. \Draft{This will have an effect in the contribution of the O$^{++}$ emission coming from the \bc\ region for ACF > 2.3~dex.} 

\begin{figure}
    \includegraphics[scale = 0.5]{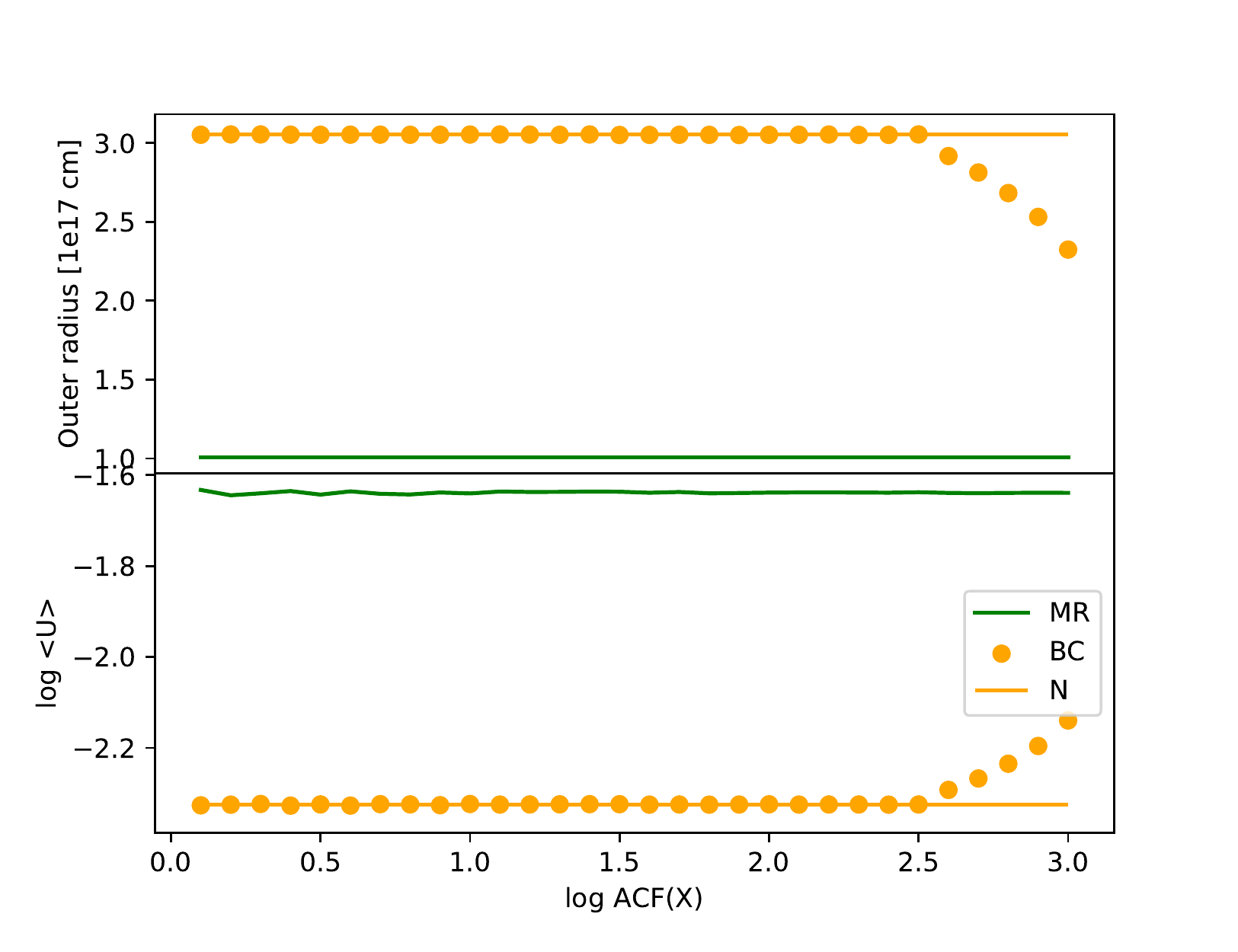}
    \caption{The top panel represents the outer radius as a function of the log~ACF(X) for: the \mr, \bc\ and \n\ components as a solid green line, orange dots and solid orange line, respectively. The bottom panel represents the log of the mean ionization parameter as a function of log~ACF(X) for: the \mr, \bc\ and \n\ components as solid green line, orange dots and solid orange line, respectively.}
    \label{fig:bc_rout_logU}
\end{figure}

The outer radius and the mean ionization parameter of the \bc\ gas are shown as a function of the ACF(X) in figure \ref{fig:bc_rout_logU}, for comparison the outer radius and the mean ionization parameter of the \n\ component are also shown. We notice that when the ACF(X) reaches values of 2.6~dex, the \bc\ gas starts to recombine at a smaller radius than the outer radius of the \n\ component. The size of the \bc\ regions starts to decrease, leading to an increase of the \Draft{mean} ionization parameter \Draft{<U>} (lower panel of fig.~\ref{fig:bc_rout_logU}). The space between the recombination front of the \bc\ region and the outer radius of the nebula is then filled by the shadow region (see Sec.~\ref{sec:shadow-model}).

\subsection{Shadow (\s) component}
\label{sec:shadow-model}
\begin{figure}
    \includegraphics[scale = 0.28]{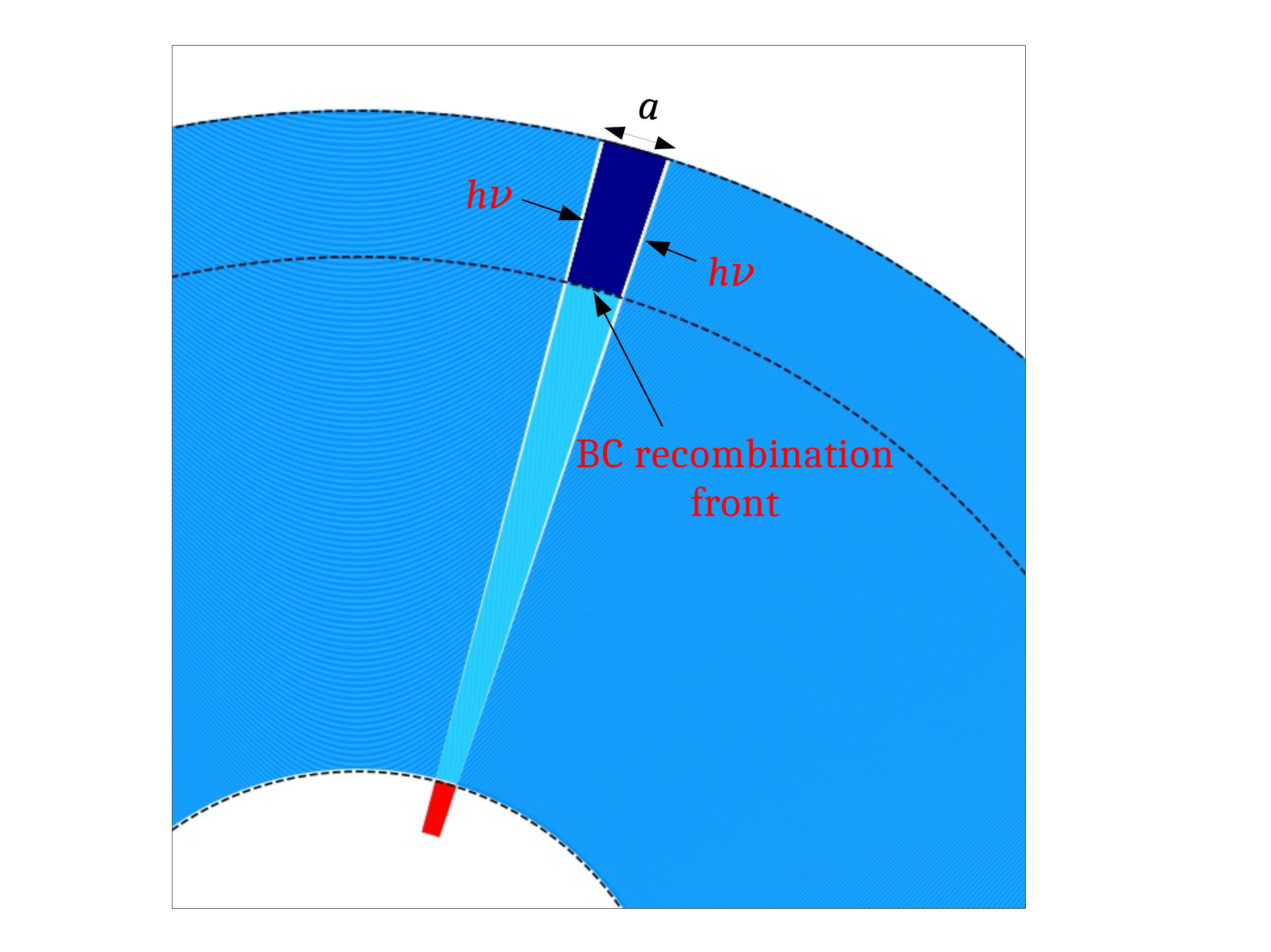}
    \caption{Colors represent the same regions as in figure \ref{fig:squeme}. Scheme characterizing the parameters of the \s\ region. The outer part of the \s\ region has an arch size of \textit{a}, the $h\nu$ arrows represent the direction of the diffuse ionizing photons coming from the \n\ region to ionize the \s\ region.}
    \label{fig:shadow_scheme}
\end{figure}

Since the models M$_1$ and M$_2$ are combined to form a spherical morphology, when the model M$_2$ recombines at a smaller radius than M$_1$, there is a shadow after M$_2$. The ionization of the shadow is due to the diffuse radiation (mainly Lyman recombination photons) coming from the \n\ gas around it. We build a model (M$_3$) to get an approximate idea of the physical conditions and the emission of the shadow (\s ) region. We use one single model for all the shadow regions obtained with different values of ACF(X), \Draft{this model has the same abundances and dust content than the \n\ component}. We use as representative ionizing flux the nebular continuum computed by Cloudy during the M$_1$ model, at the radius corresponding to half of the nebula. The figure \ref{fig:shadow_scheme} gives a schematic representation of the shadow region in dark blue. The shadow region is what follows the recombination front of the \bc\ region. We are actually computing a model where the radial direction of the Cloudy 1D model corresponds to the tangential direction of our object (illustrated by the "$h\nu$" arrows in Fig.~\ref{fig:shadow_scheme}).
The radial size of the \s\ regions are determined by the difference in radius between the recombination front of the \bc\ region and the outer radius of the \n\ region (see fig. \ref{fig:bc_rout_logU}). On the other side, the tangential size ($a$ in Fig.~\ref{fig:shadow_scheme}) of the clumps (the depth of the Cloudy model) needs to be defined. 

As we are only interested in the representative emission produced by the \s\ ionized region, we will adopt a value for this size $a$ that avoid a neutral tube inside the \s\ region. To obtain this result, we cut the Cloudy model at the half of its Str{\"o}mgren size. The electron temperature of this region is almost constant and close to 6300~K. The electron temperature of the surrounding material (\n\ region) is 30\% higher than this value; this could imply a higher density for the \s\ region in case of pressure equilibrium. We did not took this effect into account, keeping the hydrogen density at the same value in all the regions. 

\subsection{Bi-abundance models}
\label{sec:bi-model}

We combine the three models M$_1$, M$_2$ and M$_3$ described in the previous sections to obtain the emission of the whole nebula. The parameter we use to define the fraction of volume each model occupies is the solid angle ($\Omega$) of the \mr\ region (and the \bc\ and \s\ regions).

We actually do not perform a 3D model, not even a pseudo-3D models {\it a la} \citet{2016Gesicki_aap585}. We rather add the intensities computed by each models using a weight determined by $\Omega$. 

In this combined model the intensity of a line is obtained with the following equation:
\begin{equation}
I(\lambda) =  I(\lambda)_{M_1} \left(1 - \dfrac{\Omega}{4\pi} \right) + I(\lambda)_{M_2} \left(\dfrac{\Omega}{4\pi}\right) + I(\lambda)_{M_3}  \left(\dfrac{\Omega}{4\pi}\right)
\label{eq:I_lambda}
\end{equation}
where $I(\lambda)_{Mi}$ is the intensity at the wavelength $\lambda$ for model M$_i$ \Draft{as computed by Cloudy for a $4\pi$ sphere.} \DraftTwo{In the case of $M_3$, the intensity $I(\lambda)_{M_3}$ already takes into account the thickness of the shadow which depends on the position of the BC recombination front.}

It is almost topologically equivalent to use a single model with ${\Omega}/{4\pi}$ or a collection of \textit{n} clumps, each one with an angular size of ${\Omega}/({4\pi n})$. This is not totally the case for the \s\ region: a large clump may lead to a neutral tube in its shadow, while a smaller clump will have a fully ionized shadow. We choose to model the shadows without neutral component, imposing an implicit upper limit of the clump sizes.  

In the following sections we will explore the effect of the metallicity enhancement (ACF) and the contribution of the \mr\ (and \bc\ and \s ) component. This is achieved by considering a grid of models where ACF ranges from 0.1 to 3.0 dex by 0.1 dex steps, and ${\Omega}/{4\pi}$ ranges from 0.01 to 0.5. 

\subsection{NGC~6153: a reference PN with high ADF}
\label{sec:obs}
In this section we make a brief summary of the observations taken from the literature that are used to compare with our models. 

NGC~6153 is an intrinsically bright PN located in the southern sky at a distance of 1.36~kpc \citep{2018Gaia_vizier1345}, it has an angular size of 32" x 30" \citep[NIR]{2006Skrutskie_aj131}. It was observed with the ESO 1.52 m telescope by \citet{2000Liu_mnra312}. They used the B\&C spectrograph, the slit was 2 arcsec wide and 3.5 arcmin long. The spatial sampling was 1.63 arcsec per pixel. A "minor-axis" of the PN was observed by placing the slit in a PA of 122.8$^\circ$ centered in the central star. The mean spectra of the whole nebula was obtained scanning across the nebula with a long-slit oriented nort-south. A high resolution spectra (FWHM = 1.5\AA) was obtained for the spectral range $\lambda\lambda$3995-4978\AA, and a low resolution spectra (FWHM = 4.5\AA) in the spectral range $\lambda\lambda$3535-7400\AA. The observations were corrected for bias, flats, cosmic rays, they were wavelength calibrated with an He-Ar calibration lamp. The flux was calibrated using standard stars, and dereddened using the galactic extinction curve of \cite{1983Howarth_mnra203} with c(H$\beta$)=1.30 and R~=~3.1. \citet{2000Liu_mnra312} determined the chemical abundances for NGC~6153 and an ADF(O$^{++}$) of 9.2.

\section{Plasma diagnostics with the model outputs}
\label{sec:results}
The emission of the bi-abundance models (see \ref{sec:bi-model}) \Draft{are used as a simulation of a real PN} (with equation \ref{eq:I_lambda}), to determine the physical parameters and chemical abundances with the same methods than for a real object. The results are shown in the following subsections. The electronic temperature and density, and the ionic abundances are determined using the code PyNeb version 1.1.9 \citep{2013Luridiana_ascl}. The atomic data used for each ion are listed in table \ref{tab:pn_atomic_data}, if the electronic temperature is under 500~K (the case for the \mr\ component at very high ACF) an extrapolation is made to the \ion{H}{i} recombination coefficients from \citet{1995Storey_mnra272}.

\begin{table*}
\centering
\caption{Atomic data \Draft{selected from PyNeb that are} used in the determination of physical parameters and chemical abundances.}
\begin{threeparttable}
\begin{tabular}{l l l}
\hline \hline
~ & \multicolumn{2}{c}{Collisionally excited lines} \T\\
Ion & Transition probabilities &  Collision strength  \B \\
\hline \hline
N$^+$  & \citet{2004Froese-Fischer_Atom87}  & \citet{2011Tayal_apjs195}  \T \\

O$^+$ & \citet{2004Froese-Fischer_Atom87} & \citet{2009Kisielius_mnra397}  \\

S$^+$ & \citet{2009Podobedova_Jour38} & \citet{2010Tayal_apjs188} \\

O$^{++}$ & \citet{2004Froese-Fischer_Atom87}$^{a}$ &  \citet{2014Storey_mnras441} \\

Cl$^{++}$ &  \citet{1986Kaufman_jpcrd15}$^{b}$ & \citet{1989Butler_aap208}\\

Ar$^{3+}$ & \citet{1982Mendoza_mnra198} & \citet{1997Ramsbottom_ADNDT66} \\
~ & \multicolumn{2}{c}{Recombination lines} \T\B\\
Ion & Recombination coefficients &  Case  \\
H$^+$ & \citet{1995Storey_mnra272}$^{c}$ & B \\
He$^+$ & \citet{1996Smits_mnra278} & B\\
He$^{++}$ & \citet{1995Storey_mnra272} & B \\
O$^{++}$ & \citet{2017Storey_mnra470} & B \\
\hline
\end{tabular}
\begin{tablenotes}
    \item $^{a}$ For transitions 4-2 and 4-3 we use \citet{2000Storey_mnra312}.
    \item $^{b}$ For transition 4-3 we use \citet{1983Mendoza_103}.
    \item $^c$ We make an extrapolation at low Te (< 500~K).
  \end{tablenotes}
\end{threeparttable}
\label{tab:pn_atomic_data}
\end{table*}
       
\subsection{Physical parameters with CELs}
\label{sec:te-cel}
\begin{figure}
\begin{center}
\includegraphics[scale = 0.55]{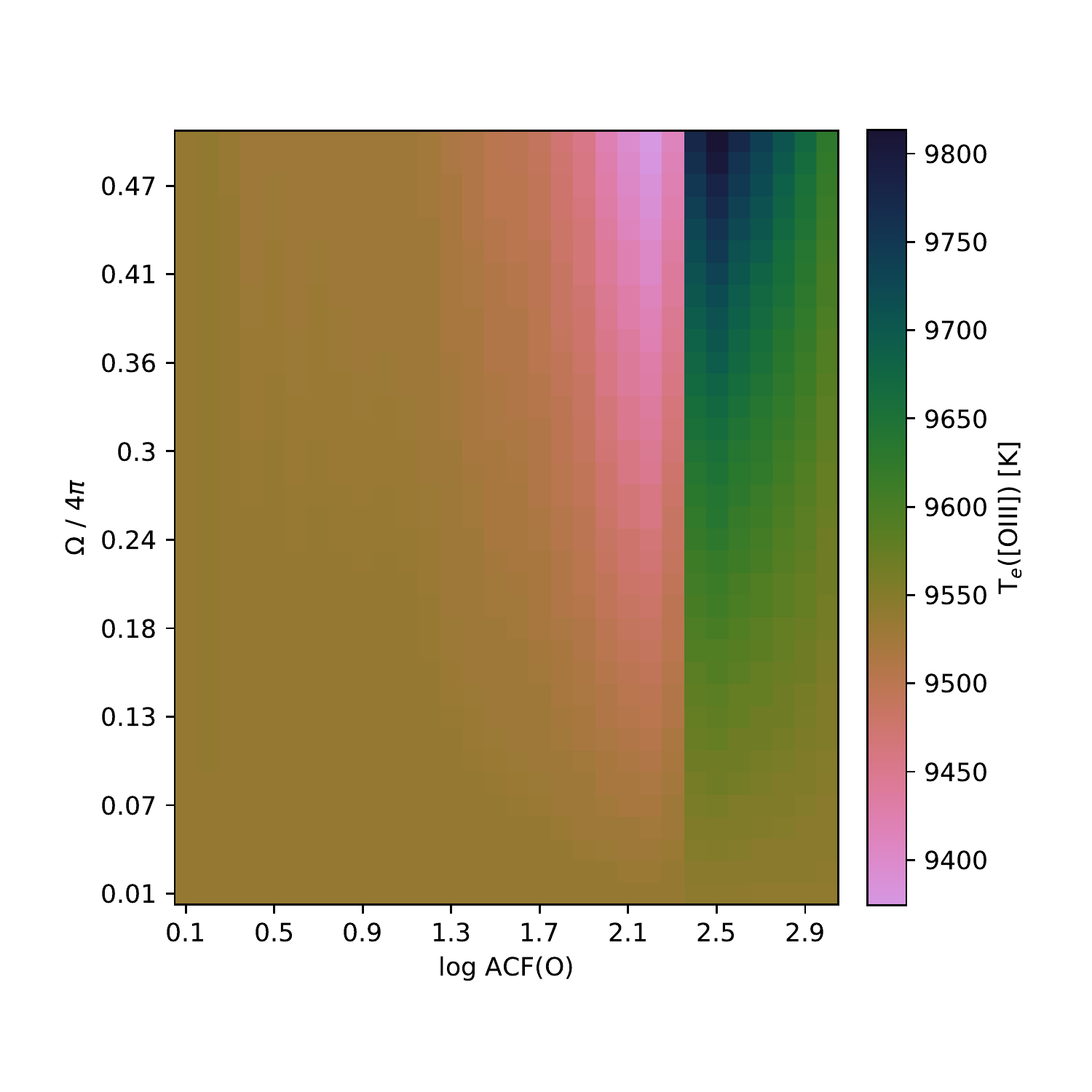}
\includegraphics[scale = 0.55]{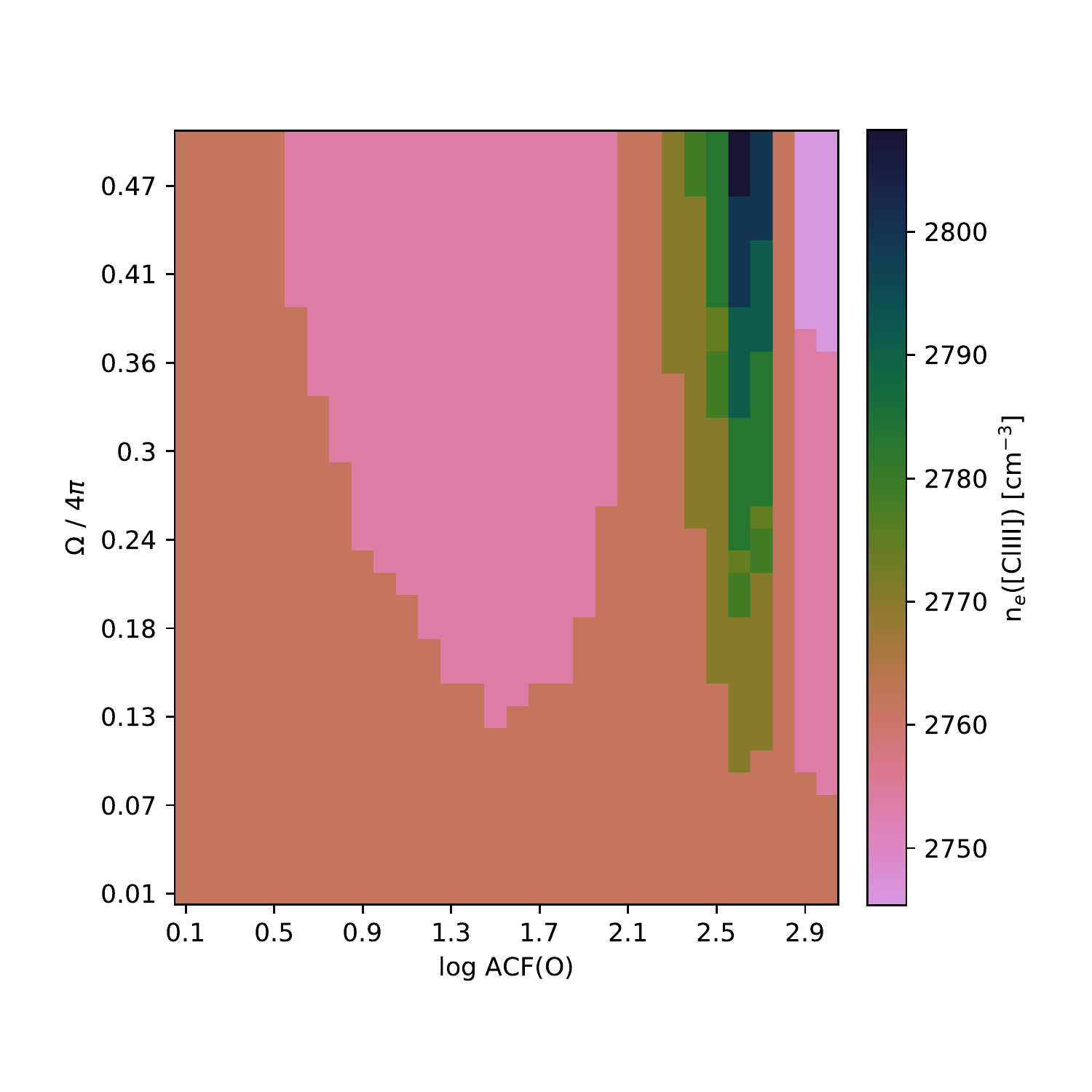}
\caption{Top (bottom) panel: the color represents the electronic temperature (density) estimation for the bi-abundance models (see sec. \ref{sec:bi-model}). The x-axis represents the variations in the ACF for oxygen (see eq. \ref{eq:ACF}) and goes from 0.1 to 3.0~dex. The y-axis represents the normalized solid angle of: the metal-rich clumps, the gas behind the clumps and the shadow, in the range of 0.01 to 0.50. Diagnostics are made with the sensitive line ratios: [\ion{O}{iii}]~4363/5007~\AA\ and [\ion{Cl}{III}]~5538/5518~\AA. The contribution of recombination in [\ion{O}{iii}]~4363~\AA\ is taken into account. Line intensities of the bi-abundance models are obtained using eq. \ref{eq:I_lambda}.}
\label{fig:high_te_ne_pc}
\end{center}
\end{figure}

 The line emissions are obtained from the bi-abundance models using eq. \ref{eq:I_lambda}, for the grid of ACF and $\Omega$/4$\pi$ explored. As we are mainly interested by the medium ionization zone (20~eV $<$ IP $<$ 45~eV), we determine the temperature and the density simultaneously using the line ratios: [\ion{O}{iii}]~4363/5007~\AA\ and [\ion{Cl}{III}]~5538/5518~\AA. The contribution of recombination to the [\ion{O}{iii}]~4363 line is naturally included in the models, no correction from this effect has been performed (but see the Sec.~\ref{sec:auroral} for a discussion of the [\ion{O}{iii}]~4363 emission). The results are shown in figure \ref{fig:high_te_ne_pc}, were the x-axis represents the changes in the ACF, the y-axis represents the changes in $\Omega$/4$\pi$ and the color represents the electronic temperature (density) in the top (bottom) panels respectively. Both parameters are found almost constant in the ACF-$\Omega$ plane, with variations of less than 5\% and 10\% for electronic density and temperature, respectively. This is due to the fact that they are determined using CEL which are mainly emitted by the close to solar components (\n, \bc\ and \s\ regions). Small variations appear in the high~ACF-high~$\Omega$ corner of the plots. In the case of the temperature, it may seems counter-intuitive that the temperature increases in the case of a higher contribution of the cold \mr\ region. What we see here is the effect of the contribution of the recombination to the [\ion{O}{iii}]~4363 emission, coming from the \mr\ region, which increases at low temperatures. The strong change in the temperature at log ACF $\sim$ 2.2 is due to the sudden vanishing of O$^{++}$ emission in the \bc\ region (see sec.~\ref{sec:behind-clumps-region}). The decrease of the temperature at constant $\Omega$ and high values of ACF is due to the increase of the contribution of the \s\ region to the emission of [\ion{O}{iii}]~5007. 

\subsection{Electronic temperature from the Balmer Jump}
\label{sec:te-bj}
\begin{figure}
\includegraphics[scale = 0.55]{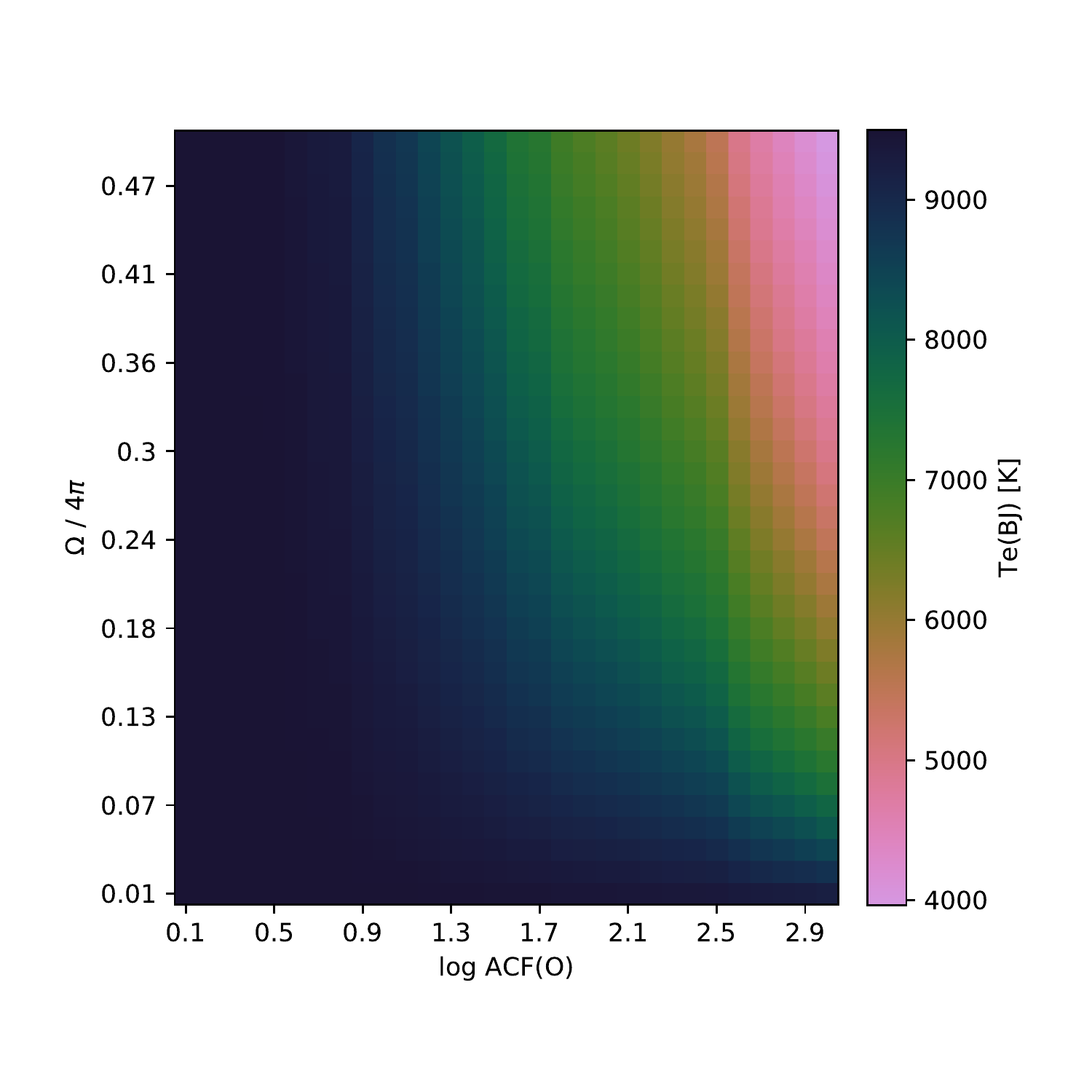}
\caption{Same axis as figure \ref{fig:high_te_ne_pc}. The color represents the Balmer jump temperature determination for the bi-abundance models.}
\label{fig:te_bj}
\end{figure}

The electronic temperature can also be estimated from the Hydrogen Balmer jump normalized to a Balmer line \citep{1967Peimbert_apj150}. For the ACF-$\Omega$ grid, we estimate the Balmer jump temperature, T(BJ), using the Continuum class from PyNeb. The method requires: the flux of the continuum (in units erg$\cdot$s$^{-1}$cm$^{-3}$\AA$^{-1}$) at a wavelength previous and latter to the Balmer jump (such wavelengths can be defined by the user, we choose 3643 and 3861~\AA), a line intensity of H (in units erg$\cdot$s$^{-1}$cm$^{-3}$, we choose H11), the electronic density and the He$^+$/H$^+$ and He$^{++}$/H$^+$ abundances ratios. As an observer would do, we use for the electron density the estimation obtained from a density diagnostic, here the [\ion{Cl}{iii}]~5538/5518\AA\ line ratio, as derived in the previous section (bottom panel of fig. \ref{fig:high_te_ne_pc}). In the range of density we are expecting for a PN (lower than 10$^5$ cm$^{-3}$), the exact adopted value does not strongly affect the derived temperature.
The He$^+$/H$^+$ and He$^{++}$/H$^+$ abundances ratios are estimated consistently by iterative process with the T(BJ) using \Draft{the following lines normalized to \hb}: \ion{He}{i}~4471, 5876 and 6678 for He$^+$/H$^+$, and \ion{He}{ii}~4686 for He$^{++}$/H$^+$. The resulting T(BJ) is shown in figure \ref{fig:te_bj}. We can see that as metallicities (or ACF) and $\Omega$ increase, the temperature starts to decrease, mainly because of the contribution of the \mr\ component that are cooler at higher metallicity and also because of the contribution of \bc\ and \s\ components that are at a lower temperature than the \n\ component.

\subsection{Estimation of O$^{++}$/H$^+$ with CELs}

\begin{figure}
	\includegraphics[scale = 0.55]{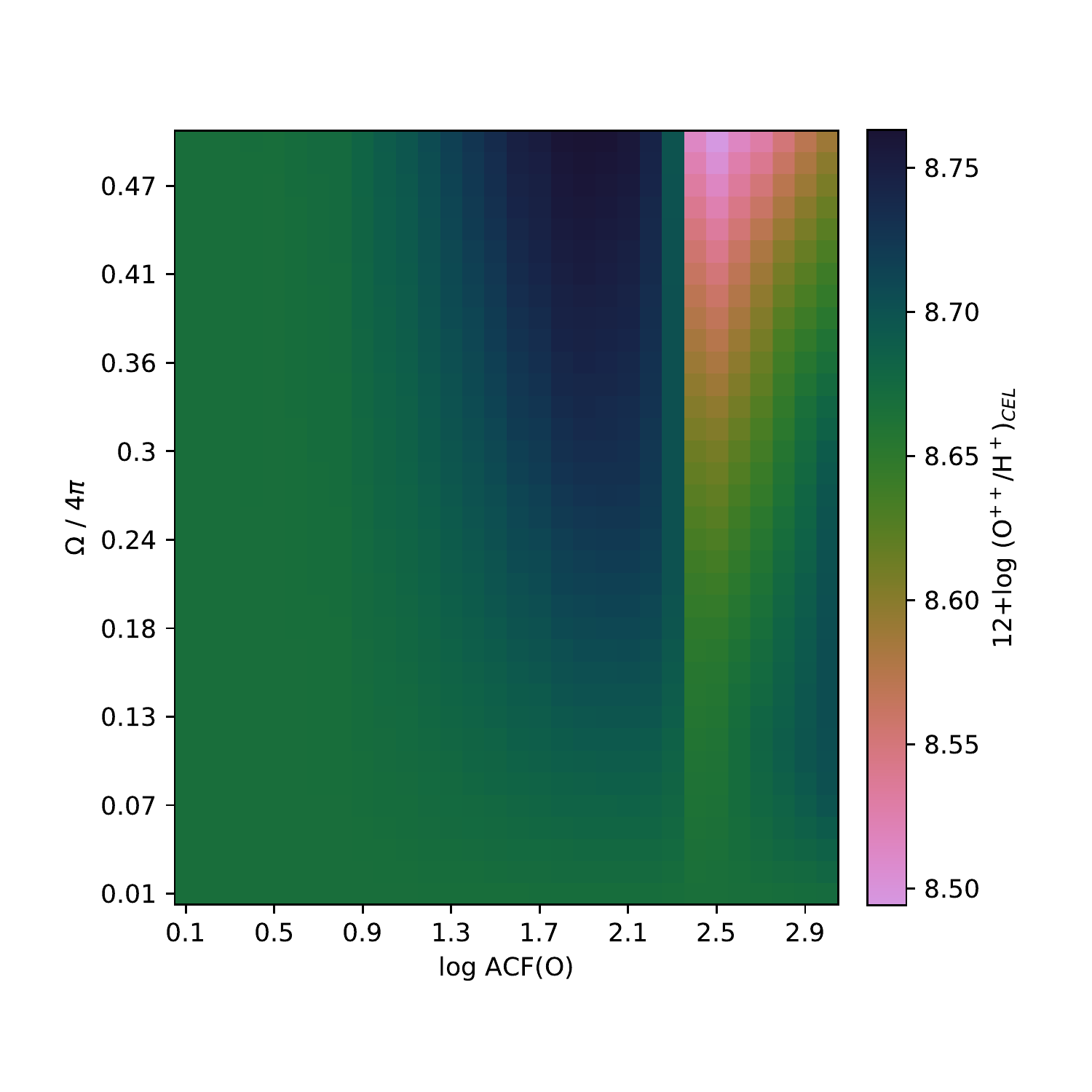}
	\caption{Same axis as figure \ref{fig:high_te_ne_pc}. The color represents the O$^{++}$/H$^{+}$ ionic abundance estimation with CELs for the bi-abundance models, using the [\ion{O}{iii}] temperature for O$^{++}$ (shown in top panel of fig. \ref{fig:high_te_ne_pc}), and the Balmer jump temperature (shown in fig. \ref{fig:te_bj}) for H$^{+}$.}
	\label{fig:opp_cel}
	\end{figure}

We estimate the ionic abundance of O$^{++}$/H$^{+}$ for the ACF-$\Omega$ map with the CEL [\ion{O}{iii}]~5007\AA\ in two ways: taking the T([\ion{O}{iii}]) for both O$^{++}$ and H$^{+}$, and taking T([\ion{O}{iii}]) for O$^{++}$ and T(BJ) for H$^{+}$. In figure \ref{fig:opp_cel} we show the determination with T(BJ) for H$^{+}$. The other determination that only uses T([\ion{O}{iii}]) differs by a factor roughly equal to T(BJ)/T([\ion{O}{iii}]) to what is shown in fig. \ref{fig:opp_cel}. From fig. \ref{fig:opp_cel} we see a 0.3~dex decrease of O$^{++}$/H$^{+}$ at high $\Omega$ and ACF larger than 2.3~dex. This is due to the abrupt decrease of O$^{++}$/O from the \bc\ region (see bottom panel of fig. \ref{fig:acf_ion_frac}). In this high ACF regime (> 2.3~dex): at constant ACF when the $\Omega$ decreases the contribution of the \bc\ region also decreases and at constant $\Omega$ when the ACF increases the outer radius of the \bc\ region decreases (see top panel of fig. \ref{fig:bc_rout_logU}). In both cases the contribution in volume of the \bc\ region to the total O$^{++}$/H$^{+}$ is smaller. Since there is less (if any) O$^{++}$ in this region (at high ACF) the total O$^{++}$/H$^{+}$ increases again.

The value of 12 + log O/H used in the close to solar components (\n, \bc\, and S) is 8.75. The value obtained from the CELs is slightly lower than this value, because of the ionic fraction O$^{++}$/O being lower than one. We also computed the "real" value of O$^{++}$/H$^{+}$ by integrating over the volume of the nebula the contribution of the \n,  \bc\, and \s\ regions (no contribution from \mr\ is taken into account, as this region does not emit [\ion{O}{iii}]~5007\AA ), which is close to both empirical determinations, being slightly closer for the T(BJ) estimation. 

\subsection{Estimation of O$^{++}$/H$^+$ with ORLs}
\begin{figure}
\includegraphics[scale = 0.55]{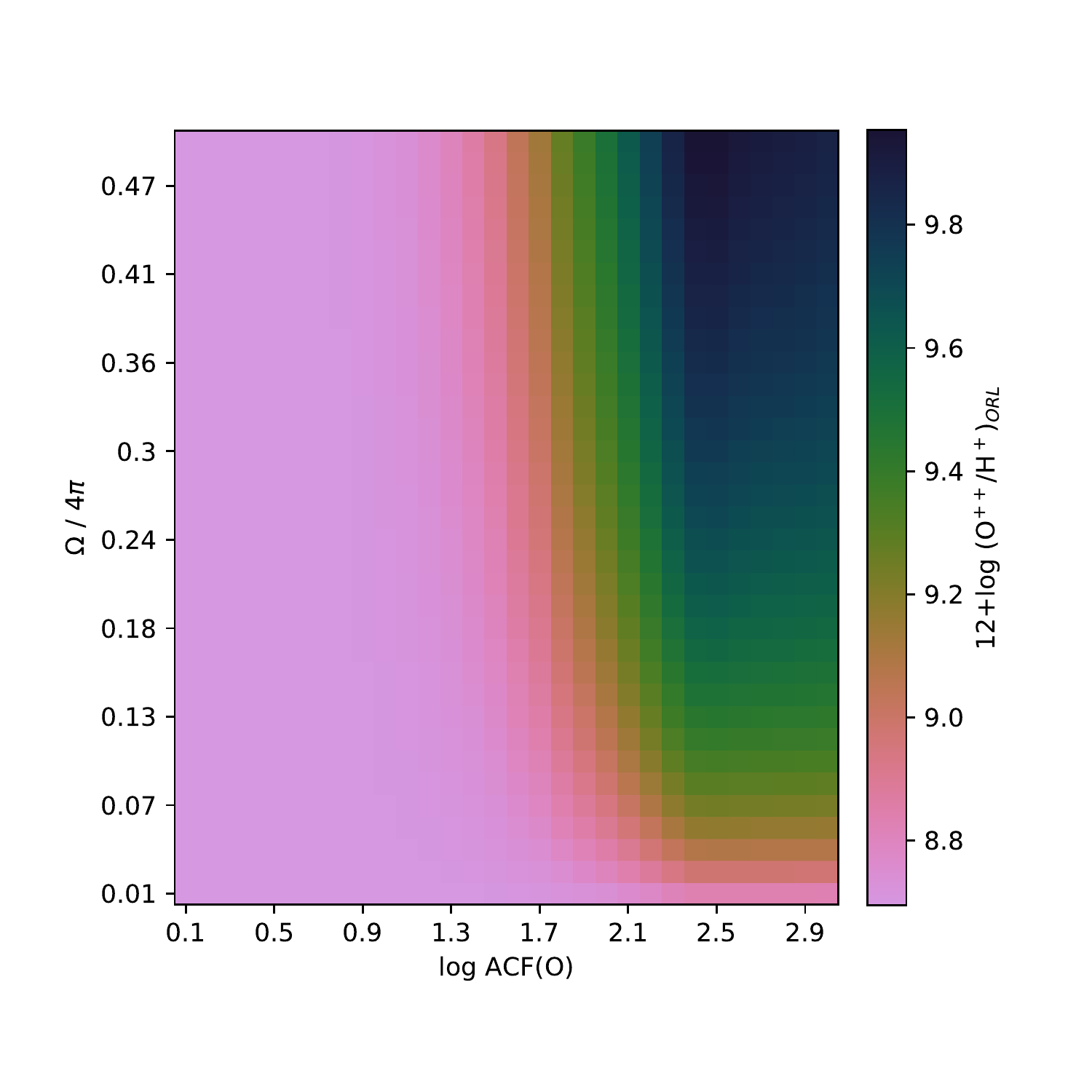}
\caption{Same axis as figure \ref{fig:high_te_ne_pc}. The color represents the O$^{++}$/H$^{+}$ ionic abundance estimation with ORLs for the bi-abundance models, using the T(BJ) (shown in fig. \ref{fig:te_bj}).}
\label{fig:opp_orl}
\end{figure}

We estimate the ionic abundances of O$^{++}$/H$^{+}$ with ORLs for the grid of ACF-$\Omega$ explored in the bi-abundance models. Given that the dependence of \ion{O}{ii} recombination lines to the electronic temperature is small, we use the Balmer jump temperature estimated in sec. \ref{sec:te-bj}. We make an average of the abundance determined with the same 3-3 recombination lines of \ion{O}{ii} than in table 14 of \citet{2000Liu_mnra312}. 
The O$^{++}$/H$^{+}$ estimation is shown in figure \ref{fig:opp_orl}. The same calculation was made with the temperature from [\ion{O}{iii}] 4363/5007~\AA, and the results were very similar. We see in fig. \ref{fig:opp_orl} changes of 1 order of magnitude from the low ACF-low $\Omega$ corner to the high ACF-high $\Omega$ opposite corner. At ACF larger than 1 dex the contribution of the \mr\ region becomes important given the low temperature of the region, favoring the recombination line emission.

The determination of O$^{++}$/H$^{+}$ from ORLs is expected to provide the actual value of the O$^{++}$/H$^{+}$ in the \mr\ region (if \hb\ is dominant in this region). We see here that this is far from being the case at high ACFs, the highest value obtained for 12 + log O$^{++}$/H$^{+}$ being 9.9 (at $\Omega$/4$\pi$ = 0.5 and ACF(O) = 2.4~dex) while for the \mr\ region it reaches 10.8.

\subsection{Estimation of the ADF(O$^{++}$)}
	
\begin{figure}
\includegraphics[scale = 0.55]{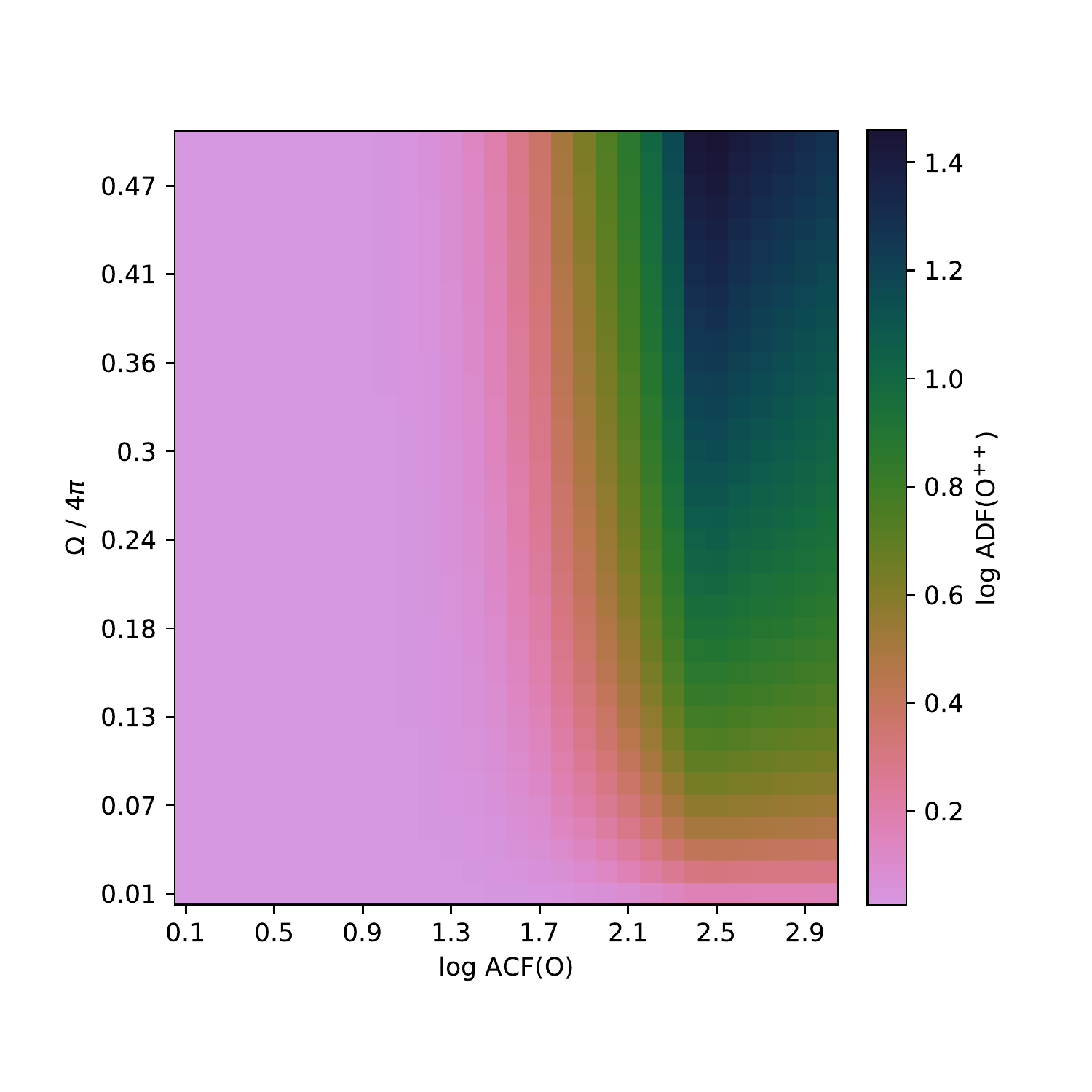}
\caption{Same axis as figure \ref{fig:high_te_ne_pc}. The color represents the log ADF(O$^{++}$) determined for the bi-abundance models. }
\label{fig:adf_opp}
\end{figure}

From the values of O$^{++}$/H$^{+}$ determined from CELs and ORLs (see previous sections) we can compute the ADF(O$^{++}$), which is shown in fig.~\ref{fig:adf_opp}. This can be compared to the ACF(O) running on the x-axis. In almost the whole ACF-$\Omega$ plane, the difference is important, reaching \Draft{a factor of 50} in the upper right corner. \Draft{For the \mr\ component, the ionic fraction O$^{++}$/O is always higher than 0.06 in the ACF-$\Omega$ plane, so even though we are comparing the discrepancy factor of O$^{++}$ to the contrast factor of O, this difference can not be completely attributed to the O$^{++}$/O ratio}. The value for the close to solar regions determined from the CELs being actually close to the "real" value used as input of the models, the discrepancy is coming from the ORLs determination that does not match the "real" value. The main issue of this determination is not the temperature but rather the estimation of the H$\beta$ proportion coming from the \ion{O}{ii} line emitting region (\mr\ component) as described in the next section. 

\begin{figure}
	\includegraphics[scale = 0.55]{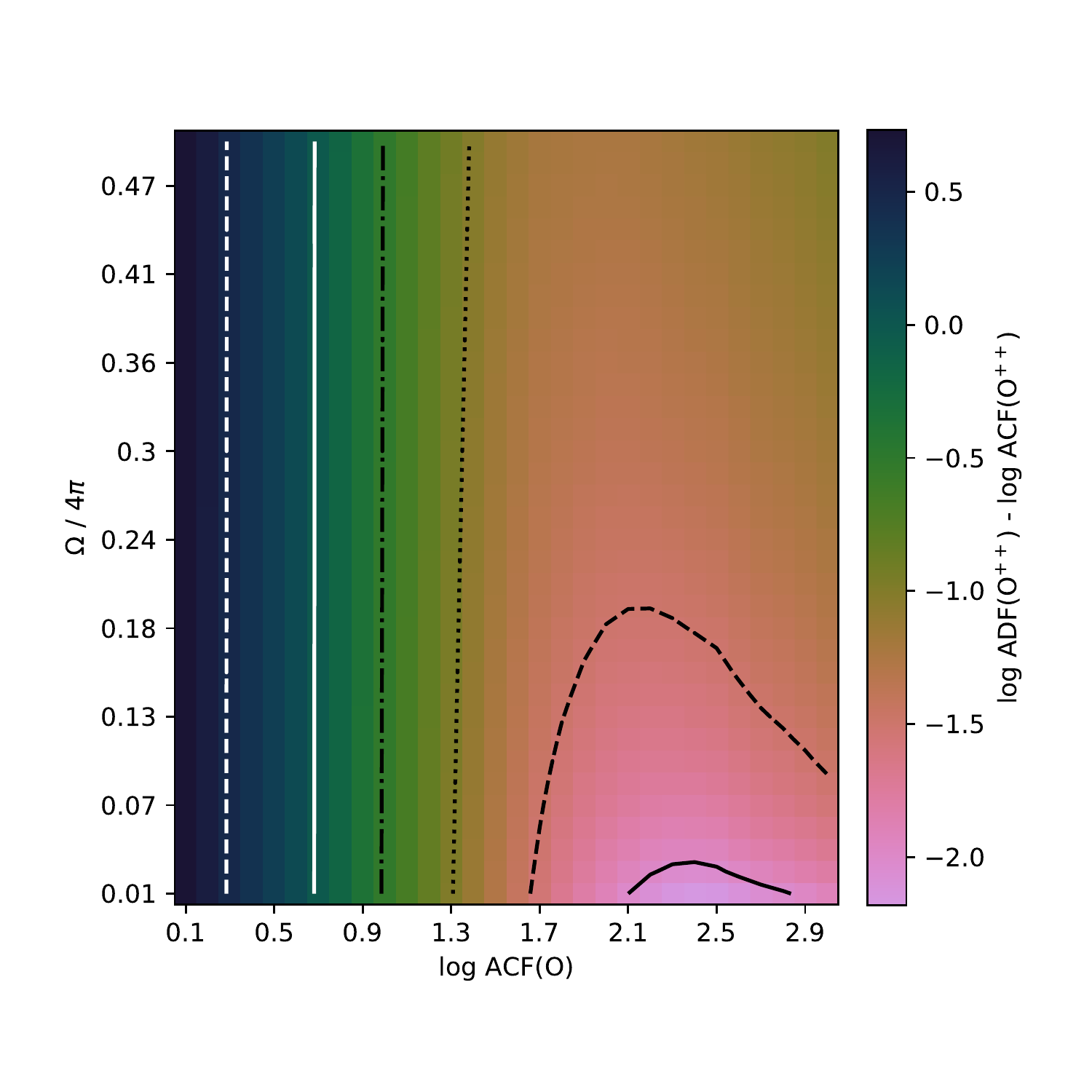}
	\caption{Same axis as figure \ref{fig:high_te_ne_pc}. The color represents the difference in log of the ADF(O$^{++}$) determined for the bi-abundance models and the ACF(\opp) (difference in \opp\ of the \mr\ and the \n,  \bc\ and \s\ components). White dashed (solid) is for log ADF(\opp) - log ACF(\opp) equal to 0.5 (0.0) and black dash-dotted, dotted, dashed and solid lines are for: -0.5, -1.0, -1.5, -2.0, respectively. }
	\label{fig:adf_acf_opp}
	\end{figure}

We compare in figure~\ref{fig:adf_acf_opp} the ADF(O$^{++}$) to the ACF(O$^{++}$), defined by the ratio of O$^{++}$/H$^{+}$ integrated over the \mr\ region and O$^{++}$/H$^{+}$ integrated over the \n +\bc +\s\ regions.
This can be seen as the measure of the error one do when determining the ADF, the ACF being the "true" value of the ionic abundance difference. Whatever the value of $\Omega$, the log ACF(O) around 0.8 corresponds to a good determination of the ADF (white solid line in fig.~\ref{fig:adf_acf_opp}). For lower values of the ACF(O), the ADF(O$^{++}$) overestimates the value of ACF(O$^{++}$) (by up to 0.7~dex). This corresponds to situations where the contribution to \oii\ coming from the close to solar regions actually dominates the total emission of the \oii\ lines. For values of ACF(O) greater than 0.8 dex, the \oii\ emission mainly comes from the \mr\ region and the ADF(O$^{++}$) underestimates the true value given by the ACF(O$^{++}$) by a factor up to $\sim$100 (dark solid line in fig.~\ref{fig:adf_acf_opp}). 

\subsection{Contributions of the \mr\ clumps to recombination lines.}
\label{sec:clumps_fraction}

\begin{figure}
	\includegraphics[scale = 0.50]{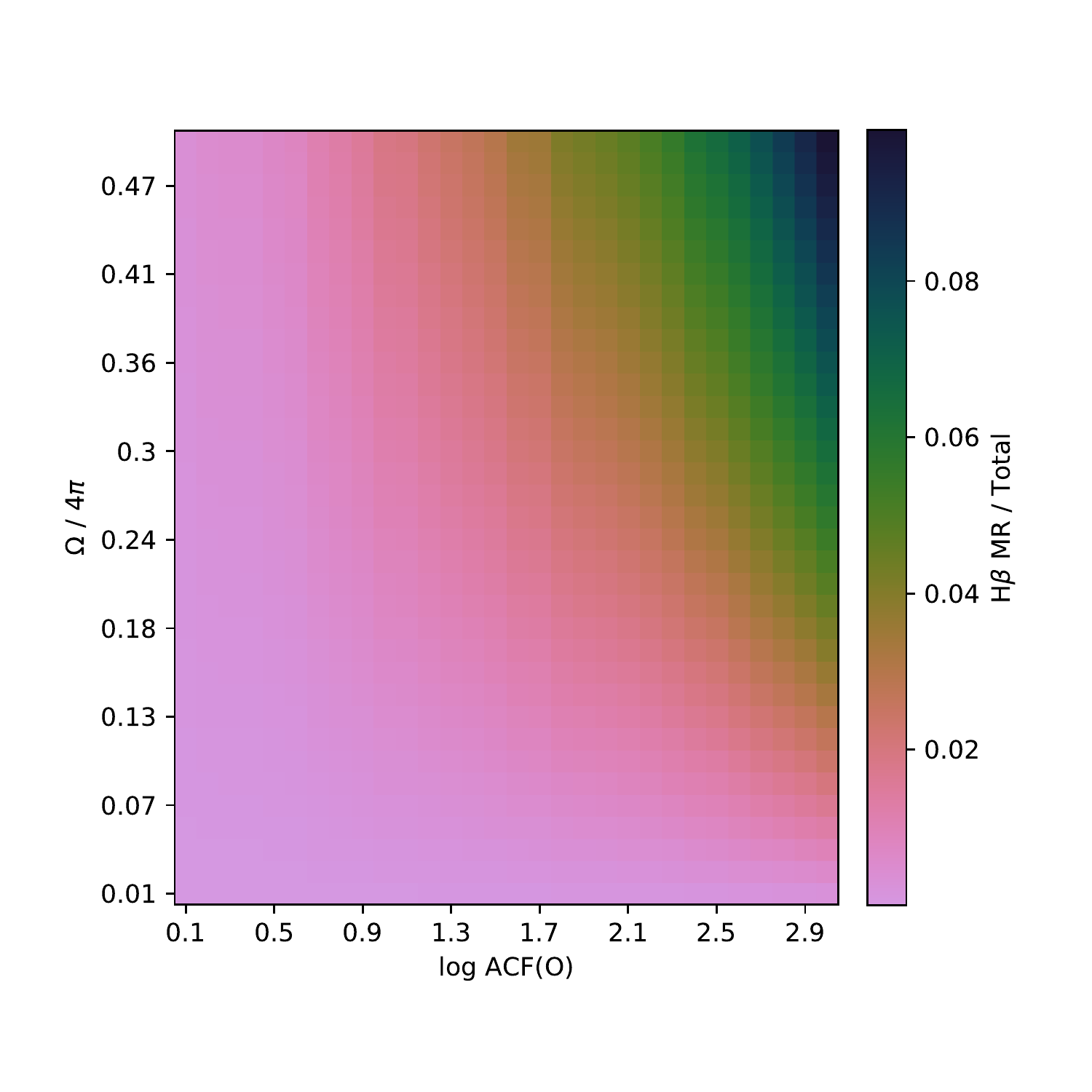}
	\caption{Same axis as figure \ref{fig:high_te_ne_pc}. The color represents the fraction of the H$\beta$ from the clumps to the total emission for the bi-abundance models.}
	\label{fig:hb_rich}
	\end{figure}

\begin{figure}
	\includegraphics[scale = 0.50]{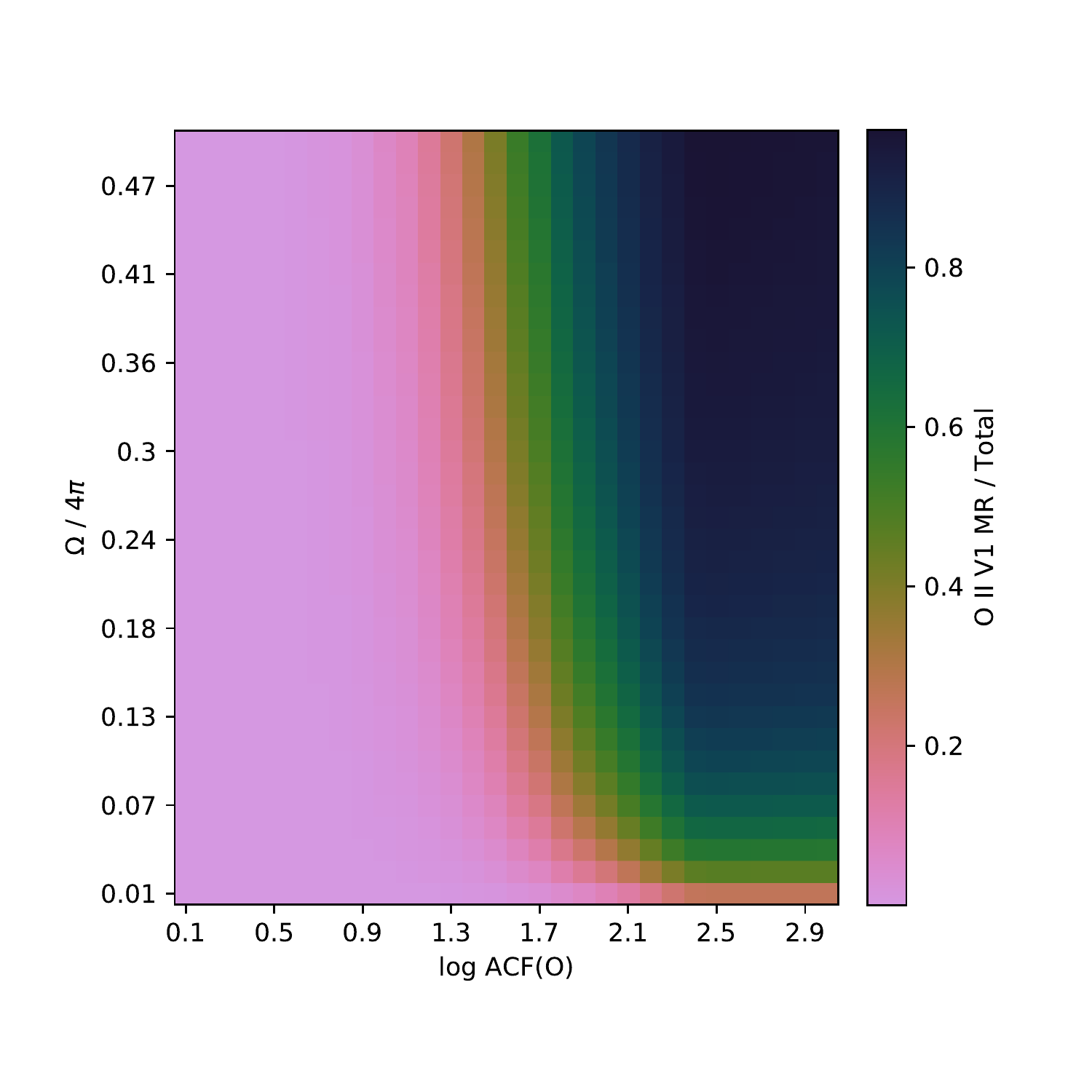}
	\caption{Same axis as figure \ref{fig:high_te_ne_pc}. The color represents the fraction of the V1 multiplet of \ion{O}{ii} from the \mr\ to the total emission for the bi-abundance models.}
	\label{fig:o2_rich}
	\end{figure}

In Figs.~\ref{fig:hb_rich} and \ref{fig:o2_rich} we draw the contribution of some recombination lines (namely for H$\beta$ and the V1 multiplet of \ion{O}{ii}) emitted by the \mr\ region relative to the total emission.
The H$\beta$ emission is mainly coming from the close to solar components, as the \mr\ contribution is never higher than 9\%. On the other side, the \ion{O}{ii} lines are well representative of the \mr\ region when this one is strongly H-poor (ACF(O) > 1.5 dex). The apparent incapacity of the \ion{O}{ii} lines to correctly predict the ACF of the nebula (see previous sections) is actually mainly due to the impossibility to only take into account the H$\beta$ emitted by the \mr\ region.

	\section{Models fitting a real object}
	\label{sec:compare_mod_object}
	\subsection{Looking for solutions for NGC~6153}
	
\begin{figure}
	    \centering
	    \includegraphics[scale = 0.55]{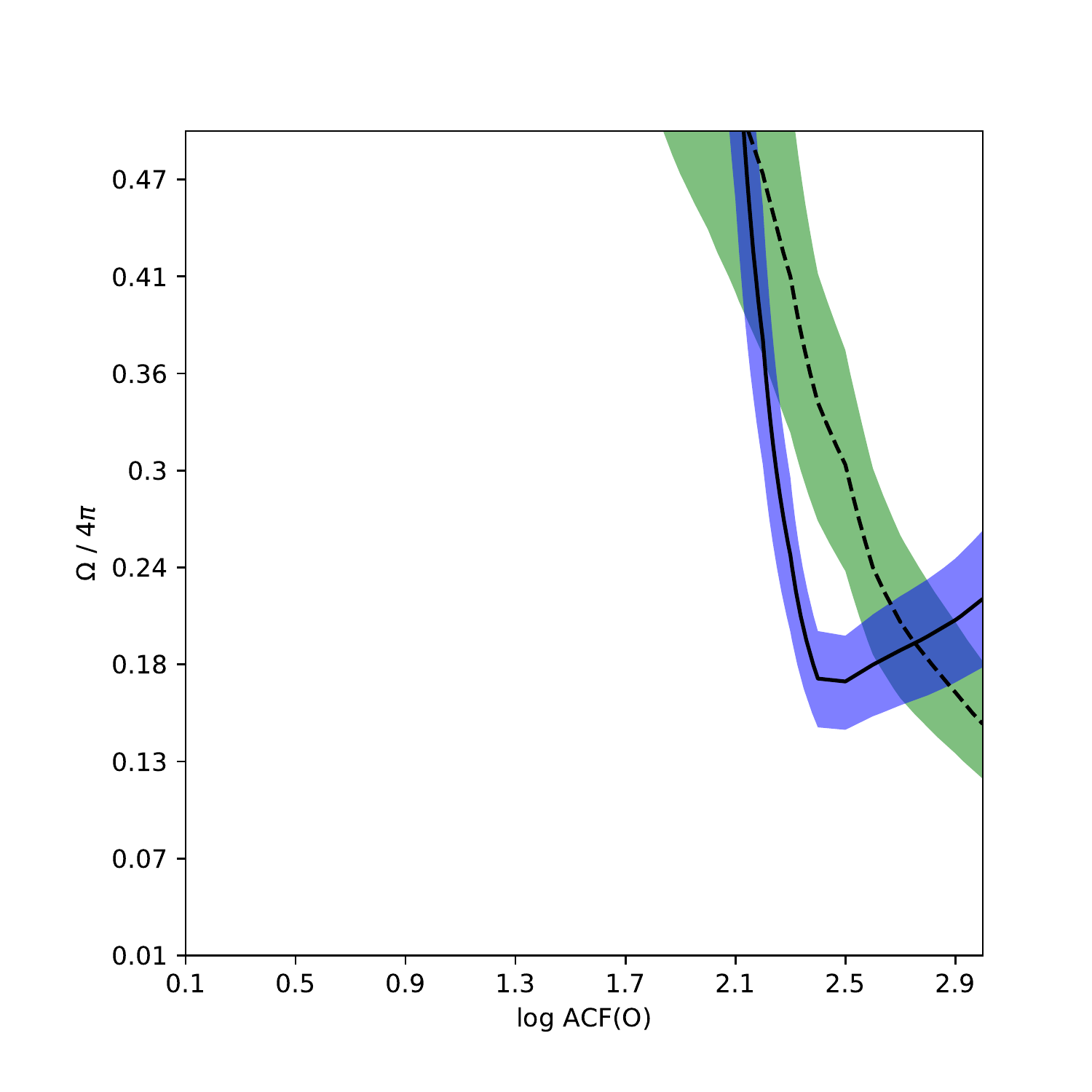}
	    \caption{Same axis as figure \ref{fig:high_te_ne_pc}. 
	    The black dashed (solid) line represents the bi-abundance models with estimations of $\Delta T$ = T([\ion{O}{iii}]) - T(BJ) (ADF(O$^{++}$)) equal to the observed value of 3025~K (8.2) for NGC~6153 \citep[our determination using the observations from][]{2000Liu_mnra312} , the green (blue) contours are for 15\% above and bellow this value.}
	    \label{fig:adf_dT}
	\end{figure}

\begin{figure*}
    \centering
	\includegraphics[scale = 0.6]{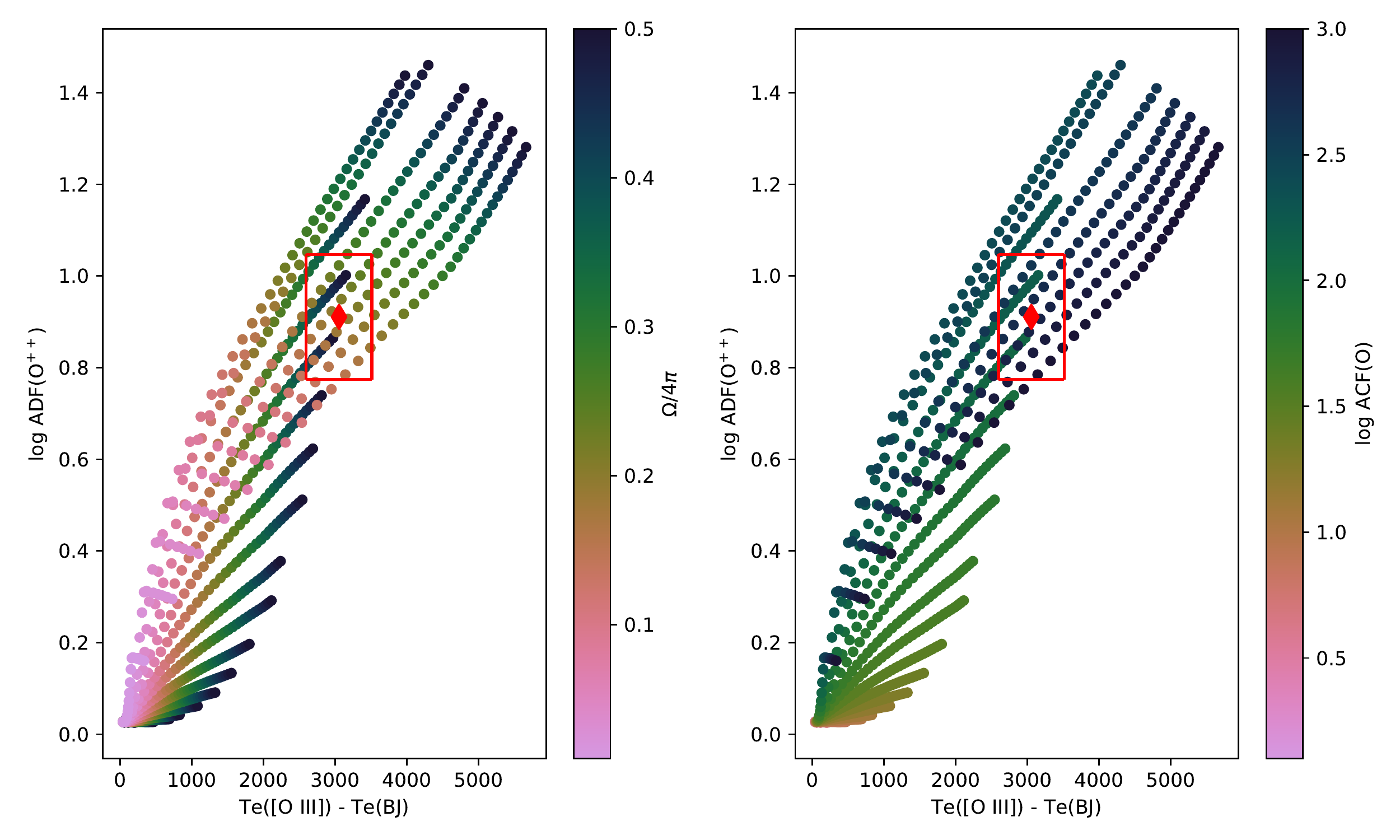}
	\caption{Left panel: color represents the normalized solid angle of: the metal-rich clumps, the gas behind the clumps and the shadow, in the range of 0.01 to 0.50. Right panel: color represents the variations in the ACF for oxygen (see eq. \ref{eq:ACF}) and goes from 0.1 to 3.0~dex for the bi-abundance models. The x-axis shows the difference in temperature estimated with: [\ion{O}{iii}]~4363/5007~\AA\ and the Balmer jump. The y-axis shows the log ADF(O$^{++}$) estimated for the bi-abundance models. The red diamond represents the value for the PN NGC~6153 taken from \citet{2000Liu_mnra312} (for the minor axis in the case of Te(BJ) and for the whole nebula in the case of Te([\ion{O}{iii}]) and ADF(O$^{++}$)).  The red square represents the region selected for the models that are closer to the observed value in NGC~6153.}
	\label{fig:adf_te}
	\end{figure*}

\begin{figure}
    \centering
    \includegraphics[scale = 0.55]{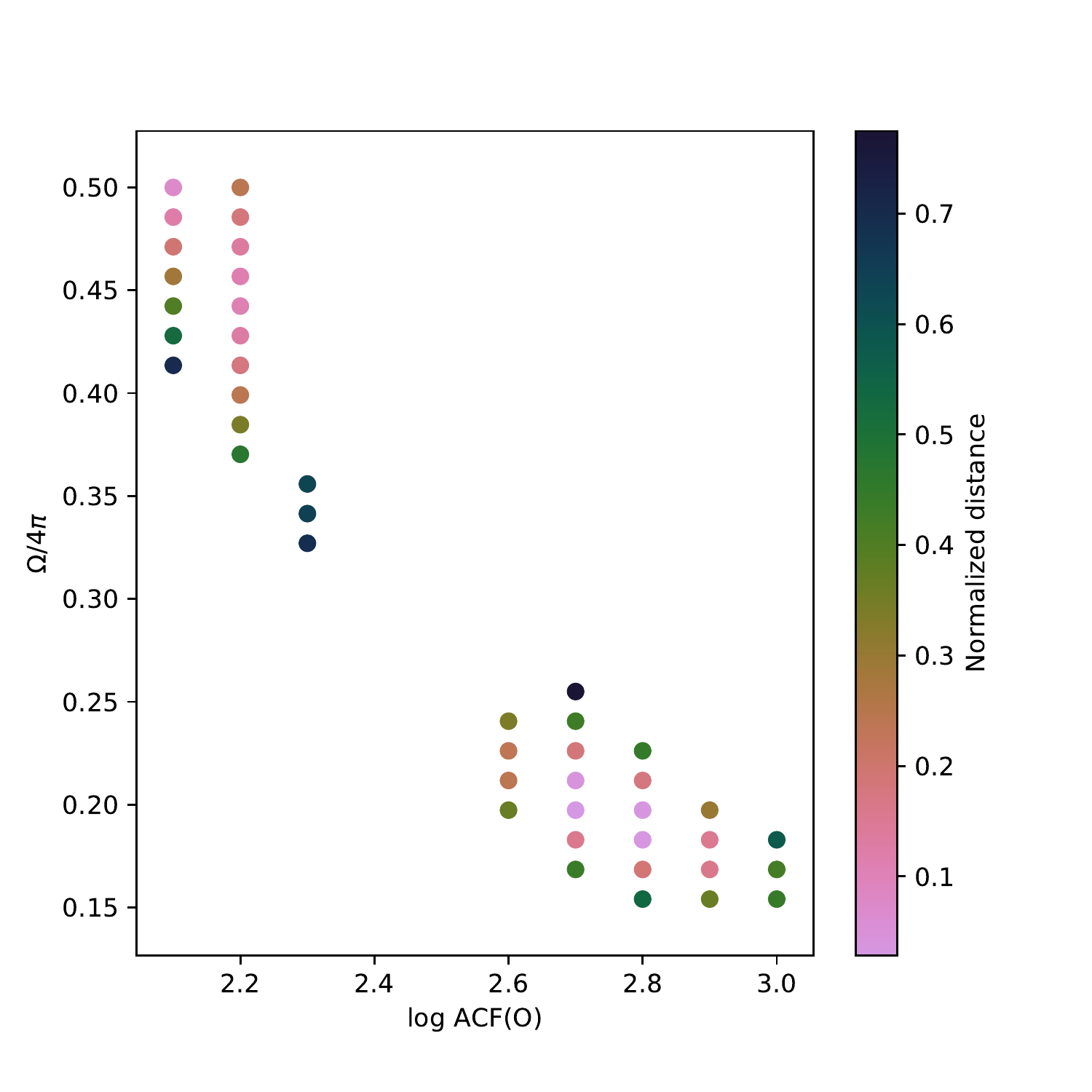}
    \caption{Selected models, as being inside the red square of figure \ref{fig:adf_te}. The color represents the distance from the model to the observed value represented as a red diamond in fig \ref{fig:adf_te}. The distance ($\sqrt{(x/x_0)^2+(y/y_0)^2}$, where x = T([\ion{O}{iii}]) - T(BJ), y = ADF(O$^{++}$)) is normalized through $x_0$ and $y_0$ so the distance from the centre of the square to a corner is 1.}
    \label{fig:sol_fam}
\end{figure}

	The two main observables related to the Abundance Discrepancy problem are the value of the ADF and $\Delta T$ the difference T([\ion{O}{iii}]) - T(BJ). We determine this parameters for NGC~6153 (with the same method than for our models), namely ADF(O$^{++}$)=8.2 (using T(BJ)) and $\Delta T$=3025~K \citep[with the observations presented in][]{2000Liu_mnra312}, to define an area in the ACF-$\Omega$ plane where the models fit these values. This area is shown in fig.~\ref{fig:adf_dT}, where the blue (green) color band shows where the model fits ADF(O$^{++}$) ($\Delta T$ resp.) within $\pm$15\%. Another way to perform a similar determination is by plotting the observables predicted by the models, as in fig.~\ref{fig:adf_te}. The color code corresponds to the values of ACF (right panel) and \om\ (left panel). The diamond corresponds to the observations for NGC~6153. Using a box of $\pm$15\% around the observed values, we can extract the models and exhibit the values of ACF-$\Omega$ combinations that reproduce the observations. This is done in fig.~\ref{fig:sol_fam}. Two families of solutions appear around [ACF(\opp), \om] = [2.1, \Draft{0.50}] and [2.8, 0.18]. The best solutions correspond to S$_1$: a lower ACF but high volume for the \mr\ region, and S$_2$: a higher ACF but smaller volume. 
	
	The characteristics of these two solutions are summarized in table~\ref{tab:phys_params}, where T$_e$, n$_e$, the ionic fractions of \hp, \op, \opp\ and \oppp\ and the fractions of volume and mass are given for each one of the 4 regions. The electron temperature of the \n\ region is close to \Draft{9500}~K, while the \bc\ and \s\ components are at slightly lower temperature. The \mr\ region is found to be as cold as 500-600~K. The high value of n$_e$ found in the \mr\ region is due to the contribution of the metals to the free electrons. The ionic fractions indicates that the two solutions are very different in the \mr\ - \bc\ - \s\ regions: the S$_1$ solution almost does not have \s\ region, and the \mr\ and \bc\ regions are mainly \opp, while the S$_2$ solution exhibits a \s\ component and show the \mr\ and \bc\ regions well recombined to \op. In both cases, the mass or volume fraction of the \mr\ region relative to the whole nebula is rather small.
	
	One interesting result is that despite the fact that both solutions differs by a factor of $\sim$ \Draft{4} in the O/H abundance, the mass of oxygen embedded in the \mr\ region only differs by a factor \Draft{of $\sim$ 1.5}. This points out that the \mr\ oxygen mass is somewhere robust against the degeneracy of the solutions relative to ACF(O) and $\Omega$, each one acting in opposite direction. We then determine the oxygen in the metal-rich region in NGC~6153 to be $\sim$40\% of the total oxygen in the nebula, this in a volume that is less than 1\%. 
	
	These two solutions S$_1$ and S$_2$ will be used in the next sections to explore the emission of the corresponding models. 
	
\begin{table*}
\centering
\caption{Physical parameters for each one of the four components, for the two solutions S$_1$, and S$_2$ in parenthesis.}
\begin{threeparttable}
\begin{tabular}{l l l l l}
\hline \hline
 & Normal & Rich clumps & Behind clumps & Shadow    \T\B\\
\hline \hline \T
\input{table2.tex}
\end{tabular}
\begin{tablenotes}
    \item $^{*}$ Fraction of solid angle for each component.
\end{tablenotes}
\end{threeparttable}
\label{tab:phys_params}
\end{table*}

\subsection{Emitted spectra }
\label{sec:em_spec}
\begin{figure*}
    \centering
    \includegraphics[scale = 0.53]{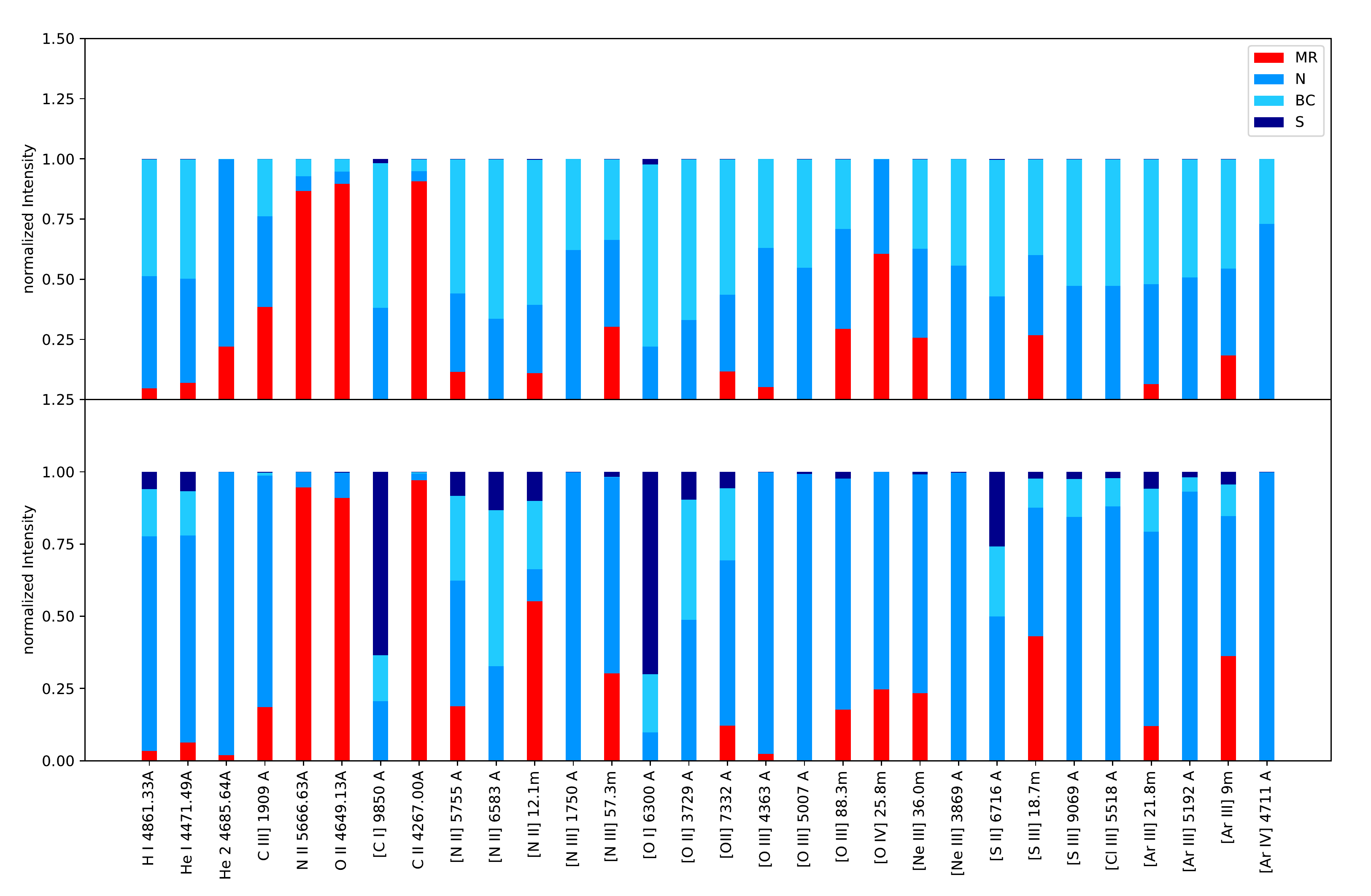}
    \caption{Normalized intensity of ORLs and CELs showing the contribution of the four components of the bi-abundance models (see sec. \ref{sec:bi-model}). Top panel represents the S$_1$ solution (log ACF(O)~=~2.1, and $\Omega$/4$\pi$~=~\Draft{0.50}), and bottom panel the S$_2$ solution (log ACF(O)~=~2.8, and $\Omega$/4$\pi$~=~0.18). The two solutions are selected from the two solution families shown in figure \ref{fig:sol_fam}. }
    \label{fig:lines_cel_orl}
\end{figure*}

We use the solutions S$_1$ and S$_2$ determined in the previous section to show in fig.~\ref{fig:lines_cel_orl} the contribution to the total emission coming from each one of the 4 regions (in different colors), for a set of 28 representative emission lines.
An extensive list of emission lines with the intensity emitted by each region, normalized to H$\beta$, for both solutions, is given in table~\ref{tab:intensities}.

\Draft{For both solutions we compare in table \ref{tab:IR_emission} the infrared emission to the observed values through the 4 IRAS bands at 12, 25, 60 and 100 $\mu$m obtained for PN NGC~6153 from \citet{1988iras7H}.}
\Draft{We can see that in both solutions the IR emission from the model is very close to the observed emission. \DraftThree{For the S$_1$, the IR band fluxes are well reproduced. For the S$_2$, the model is higher than the observed fluxes at 12, 25, and 60 $\mu$m, and a good fit for 100~$\mu$m is obtained. The emission at the shorter wavelengths bands is already higher than what is observed, adding dust to the \mr\ component will increase the emission in this range.} Thus there is no evidence of the presence of dust in the \mr\ region, \DraftThree{thus we don't include dust in the \mr\ component.} This result agrees with what \citet[][]{2003Ercolano_mnra344} found for the polar knots of Abell~30, but is in contradiction with what \citet[][]{1994Borkowski_apj435} found in the case of the equatorial H-poor ring of the same object. The interpretation of this lack of dust in the metal rich region, in terms of creation and destruction of dust, is out of the scope of this paper, \DraftThree{nonetheless, the extreme opposite case where all the dust is in the \mr\ component is discussed in sec. \ref{sec:dust_mr}.} }

\begin{table}
    \centering
    \caption{Emission from IRAS bands for PN NGC~6153 \citep{1988iras7H} and the two solutions S$_1$ and S$_2$ in Jansky units.}
    \begin{tabular}{c c c c}
    \hline \hline
    Wavelength & Observations & S$_1$  & S$_2$  \\ \hline \hline
    12~$\mu$m   &  6.9 &  7.2  &  10.3 \\
    25~$\mu$m   &  52.1 &  72.8 & 107.0  \\
    60~$\mu$m & 120.0 &  127.7  & 173.6 \\
    100~$\mu$m  & 52.1 &  40.3  & 52.0 \\
    
     \hline \B
    \end{tabular}
    \label{tab:IR_emission}
\end{table}

We see again that the contribution of H$\beta$ coming from the \mr\ region is very small. This is the key problem in the determination of the true value of the \mr\ region metallicity, translated in the difference between the ADF (observed) and ACF (real).

The important contribution of the \mr\ region coming from the metal recombination lines is clearly seen, as well as from the infrared lines. The contribution of the \bc\ and \s\ regions are different in the two solutions.

From the table~\ref{tab:intensities} we can see that the total intensities of some lines (latest column) change between the 2 solutions: the \ion{C}{ii} (\ion{N}{ii}) recombination lines increase by a factor of $\sim$\Draft{2.8 (1.7)}, while \Draft{the \ion{Ne}{ii} recombination lines decrease by $\sim$ 25\%}, between the S$_1$ and the S$_2$ solutions. 
This is mainly due to a change in the ionization of the \mr\ region. For the same reason we see a difference in the emission of the infrared lines of [\ion{N}{ii}] and [\ion{Ne}{ii}]. The other IR lines are from higher charged ions, less affected by the ionization changes in the \mr\ regions. The \oii\ recombination lines are not changing, by construction of our solutions based on fitting the observed ADF(\opp). 

The optical CELs are mainly unchanged between the two solutions.

We use a simplistic relation in the way metals are enhanced in the \mr\ region (same ACF for all the metals). The real situation may certainly be more complex and these differences in the \ion{C}{ii} and \ion{N}{ii} emission between the two solutions \Draft{can not be used to derive properties related to the oxygen abundance like ACF(O).} \DraftTwo{Nevertheless, the \ion{C}{ii}, \ion{O}{ii}, and \ion{N}{ii} intensities predicted by both models are close to the observed values given by \citet[][]{2011Yuan_mnra411}. This could be seen as an indication that the C:N:O relative abundances used in \mr\ region are adequate.}

\section{Discussion}
\label{sec:discussion}

\subsection{Helium abundance in the \mr\ component}

The material that is responsible for the observed high ADF ($>$~10) in some PNe is unique and original. It is not a common ISM component. It is very hard to observe as it is blended with "normal" PN gas in the observations. It mainly emits metal recombination lines, and some H recombination lines. The contribution from this component to the total He recombination lines is almost negligible (as well as in the case of HI lines), and barely distinguishable from the "normal" component emission. The exact He abundance in this \mr\ component is then very hard to determine. On the other side, the fact that this component is cold points to a small heating and strong cooling. We explore a variant of the solutions obtained above, where the He/H abundance is also enhanced. The results are presented in details in the Appendix~\ref{sec:App-He}. An acceptable solution can be found for He/H as high as 6.3, leading to a slight overprediction of the \ion{He}{I} line intensities compared to values reported by \citet[][]{2011Yuan_mnra411}. The total amount of oxygen in the whole nebula derived for this solution does not significantly changed compared to what is found in solutions S$_1$ and S$_2$.

This result is similar to that of \citet{2011Yuan_mnra411}, where the He abundance is less enhanced than metals in their metal rich region (He/H (MR/NR) $\simeq$ 5, while O/H (MR/NR) $\simeq$ 80, from their Tab.2.). 

\DraftThree{On the other hand, in the case of the model of Abell~30 presented by \citet[][]{2003Ercolano_mnra344}, the He/H abundance is reaching a value as high as 40, while O was determined to be almost in the same amount than H. We have here lower values for O/H (between 0.07 and 0.35) and for He/H (lower than 6.3). The process leading to the H-poor components in these two object seems to be different.

Finally, if we compare with the abundances determined in some nova shells (nova event is not excluded by \citet[][]{2018Wesson_mnra480} as a proxy for a scenario to these H-poor clumps), one can see that the strong enhancement determined for the C:N:O elements is not associated to an equivalent enhancement in Helium \citep[See e.g.][]{1996Morisset_aap312}.}

\subsection{Dust in the \mr\ component}
\label{sec:dust_mr}
\DraftThree{The creation and destruction of dust in the \mr\ component is unknown, and due to the lack of spatially resolved IR observations, we cannot constrain the radial distribution of dust.
In the previous sections we have assumed that no dust is in the \mr\ component, and all the dust necessary to reproduce the IRAS observations is located in the close-to-solar components. Here we explore the opposite case where the dust is only present in the \mr component, although the reality may be an intermediate solution between both hypothesis. We generate grids of models including dust of graphite and silicate types with two sizes for each: 1~$\mu$m and 0.01$\mu$m, we search for the D/G values that have the closest fit to the IRAS observations. We define the D/G for each model to be proportional to the O/H of the \mr\ component. We follow the same procedure searching for solutions that reproduce $\Delta$T and the ACF(O$^{++}$), like in the figures \ref{fig:adf_te} and \ref{fig:sol_fam}. In these models, at some point when the ACF increases the dust becomes optically thick (ACF > 2.7~dex) and the gas behind the clumps is no longer ionized by the star, becoming a shadow ionized by the Lyman continuum radiation from the \n\ component. The solution at "lower" ACF and higher $\Omega$/4$\pi$ is lost, and only one solution with ACF = 2.8~dex and $\Omega$/4$\pi$ = 0.27, marginally reproduces simultaneously $\Delta$T and  ACF(O$^{++}$). For this solution the IR emission at 12, 25, 60 and 100~$\mu$m is 10.2, 44.6, 134.8 and 71.4~Jy, close to the observed values (See Tab.~\ref{tab:IR_emission}). For this solution, the D/G by mass is 1.04, which is considerably higher than the canonical value in the ISM of 6.3$\times$10$^{-3}$. The abundances of the elements trapped in the dust (in log, by number) are: C = -1.21, O = -0.4, Mg = Si = Fe = -1.0. These values of abundances of elements in dusty phase are even higher than the corresponding abundances in the gaseous phase for Mg, Si and Fe, for O is roughly the same amount and for C is about 30\% of the gaseous phase. A more detailed modeling of the distribution of dust could be of interest as further work, but additional observations are needed to constrain the free parameters.}

\subsection{Recombination contribution to auroral lines}
\label{sec:auroral}

\DraftThree{The recombination contribution to the auroral lines for [\ion{N}{ii}]~5755 is taken from Cloudy based on \citet{1984Nussbaumer_Astr56}, for [\oii]~7332 from Cloudy based on \citet{2001Liu_mnra323}, for [\oiii]~4363 from Cloudy (version 17.02) using \citet{1991Pequignot_aap251} and \citet{1984Nussbaumer_Astr56}}. We see in Fig.~\ref{fig:lines_cel_orl} that the contribution of the \mr\ component to the auroral lines [\ion{N}{ii}]~5755 and \Draft{[\oii]~7332} emission is not negligible. This is due to the recombination of N$^{++}$ and \opp, favored by the low temperature of the \mr\ region. In the grid we explored, the most important contributions reach 1/3 of the total auroral emission, for a model with a determined ADF(\opp) of $\sim$30. In case of more extreme ADFs, the recombination contribution may even be dominant. \Draft{In case of higher excitation nebula the recombination of [\oiii]~4363 may become significant.} This is in total coherence with observations from \citet{2016Jones_mnra455} and \citet{2016Garcia-Rojas_apjl824} who found the spacial location of the [\oiii]~4363 emission in coincidence with the \opp\ recombination lines and not with the [\oiii]~5007 emission. This effect can be amplified when the observation is obtained in the direction of the metal rich region rather than for the whole nebula, which is the case in high spatial resolution observations. Without taking the effect of the recombination into account, one can determine a gradient of temperature increasing toward the central part of the nebula, where the \mr\ regions are located. On the contrary, the increase of [\oiii]~4363 emission is actually related to a very strong decrease of the electron temperature! \DraftTwo{We will explore the contribution of the recombination to [\oiii]~4363 in a forthcoming paper \citep{2020Gomez-Llanos_arxiv}}

\subsection{Ionization correction factor for the \mr\ region}
\label{sec:ICF-MR}

The results presented in this work are about the ADF(\opp). If one is interested in the ADF(O), an Ionization Correction Factor (ICF) needs to be applied. Even in the case of a low ionization nebula where only \op\ and \opp\ are present, these ionic abundances are determined from the [\oii] and [\oiii] lines that are mainly emitted by the close to solar regions (\n, \bc\ and \s). For the \mr\ region where only \opp\ is observed through the \oii\ lines, the ICF(\opp) needs to be used. Since the ICF(\opp) for ORLs can't be determined, the one derived for the forbidden lines has to be used. From table ~\ref{tab:phys_params}, we see that the ICF(\opp) - which is O/\opp\ - is almost the same for the \n, \mr\ and \bc\ regions in the S$_1$ solution (\Draft{between 1.1 and 1.2}). But in the case of the S$_2$ solution, the ICF(\opp) for the close to solar region is close to 1.1, while the ICF(\opp) that needs to be applied to \opp\ in the \mr\ region is 4.8.

If the density of the \mr\ region, or its radial size, is changed, the ionization of the region will change, and so will the ICF(\opp).

The \ion{He}{ii}~4686 emission is classically used to determine the ICF(\opp) taking into account the presence of \oppp\ \citep{2014Delgado-Inglada_mnra440}. But we see that in the case of the S$_1$ solution, 20\% of the intensity of this line comes from the \mr\ region and should be removed before computing the ICF.

\subsection{Effects of density and size of the \mr\ region}
\label{sec:density-MR}

We explored in the previous sections the behaviour of a grid of models. In this grid some parameters have been fixed to reasonable values. This is the case of the \mr\ region density, its distance to the central star and its radial size. 

The electron density of the \mr\ region is hard to constrain by observations: recombination lines (the only ones emitted by this cold region) have quite low dependency on the density. There is no indication that a pressure equilibrium must exist between the warm and the cold regions, as they are not supposed to be in contact. We adopted then the same hydrogen density as for the \n,  \bc\ and \s\ regions. Increasing for example the density of the \mr\ region would increase its optical depth, decreasing its global ionization and possibly leading to a vanishing of the \bc\ region, replaced by a pure shadow region. 

The effect of changing the radial size of the \mr\ region, decreasing for example its inner radius, would lead to a very similar effect: increasing its optical depth. 

In both cases, the global emission of the \mr\ region, for a given $\Omega$, will increase. To recover the same ADF and $\Delta T$, a lower ACF will be needed for a given $\Omega$.

We performed test cases to explore the effect of changing the hydrogen density and the inner radius of the \mr\ component. 
In a first test, we explore the density range from 1$\times$10$^3$~cm$^{-3}$ to 4$\times 10^3$~cm$^{-3}$. At lower densities the models fitting the observed ADF(\opp) and $\Delta T$, as shown in fig.~\ref{fig:adf_te}, have higher $\Omega$ eventually reaching impossible values higher than 1. On the other side, at higher densities no solution is found that reproduces simultaneously the ADF(\opp) and $\Delta T$. 

The other test varying the inner radius between 4.3\arcsec\ and 4.7\arcsec, leads to a similar behaviour. A smaller inner radius needs less $\Omega$ to fit the two observables.

These tests show that the main result obtained in the previous section is still valid when changing some of the parameters we fixed {\it a priori} in our grid of models: the ACF (the real metal enhancement of the rich region) is higher than the ADF determined from observations. 

From the exploration of this grid, we determine the mass of oxygen embedded in the metal rich clumps to be between 25\% and 60\% of the total mass of the nebula.

Another limitation of our study is the reality of the \bc\ and \s\ regions. The very simplistic model presented here assumes a very sharp separation between the \n\ region and what is happening behind the \mr\ region. Some observations exhibit cometary tails \citep[see e.g.][in the Helix nebula]{2005AJ....130..172O} that are very aligned in the direction of the central star. On the other side, the image of NGC~6153 does not show clear evidences of this kind of structure. 

\section{Conclusions}
\label{sec:conclusions}

In the previous sections we described how a grid of models is obtained considering two metallicity models when varying the abundance of a metal-rich (\mr) component and its volume. The ratio between the two metallicities is called the Abundance Contrast Factor (ACF). We explored how the physical parameters and chemical abundances can be determined from the observables produced by the models, especially the [\oiii] and \oii\ emission lines leading to the determination of \opp/\hp\ by two methods. These two values are classically used to define the Abundance Discrepancy Factor (ADF(\opp)). 

We show that the determination of the ADF(\opp) does not represent an good estimation of the ACF(O) given as input of the models, even taking into account that the ACF is defined for the oxygen element and the ADF only for the \opp\ ion. The difference between the ACF and the ADF can be as high as two orders of magnitude. The main limitation is coming from the fact that the ionic abundance \opp/\hp\ determined from the \oii\ region is computed using the \ion{H}{i} intensity resulting from the emission of the whole nebula, when only a small fraction is actually coming from the \mr\ region (a maximum of 8\% is found in the models we present here). 

It is today almost impossible to isolate the contribution of the \mr\ region from the total \ion{H}{i} emission by observations. Very high spatial and/or spectral resolution observation may solve this important issue. 

Nevertheless, we show that it is possible to determine an estimation of the mass of metals (at least oxygen in our case) embedded in the rich component, from the optical recombination lines.

\DraftTwo{The models reproduce the Infrared observations without needs of an enhanced dust content in the metal rich region.}

\section*{Acknowledgements}

\DraftTwo{We thank the anonymous referee for helping to improve the quality of this research, especially on the discussion of the presence of dust in the models, and the He/H abundance in the \mr\ region.}
We are very grateful to Grazyna Stasi\'nska and Jorge Garc\'ia-Rojas for reading and helpfully commenting the draft of this paper.
This work was supported by CONACyT-CB2015 254132, DGAPA/PAPIIT-107215, and DGAPA/PAPIIT-101220 projects. VGL acknowledges the support of CONACyT and PAEP-UNAM grants. The manuscript of this paper has been written in the Overleaf environment. 

\section*{Data availability}
The models presented in the paper are available under request to the authors.


\appendix
\onecolumn

\section{Emission lines intensities of the two solutions}
\label{sec:App-intensities}

\DraftTwo{In this appendix we present the line intensities obtained for the two solutions presented in Sec.~\ref{sec:compare_mod_object}, the second solution being in parenthesis. The intensities emitted by the 4 regions considered in each model are given, as well as the total intensities. Some recombination lines are not computed by Cloudy, we obtained them by calling PyNeb from pyCloudy through the add\_emis\_from\_pyneb method (See Sec.~\ref{sec:models} for details). These lines are identified with an asterisk. The recombination contribution to the [\ion{O}{III}] 4363 emission line is not taken from Cloudy, but rather obtained by adding the radiative recombination and the dielectronic recombination, following \citet[][]{1991Pequignot_aap251} and \citet[][]{1984Nussbaumer_Astr56} respectively (See Sec.~\ref{sec:auroral}).
}

\begin{longtable}[t]{l l l l l l l}
\caption{Line intensities emitted by each of the four components (\n, \mr, \bc, and \s ) and the total emission of our modeled bi-abundance nebula, for the S$_1$, and S$_2$ solutions in parenthesis. Values are normalized to H$\beta$.\\
* Line intensities obtained using \textit{PyNeb} (see sec. \ref{sec:models}).\\
** The collisional contribution of this line is taken from \textit{Cloudy}, the recombination contribution is obtained adding the radiative and dielectronic recombination from \citet[][]{1991Pequignot_aap251} and \citet[][]{1984Nussbaumer_Astr56}, respectively (See Sec.~\ref{sec:auroral}) .}
\label{tab:intensities}\\

\hline
Line & Normal & Rich clumps & Behind clumps & Shadow & Total  \T\B\\
\hline \hline 
\endfirsthead
\caption{Continues}\\
\hline
Line & Normal & Rich clumps & Behind clumps & Shadow & Total \T\B\\
\hline \hline 
\endhead
\input{table.tex}
\end{longtable}

\bsp	
\label{lastpage}

\section{The helium abundance in \mr\ component}
\label{sec:App-He}

\begin{figure}
    \centering
    \includegraphics[scale = 0.65]{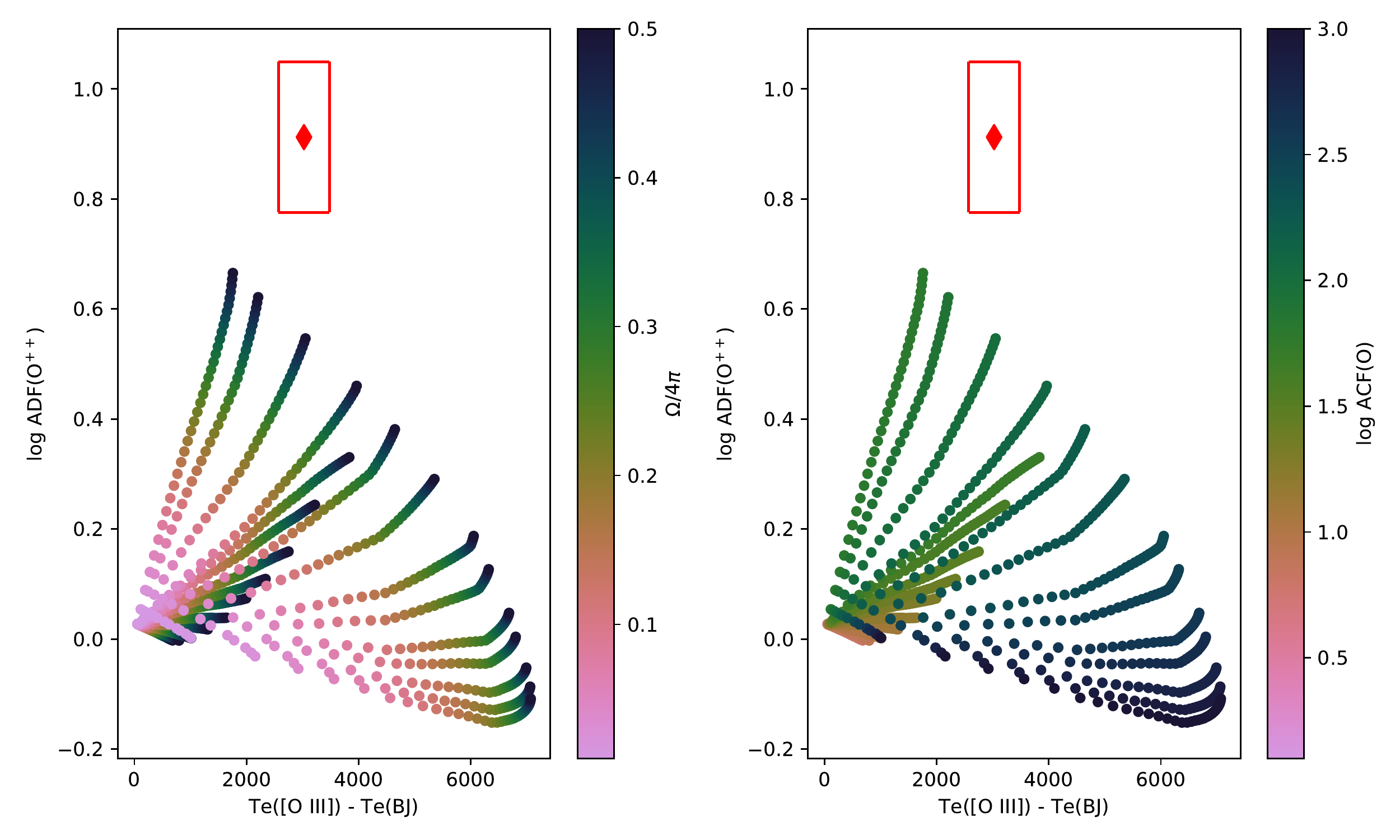}
    \caption{Relation between ADF(\opp) and T([\oiii]) - Te(BJ) for a sample of models where the helium in the \mr\ component is enhanced in the same way as the metals.}
    \label{fig:adf_te_he_rich}
\end{figure}

\begin{figure}
    \centering
    \includegraphics[scale = 0.65]{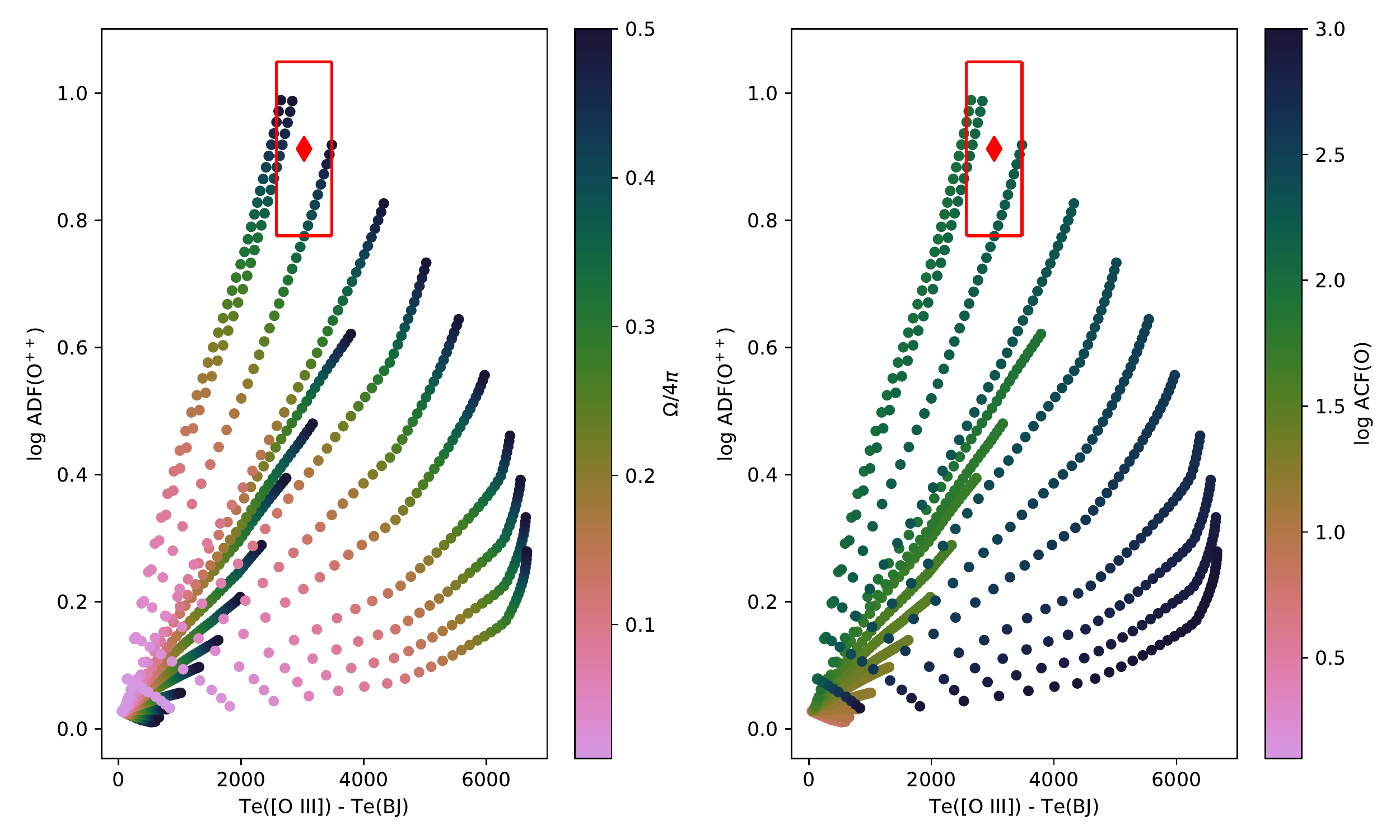}
    \caption{Relation between ADF(\opp) and T([\oiii]) - Te(BJ) for a sample of models where the helium in the \mr\ component is enhanced following this relation: He/H$_{MR}$ = He/H$_N$ x 0.5 (1 + ACF(O) ).}
    \label{fig:adf_te_he_50p}
\end{figure}

\DraftTwo{In the main part of this paper, we present models where the helium abundance is locked to the hydrogen one. 
Here we explore an extreme case where the \mr\ region is only H-poor and where helium is enhanced in the same way as the metals. We found that no solution can be determined in the ADF(\opp) vs $\Delta$T observable space, as shown in fig. \ref{fig:adf_te_he_rich}. A value of T([\oiii]) - Te(BJ) as high as the observed one can be obtained, but no high value for the ADF(\opp) can be reached (the maximum reached by the models is 5, while the observation is 8) . This is mainly due to the fact that the \mr\ region does not emit anymore the \ion{O}{II} recombination lines: in the inner part of the \mr\ region, where He is ionized (once or twice), the electron temperature is rather high due to the heating from this He ionizations (around 5,000K) and the recombination of \opp is strongly reduced. On the other side of the \mr\ region, when He is recombined and does not heat anymore the gas, there is no photons anymore to ionize \op into \opp, and no \ion{O}{II} recombination lines can be produced. Nevertheless this cold region emits some continuum, leading to a low value for Te(BJ) and {\it in fine} to the observed T([\oiii]) - Te(BJ).

We explored another set of models where the He abundances is only enhanced by a fraction of the metal enhancement. We found that the maximal value for He/H that still leads to a marginal solution S3 that reproduces the observed ADF(\opp) and $\Delta$T is: He/H = 6.3 (no solution is found for He/H higher than 6.3). The solution we obtain in this case corresponds to log~ACF = 2.1 and $\Omega$/4$\pi$ = 0.5. The corresponding plot is shown in fig. \ref{fig:adf_te_he_50p}.  Intensities of He lines for this solution S3 are presented in \ref{tab:he_lines}, as well as the values for S$_1$ and S$_2$ presented in Sec.~\ref{sec:compare_mod_object} and the observation from \citet[][]{2011Yuan_mnra411}. We can see that for this high He content solution, the predictions of the \ion{He}{I} lines are more or less 4 times higher than for the S$_1$ and S$_2$ solutions and are above the observed values. Regarding the \ion{He}{II} line, this S$_3$ solution is closer to the observation than the S1 or S2 values. We can conclude that He/H in the \mr\ region is not well constraint by the model, being between 0.1 and 6, reminding that the enhancement in metals is of order of 125 for the S$_3$ solution. We should also keep in mind that the ionized SED used in our models is a simple Planck function, adopting a more detailed atmosphere model being out of the scope of this paper.
}

\begin{table}
    \centering
    \caption{Emission line intensities of He lines for the solution S3 with He/H = 6.3, log~ACF = 2.1, and $\Omega$/4$\pi$ = 0.5. The observations for PN NGC~6153 from \citet[][]{2011Yuan_mnra411} and the solutions S1 and S2 of the paper are also shown.}
    \begin{tabular}{c c c c c}
    \hline \hline
Line &  S1 &  S2 & S3 &  Observation \\ \hline \hline
He I 4471.49 & 5.0e-02 & 4.5e-02 & 1.9e-01 & 6.5e-02 \\
He I 5875.64 & 1.4e-01 & 1.3e-01 & 5.7e-01 & 1.9e-01 \\
He I 6678.15 & 4.1e-02 & 3.7e-02 & 1.6e-01 & 4.8e-02 \\
He I 7065.22 & 3.2e-02 & 2.8e-02 & 8.4e-02 & 4.3e-02 \\
He II 4685.64 & 6.8e-02 & 8.2e-02 & 1.0e-01 & 1.3e-01 \\
     \hline \B
    \end{tabular}
    \label{tab:he_lines}
\end{table}

\begin{table*}
\centering
\caption{Physical parameters for each one of the four components, for the solution S$_3$.}
\begin{threeparttable}
\begin{tabular}{l l l l l}
\hline \hline
 & Normal & Rich clumps & Behind clumps & Shadow    \T\B\\
\hline \hline \T
\input{table2_s3.tex}
\end{tabular}
\begin{tablenotes}
    \item $^{*}$ Fraction of solid angle for each component.
\end{tablenotes}
\end{threeparttable}
\label{tab:phys_params_s3}
\end{table*}

\end{document}

%% file: abunds.tex
He = 11.00 & Al = 6.45 & Ti = 4.95 \\ 
B = 2.70 & Si = 7.54 & V = 3.93 \\ 
C = 8.50 & P = 5.41 & Cr = 5.64 \\ 
N = 8.58 & S = 7.24 & Mn = 5.43 \\ 
O = 8.75 & Cl = 5.37 & Fe = 6.18 \\ 
F = 4.56 & Ar = 6.46 & Co = 4.99 \\ 
Ne = 8.25 & K = 5.03 & Ni = 6.22 \\ 
Na = 6.24 & Ca = 6.34 & Cu = 4.19 \\ 
Mg = 7.58 & Sc = 3.15 & Zn = 4.56 \\ 
\hline \hline

%% file: table2.tex
<T$_e$> [K] & 9536 (9536) & 629                           (527) & 8906 (7252) & 7117 (7117) \\ 
<n$_e$> [cm$^{{-3}}$] & 2770 (2770) & 3817                           (5728) & 2738 (2610) & 2056 (2056) \\ 
12+log(O/H) & 8.75 (8.75) & 10.85                           (11.45) & 8.75 (8.75) & 8.75 (8.75) \\ 
$\Omega$/4$\pi$* & 0.50 (0.80) & 0.50                           (0.20) & 0.50 (0.20) & 0.50 (0.20) \\ 
H$^+$/H & 0.99 (0.99) & 1.00                           (1.00) & 0.99 (0.96) & 0.74 (0.74) \\ 
O$^+$/O & 0.06 (0.06) & 0.04                           (0.70) & 0.15 (0.96) & 0.46 (0.46) \\ 
O$^{{++}}$/O & 0.86 (0.86) & 0.84                           (0.28) & 0.84 (0.00) & 0.26 (0.26) \\ 
O$^{{+3}}$/O & 0.07 (0.07) & 0.11                           (0.02) & 0.00 (0.00) & 0.00 (0.00) \\ 
O mass (10$^{{-4}}$) [M$_\odot$] & 6.6 (10.5) & 7.5                           (12.3) & 6.6 (2.0) & 0.0 (1.0) \\ 
Volume [10$^{{50}}$cm$^{{-3}}$] & 345.1 (554.2) & 3.1                           (1.3) & 345.0 (105.0) & 0.5 (51.3) \\ 
\% mass & 49.3 (76.7) & 1.4                           (1.7) & 49.3 (14.5) & 0.1 (7.1) \\ 
\hline

%% file: table.tex
$\ion{C}{ii}$~1335.0~\AA\** & 4.8e-3 (7.6e-3) & 5.6e-2 (1.7e-1) & 5.6e-3 (1.3e-3) & 3.6e-6 (4.e-4) & 6.7e-2 (1.8e-1) \\ 
$[\ion{N}{iii}]$~1746.82~\AA\ & 2.6e-3 (4.1e-3) & 9.8e-8 (1.2e-7) & 1.6e-3 (2.5e-7) & 2.0e-8 (2.2e-6) & 4.1e-3 (4.1e-3) \\ 
$[\ion{N}{iii}]$~1748.65~\AA\ & 5.4e-3 (8.5e-3) & 2.2e-6 (2.5e-6) & 3.2e-3 (5.2e-7) & 4.2e-8 (4.6e-6) & 8.5e-3 (8.5e-3) \\ 
$[\ion{N}{iii}]$~1749.67~\AA\ & 3.0e-2 (4.9e-2) & 3.9e-6 (4.3e-6) & 1.9e-2 (3.0e-6) & 2.4e-7 (2.6e-5) & 4.9e-2 (4.9e-2) \\ 
$[\ion{N}{iii}]$~1750.0~\AA\ & 6.2e-2 (9.8e-2) & 9.2e-6 (1.0e-5) & 3.7e-2 (6.1e-6) & 4.8e-7 (5.3e-5) & 9.9e-2 (9.8e-2) \\ 
$[\ion{N}{iii}]$~1752.16~\AA\ & 1.8e-2 (2.8e-2) & 6.7e-7 (8.0e-7) & 1.1e-2 (1.7e-6) & 1.4e-7 (1.5e-5) & 2.8e-2 (2.8e-2) \\ 
$[\ion{N}{iii}]$~1753.99~\AA\ & 5.6e-3 (8.8e-3) & 2.4e-6 (2.7e-6) & 3.4e-3 (5.5e-7) & 4.4e-8 (4.8e-6) & 8.9e-3 (8.9e-3) \\ 
$\ion{C}{ii}$~1761.0~\AA\** & 1.1e-2 (1.8e-2) & 1.2e-1 (3.5e-1) & 1.3e-2 (3.e-3) & 8.3e-6 (9.1e-4) & 1.4e-1 (3.7e-1) \\ 
$\ion{C}{iii}]$~1909.0~\AA\ & 2.4e-1 (3.9e-1) & 2.5e-1 (9.e-2) & 1.5e-1 (4.9e-3) & 1.2e-5 (1.3e-3) & 6.5e-1 (4.9e-1) \\ 
$\ion{C}{ii}$~2837.0~\AA\** & 1.8e-3 (2.9e-3) & 2.1e-2 (6.4e-2) & 2.1e-3 (4.9e-4) & 1.4e-6 (1.5e-4) & 2.5e-2 (6.7e-2) \\ 
$\ion{Ne}{ii}$~3218.19~\AA\** & 3.1e-4 (4.9e-4) & 4.7e-3 (3.7e-3) & 3.2e-4 (----) & 5.9e-8 (6.5e-6) & 5.3e-3 (4.2e-3) \\ 
$\ion{Ne}{ii}$~3244.09~\AA\** & 2.1e-4 (3.3e-4) & 3.2e-3 (2.5e-3) & 2.2e-4 (----) & 4.0e-8 (4.4e-6) & 3.6e-3 (2.9e-3) \\ 
$\ion{Ne}{ii}$~3334.83~\AA\** & 5.e-4 (8.e-4) & 7.1e-3 (5.6e-3) & 5.2e-4 (----) & 9.4e-8 (1.0e-5) & 8.1e-3 (6.4e-3) \\ 
$\ion{Ne}{ii}$~3355.01~\AA\** & 2.6e-4 (4.2e-4) & 3.7e-3 (2.9e-3) & 2.7e-4 (----) & 4.9e-8 (5.4e-6) & 4.2e-3 (3.3e-3) \\ 
$\ion{Ne}{ii}$~3360.59~\AA\** & 1.0e-4 (1.6e-4) & 1.5e-3 (1.2e-3) & 1.1e-4 (----) & 1.9e-8 (2.1e-6) & 1.7e-3 (1.3e-3) \\ 
$\ion{Ne}{ii}$~3367.21~\AA\** & 2.4e-4 (3.8e-4) & 3.6e-3 (2.8e-3) & 2.5e-4 (----) & 4.5e-8 (5.e-6) & 4.1e-3 (3.2e-3) \\ 
$\ion{Ne}{ii}$~3388.41~\AA\** & 1.6e-4 (2.6e-4) & 2.5e-3 (2.e-3) & 1.7e-4 (----) & 3.1e-8 (3.5e-6) & 2.8e-3 (2.2e-3) \\ 
$\ion{Ne}{ii}$~3694.21~\AA\** & 2.5e-4 (4.e-4) & 3.1e-3 (2.5e-3) & 2.6e-4 (----) & 4.6e-8 (5.1e-6) & 3.7e-3 (2.9e-3) \\ 
$\ion{Ne}{ii}$~3709.62~\AA\** & 9.8e-5 (1.6e-4) & 1.2e-3 (9.9e-4) & 1.0e-4 (----) & 1.8e-8 (2.e-6) & 1.4e-3 (1.1e-3) \\ 
$[\ion{O}{ii}]$~3726.03~\AA\ & 1.6e-1 (2.6e-1) & ---- (----) & 3.3e-1 (2.3e-1) & 4.4e-4 (4.9e-2) & 4.9e-1 (5.4e-1) \\ 
$[\ion{O}{ii}]$~3728.81~\AA\ & 9.9e-2 (1.6e-1) & ---- (----) & 2.e-1 (1.3e-1) & 2.8e-4 (3.1e-2) & 3.e-1 (3.2e-1) \\ 
$\ion{H}{i}$~3770.63~\AA\ & 1.9e-2 (3.e-2) & 1.7e-3 (1.2e-3) & 1.9e-2 (6.5e-3) & 2.1e-5 (2.4e-3) & 4.e-2 (4.e-2) \\ 
$\ion{Ne}{ii}$~3777.13~\AA\** & 9.6e-5 (1.5e-4) & 1.2e-3 (9.7e-4) & 9.9e-5 (----) & 1.8e-8 (2.e-6) & 1.4e-3 (1.1e-3) \\ 
$\ion{O}{ii}$~3856.13~\AA\** & 1.2e-5 (2.e-5) & 2.5e-4 (2.4e-4) & 1.3e-5 (----) & ---- (7.9e-7) & 2.7e-4 (2.6e-4) \\ 
$[\ion{Ne}{iii}]$~3868.76~\AA\ & 4.4e-1 (7.e-1) & 1.7e-4 (9.1e-5) & 3.5e-1 (----) & 1.8e-5 (2.e-3) & 7.9e-1 (7.0e-1) \\ 
$\ion{Ne}{iii}$~3868.8~\AA\** & 5.4e-1 (8.6e-1) & ---- (----) & 4.3e-1 (----) & 2.2e-5 (2.4e-3) & 9.6e-1 (8.6e-1) \\ 
$\ion{O}{ii}$~3882.19~\AA\** & 7.0e-5 (1.1e-4) & 1.4e-3 (1.3e-3) & 7.2e-5 (----) & 4.0e-8 (4.4e-6) & 1.5e-3 (1.4e-3) \\ 
$\ion{He}{i}$~3888.63~\AA\ & 5.2e-2 (8.2e-2) & 6.8e-3 (5.8e-3) & 5.8e-2 (1.6e-2) & 6.2e-5 (6.8e-3) & 1.2e-1 (1.1e-1) \\ 
$\ion{O}{ii}$~3907.45~\AA\** & 3.6e-5 (5.7e-5) & 5.7e-4 (5.2e-4) & 3.6e-5 (----) & 2.e-8 (2.2e-6) & 6.4e-4 (5.8e-4) \\ 
$\ion{Ne}{iii}$~3967.5~\AA\** & 1.6e-1 (2.6e-1) & ---- (----) & 1.3e-1 (----) & 6.6e-6 (7.3e-4) & 2.9e-1 (2.6e-1) \\ 
$\ion{N}{ii}$~4041.0~\AA\ & 4.8e-4 (7.7e-4) & 9.7e-3 (2.e-2) & 5.7e-4 (8.0e-7) & 2.6e-7 (2.8e-5) & 1.1e-2 (2.0e-2) \\ 
$\ion{N}{ii}$~4041.31~\AA\** & 1.8e-4 (2.9e-4) & 5.1e-3 (1.0e-2) & 2.2e-4 (3.1e-7) & 1.0e-7 (1.1e-5) & 5.5e-3 (1.1e-2) \\ 
$[\ion{S}{ii}]$~4068.6~\AA\ & 1.0e-2 (1.6e-2) & 1.4e-6 (7.9e-6) & 1.3e-2 (5.3e-3) & 4.4e-5 (4.9e-3) & 2.3e-2 (2.7e-2) \\ 
$\ion{O}{ii}$~4069.88~\AA\** & 4.9e-4 (7.8e-4) & 7.2e-3 (6.7e-3) & 5.e-4 (----) & 2.7e-7 (3.e-5) & 8.2e-3 (7.5e-3) \\ 
$\ion{O}{ii}$~4072.15~\AA\** & 6.1e-4 (9.7e-4) & 9.7e-3 (9.0e-3) & 6.2e-4 (----) & 3.4e-7 (3.7e-5) & 1.1e-2 (1.0e-2) \\ 
$[\ion{S}{ii}]$~4076.35~\AA\ & 3.3e-3 (5.3e-3) & 2.1e-7 (1.2e-6) & 4.2e-3 (1.7e-3) & 1.4e-5 (1.6e-3) & 7.5e-3 (8.5e-3) \\ 
$\ion{O}{ii}$~4078.84~\AA\** & 1.3e-4 (2.0e-4) & 1.9e-3 (1.8e-3) & 1.3e-4 (----) & 7.0e-8 (7.7e-6) & 2.2e-3 (2.e-3) \\ 
$\ion{O}{ii}$~4085.11~\AA\** & 1.4e-4 (2.2e-4) & 2.1e-3 (1.9e-3) & 1.4e-4 (----) & 7.6e-8 (8.4e-6) & 2.3e-3 (2.1e-3) \\ 
$\ion{O}{ii}$~4087.15~\AA\** & 1.0e-4 (1.6e-4) & 2.1e-3 (1.9e-3) & 1.0e-4 (----) & 6.e-8 (6.6e-6) & 2.3e-3 (2.1e-3) \\ 
$\ion{O}{ii}$~4089.29~\AA\** & 2.e-4 (3.1e-4) & 5.6e-3 (5.4e-3) & 2.0e-4 (----) & 1.2e-7 (1.3e-5) & 6.0e-3 (5.8e-3) \\ 
$\ion{O}{ii}$~4092.93~\AA\** & 8.3e-5 (1.3e-4) & 1.3e-3 (1.2e-3) & 8.5e-5 (----) & 4.6e-8 (5.1e-6) & 1.5e-3 (1.4e-3) \\ 
$\ion{O}{ii}$~4103.0~\AA\** & 2.4e-5 (3.9e-5) & 4.9e-4 (4.6e-4) & 2.5e-5 (----) & 1.4e-8 (1.5e-6) & 5.3e-4 (5.0e-4) \\ 
$\ion{O}{ii}$~4104.99~\AA\** & 1.0e-4 (1.7e-4) & 2.2e-3 (2.1e-3) & 1.1e-4 (----) & 6.0e-8 (6.7e-6) & 2.4e-3 (2.3e-3) \\ 
$\ion{O}{ii}$~4110.79~\AA\** & 5.2e-5 (8.2e-5) & 1.0e-3 (9.8e-4) & 5.3e-5 (----) & 3.e-8 (3.3e-6) & 1.1e-3 (1.1e-3) \\ 
$\ion{O}{ii}$~4119.22~\AA\** & 1.8e-4 (2.9e-4) & 3.5e-3 (3.3e-3) & 1.9e-4 (----) & 1.1e-7 (1.2e-5) & 3.9e-3 (3.6e-3) \\ 
$\ion{O}{ii}$~4120.28~\AA\** & 1.7e-5 (2.8e-5) & 3.1e-4 (2.9e-4) & 1.8e-5 (----) & 9.8e-9 (1.1e-6) & 3.5e-4 (3.2e-4) \\ 
$\ion{O}{ii}$~4120.55~\AA\** & 4.0e-5 (6.5e-5) & 8.6e-4 (8.1e-4) & 4.1e-5 (----) & 2.3e-8 (2.6e-6) & 9.4e-4 (8.8e-4) \\ 
$\ion{O}{ii}$~4121.46~\AA\** & 7.6e-5 (1.2e-4) & 1.3e-3 (1.3e-3) & 7.8e-5 (----) & 4.3e-8 (4.7e-6) & 1.5e-3 (1.4e-3) \\ 
$\ion{O}{ii}$~4132.8~\AA\** & 1.7e-4 (2.6e-4) & 2.7e-3 (2.5e-3) & 1.7e-4 (----) & 9.3e-8 (1.0e-5) & 3.1e-3 (2.8e-3) \\ 
$\ion{O}{ii}$~4140.7~\AA\** & 6.1e-6 (9.7e-6) & 1.0e-4 (9.3e-5) & 6.2e-6 (----) & ---- (3.8e-7) & 1.1e-4 (1.0e-4) \\ 
$\ion{O}{ii}$~4153.3~\AA\** & 2.6e-4 (4.1e-4) & 4.1e-3 (3.8e-3) & 2.7e-4 (----) & 1.4e-7 (1.6e-5) & 4.7e-3 (4.2e-3) \\ 
$\ion{O}{ii}$~4156.53~\AA\** & 3.2e-5 (5.2e-5) & 5.4e-4 (5.e-4) & 3.3e-5 (----) & 1.8e-8 (2.0e-6) & 6.0e-4 (5.5e-4) \\ 
$\ion{O}{ii}$~4169.22~\AA\** & 7.5e-5 (1.2e-4) & 1.2e-3 (1.1e-3) & 7.7e-5 (----) & 4.2e-8 (4.6e-6) & 1.3e-3 (1.2e-3) \\ 
$\ion{C}{iii}$~4187.0~\AA\ & 1.3e-4 (2.2e-4) & 2.8e-3 (1.0e-3) & 1.1e-5 (----) & ---- (2.2e-8) & 3.e-3 (1.2e-3) \\ 
$\ion{O}{ii}$~4189.79~\AA\** & 2.1e-4 (3.3e-4) & 7.2e-5 (6.8e-5) & 2.1e-4 (----) & 1.0e-7 (1.1e-5) & 4.9e-4 (4.1e-4) \\ 
$\ion{C}{ii}$~4267.0~\AA\ & 1.3e-3 (2.1e-3) & 2.7e-2 (8.4e-2) & 1.5e-3 (3.8e-4) & 1.1e-6 (1.2e-4) & 3.0e-2 (8.7e-2) \\ 
$\ion{O}{ii}$~4317.14~\AA\** & 1.4e-4 (2.3e-4) & 2.3e-3 (2.1e-3) & 1.5e-4 (----) & 7.8e-8 (8.6e-6) & 2.6e-3 (2.4e-3) \\ 
$\ion{O}{ii}$~4319.63~\AA\** & 8.3e-5 (1.3e-4) & 1.3e-3 (1.3e-3) & 8.4e-5 (----) & 4.5e-8 (5.e-6) & 1.5e-3 (1.4e-3) \\ 
$\ion{O}{ii}$~4325.76~\AA\** & 3.2e-5 (5.1e-5) & 5.1e-4 (4.7e-4) & 3.3e-5 (----) & 1.8e-8 (1.9e-6) & 5.7e-4 (5.3e-4) \\ 
$\ion{O}{ii}$~4336.86~\AA\** & 5.9e-5 (9.5e-5) & 9.5e-4 (8.8e-4) & 6.0e-5 (----) & 3.2e-8 (3.6e-6) & 1.1e-3 (9.8e-4) \\ 
$\ion{H}{i}$~4340.46~\AA\ & 2.2e-1 (3.5e-1) & 2.0e-2 (7.6e-3) & 2.3e-1 (7.6e-2) & 2.5e-4 (2.8e-2) & 4.7e-1 (4.6e-1) \\ 
$\ion{O}{ii}$~4345.56~\AA\** & 1.8e-4 (2.8e-4) & 2.8e-3 (2.6e-3) & 1.8e-4 (----) & 9.8e-8 (1.1e-5) & 3.2e-3 (2.9e-3) \\ 
$\ion{O}{ii}$~4349.43~\AA\** & 2.3e-4 (3.6e-4) & 3.7e-3 (3.5e-3) & 2.3e-4 (----) & 1.3e-7 (1.4e-5) & 4.2e-3 (3.9e-3) \\ 
$[\ion{O}{iii}]$~4363.0~\AA\*** & 3.e-2 (4.7e-2) & 2.7e-3 (1.2e-3) & 1.9e-2 (----) & 6.6e-7 (7.3e-5) & 5.1e-2 (4.9e-2) \\ 
$\ion{O}{ii}$~4366.89~\AA\** & 1.5e-4 (2.4e-4) & 2.4e-3 (2.2e-3) & 1.5e-4 (----) & 8.2e-8 (9.0e-6) & 2.7e-3 (2.5e-3) \\ 
$[\ion{N}{iii}]$~4379.0~\AA\ & 7.5e-4 (1.2e-3) & 2.1e-2 (8.e-3) & 9.2e-5 (----) & ---- (1.5e-7) & 2.2e-2 (9.2e-3) \\ 
$\ion{O}{ii}$~4414.9~\AA\** & 1.2e-4 (1.9e-4) & 1.5e-3 (1.4e-3) & 1.2e-4 (----) & 6.1e-8 (6.7e-6) & 1.7e-3 (1.6e-3) \\ 
$\ion{O}{ii}$~4416.97~\AA\** & 1.1e-4 (1.8e-4) & 1.1e-3 (1.0e-3) & 1.1e-4 (----) & 5.6e-8 (6.1e-6) & 1.3e-3 (1.2e-3) \\ 
$\ion{He}{i}$~4471.49~\AA\ & 2.2e-2 (3.4e-2) & 3.5e-3 (3.e-3) & 2.5e-2 (7.3e-3) & 2.9e-5 (3.2e-3) & 5.e-2 (4.8e-2) \\ 
$\ion{O}{ii}$~4590.97~\AA\** & 2.1e-4 (3.4e-4) & 1.1e-4 (1.1e-4) & 2.1e-4 (----) & 1.0e-7 (1.1e-5) & 5.4e-4 (4.5e-4) \\ 
$\ion{N}{ii}$~4607.16~\AA\** & 7.3e-5 (1.2e-4) & 9.3e-4 (1.9e-3) & 8.5e-5 (1.1e-7) & 3.4e-8 (3.7e-6) & 1.1e-3 (2.0e-3) \\ 
$\ion{N}{ii}$~4613.87~\AA\** & 4.6e-5 (7.4e-5) & 5.9e-4 (1.2e-3) & 5.4e-5 (7.0e-8) & 2.2e-8 (2.4e-6) & 6.9e-4 (1.3e-3) \\ 
$\ion{N}{ii}$~4621.39~\AA\** & 7.7e-5 (1.2e-4) & 9.5e-4 (1.9e-3) & 9.e-5 (1.2e-7) & 3.6e-8 (4.e-6) & 1.1e-3 (2.1e-3) \\ 
$\ion{N}{ii}$~4630.54~\AA\** & 2.8e-4 (4.5e-4) & 3.8e-3 (7.8e-3) & 3.3e-4 (4.3e-7) & 1.3e-7 (1.5e-5) & 4.4e-3 (8.3e-3) \\ 
$\ion{O}{ii}$~4638.86~\AA\** & 3.2e-4 (5.1e-4) & 4.3e-3 (4.e-3) & 3.2e-4 (----) & 1.7e-7 (1.9e-5) & 5.e-3 (4.5e-3) \\ 
$\ion{O}{ii}$~4641.81~\AA\** & 6.4e-4 (1.0e-3) & 9.2e-3 (8.5e-3) & 6.5e-4 (----) & 3.5e-7 (3.8e-5) & 1.0e-2 (9.6e-3) \\ 
$\ion{N}{ii}$~4643.09~\AA\** & 1.e-4 (1.6e-4) & 1.3e-3 (2.6e-3) & 1.2e-4 (1.5e-7) & 4.7e-8 (5.1e-6) & 1.5e-3 (2.7e-3) \\ 
$\ion{O}{ii}$~4649.13~\AA\** & 7.7e-4 (1.2e-3) & 1.3e-2 (1.3e-2) & 7.9e-4 (----) & 4.3e-7 (4.8e-5) & 1.5e-2 (1.4e-2) \\ 
$\ion{C}{iii}$~4650.25~\AA\ & 1.8e-4 (2.8e-4) & 7.6e-5 (3.4e-5) & 7.6e-7 (----) & ---- (9.2e-7) & 2.5e-4 (3.2e-4) \\ 
$\ion{O}{ii}$~4650.84~\AA\** & 3.4e-4 (5.4e-4) & 4.6e-3 (4.3e-3) & 3.4e-4 (----) & 1.8e-7 (2.e-5) & 5.3e-3 (4.8e-3) \\ 
$\ion{O}{ii}$~4661.63~\AA\** & 3.6e-4 (5.7e-4) & 4.8e-3 (4.5e-3) & 3.6e-4 (----) & 1.9e-7 (2.1e-5) & 5.6e-3 (5.1e-3) \\ 
$\ion{O}{ii}$~4673.73~\AA\** & 5.9e-5 (9.4e-5) & 8.1e-4 (7.4e-4) & 6.e-5 (----) & 3.2e-8 (3.5e-6) & 9.3e-4 (8.4e-4) \\ 
$\ion{O}{ii}$~4676.23~\AA\** & 2.2e-4 (3.5e-4) & 3.2e-3 (2.9e-3) & 2.3e-4 (----) & 1.2e-7 (1.3e-5) & 3.6e-3 (3.3e-3) \\ 
$\ion{He}{ii}$~4685.64~\AA\ & 5.3e-2 (8.4e-2) & 1.5e-2 (1.6e-3) & 8.1e-5 (----) & 1.9e-8 (2.1e-6) & 6.7e-2 (8.5e-2) \\ 
$\ion{O}{ii}$~4696.35~\AA\** & 2.7e-5 (4.3e-5) & 3.6e-4 (3.3e-4) & 2.7e-5 (----) & 1.4e-8 (1.6e-6) & 4.2e-4 (3.8e-4) \\ 
$\ion{O}{ii}$~4699.22~\AA\** & 1.7e-5 (2.7e-5) & 3.e-4 (2.8e-4) & 1.7e-5 (----) & 9.5e-9 (1.0e-6) & 3.3e-4 (3.1e-4) \\ 
$\ion{O}{ii}$~4705.35~\AA\** & 1.7e-5 (2.7e-5) & 3.7e-4 (3.5e-4) & 1.7e-5 (----) & 1.e-8 (1.1e-6) & 4.0e-4 (3.8e-4) \\ 
$[\ion{Ar}{iv}]$~4711.26~\AA\ & 1.5e-2 (2.4e-2) & ---- (----) & 5.5e-3 (----) & 2.e-7 (2.2e-5) & 2.1e-2 (2.4e-2) \\ 
$[\ion{Ar}{iv}]$~4740.12~\AA\ & 1.1e-2 (1.8e-2) & ---- (----) & 4.2e-3 (----) & 1.5e-7 (1.6e-5) & 1.5e-2 (1.8e-2) \\ 
$\ion{N}{ii}$~4779.72~\AA\** & 5.2e-5 (8.2e-5) & 8.3e-4 (1.7e-3) & 6.1e-5 (8.2e-8) & 2.6e-8 (2.9e-6) & 9.5e-4 (1.8e-3) \\ 
$\ion{N}{ii}$~4788.13~\AA\** & 6.9e-5 (1.1e-4) & 1.2e-3 (2.4e-3) & 8.2e-5 (1.1e-7) & 3.5e-8 (3.9e-6) & 1.3e-3 (2.5e-3) \\ 
$\ion{N}{ii}$~4803.29~\AA\** & 1.3e-4 (2.1e-4) & 2.3e-3 (4.7e-3) & 1.6e-4 (2.1e-7) & 6.7e-8 (7.4e-6) & 2.6e-3 (5.e-3) \\ 
$\ion{H}{i}$~4861.33~\AA\ & 4.7e-1 (7.4e-1) & 4.8e-2 (3.4e-2) & 4.9e-1 (1.6e-1) & 5.4e-4 (6.e-2) & 1.0e+0 (1.0e+0) \\ 
$\ion{O}{ii}$~4890.86~\AA\** & 2.6e-5 (4.2e-5) & 4.6e-4 (4.3e-4) & 2.7e-5 (----) & 1.5e-8 (1.6e-6) & 5.1e-4 (4.7e-4) \\ 
$\ion{O}{ii}$~4924.53~\AA\** & 1.4e-4 (2.2e-4) & 2.1e-3 (2.e-3) & 1.4e-4 (----) & 7.5e-8 (8.3e-6) & 2.4e-3 (2.2e-3) \\ 
$\ion{O}{ii}$~4943.0~\AA\** & 8.6e-6 (1.4e-5) & 2.e-4 (1.9e-4) & 8.9e-6 (----) & ---- (5.6e-7) & 2.2e-4 (2.e-4) \\ 
$[\ion{O}{iii}]$~4958.91~\AA\ & 1.8e+0 (2.8e+0) & 9.7e-4 (5.e-4) & 1.5e+0 (----) & 1.9e-4 (2.1e-2) & 3.2e+0 (2.8e+0) \\ 
$[\ion{O}{iii}]$~5006.84~\AA\ & 5.3e+0 (8.4e+0) & 2.9e-3 (1.5e-3) & 4.4e+0 (----) & 5.6e-4 (6.2e-2) & 9.6e+0 (8.5e+0) \\ 
$[\ion{Ar}{iii}]$~5191.82~\AA\ & 6.e-4 (9.5e-4) & ---- (----) & 5.8e-4 (5.1e-5) & 1.7e-7 (1.9e-5) & 1.2e-3 (1.0e-3) \\ 
$[\ion{N}{i}]$~5197.9~\AA\ & 2.3e-4 (3.7e-4) & 1.8e-6 (1.2e-5) & 7.1e-4 (1.2e-3) & 5.1e-5 (5.6e-3) & 9.9e-4 (7.2e-3) \\ 
$[\ion{N}{i}]$~5200.26~\AA\ & 1.6e-4 (2.6e-4) & 2.5e-7 (3.2e-6) & 4.9e-4 (9.4e-4) & 4.3e-5 (4.8e-3) & 7.e-4 (6.e-3) \\ 
$[\ion{Cl}{iii}]$~5517.71~\AA\ & 4.2e-3 (6.6e-3) & ---- (----) & 4.6e-3 (7.4e-4) & 1.5e-6 (1.6e-4) & 8.8e-3 (7.5e-3) \\ 
$[\ion{Cl}{iii}]$~5537.87~\AA\ & 4.7e-3 (7.6e-3) & ---- (----) & 5.3e-3 (8.6e-4) & 1.6e-6 (1.8e-4) & 1.0e-2 (8.6e-3) \\ 
$\ion{N}{ii}$~5666.63~\AA\** & 2.6e-4 (4.2e-4) & 3.7e-3 (7.6e-3) & 3.1e-4 (4.1e-7) & 1.3e-7 (1.4e-5) & 4.3e-3 (8.0e-3) \\ 
$\ion{N}{ii}$~5676.02~\AA\** & 1.3e-4 (2.1e-4) & 1.7e-3 (3.5e-3) & 1.5e-4 (2.0e-7) & 6.2e-8 (6.8e-6) & 2.0e-3 (3.7e-3) \\ 
$\ion{N}{ii}$~5679.0~\AA\ & 4.4e-4 (7.1e-4) & 5.3e-3 (1.1e-2) & 5.2e-4 (6.9e-7) & 2.2e-7 (2.4e-5) & 6.2e-3 (1.1e-2) \\ 
$\ion{N}{ii}$~5679.56~\AA\** & 4.7e-4 (7.5e-4) & 7.7e-3 (1.6e-2) & 5.5e-4 (7.4e-7) & 2.3e-7 (2.5e-5) & 8.7e-3 (1.7e-2) \\ 
$\ion{N}{ii}$~5686.21~\AA\** & 7.8e-5 (1.2e-4) & 1.0e-3 (2.1e-3) & 9.1e-5 (1.2e-7) & 3.7e-8 (4.1e-6) & 1.2e-3 (2.2e-3) \\ 
$\ion{N}{ii}$~5710.77~\AA\** & 9.4e-5 (1.5e-4) & 1.3e-3 (2.7e-3) & 1.1e-4 (1.5e-7) & 4.5e-8 (5.e-6) & 1.5e-3 (2.9e-3) \\ 
$[\ion{N}{ii}]$~5755.0~\AA\ & 6.9e-3 (1.1e-2) & 2.4e-3 (4.7e-3) & 1.2e-2 (7.4e-3) & 1.9e-5 (2.1e-3) & 2.1e-2 (2.5e-2) \\ 
$\ion{He}{i}$~5875.64~\AA\ & 6.2e-2 (9.9e-2) & 1.0e-2 (8.9e-3) & 7.2e-2 (2.1e-2) & 8.4e-5 (9.3e-3) & 1.4e-1 (1.4e-1) \\ 
$\ion{N}{ii}$~5927.81~\AA\** & 5.3e-5 (8.4e-5) & 8.5e-4 (1.7e-3) & 6.2e-5 (8.4e-8) & 2.7e-8 (2.9e-6) & 9.7e-4 (1.8e-3) \\ 
$\ion{N}{ii}$~5931.78~\AA\** & 9.4e-5 (1.5e-4) & 1.6e-3 (3.2e-3) & 1.1e-4 (1.5e-7) & 4.8e-8 (5.3e-6) & 1.8e-3 (3.4e-3) \\ 
$\ion{N}{ii}$~5941.65~\AA\** & 1.8e-4 (2.9e-4) & 3.2e-3 (6.6e-3) & 2.2e-4 (3.e-7) & 9.4e-8 (1.0e-5) & 3.6e-3 (6.9e-3) \\ 
$\ion{N}{ii}$~5952.39~\AA\** & 2.8e-5 (4.4e-5) & 4.7e-4 (9.5e-4) & 3.3e-5 (4.5e-8) & 1.4e-8 (1.5e-6) & 5.3e-4 (1.e-3) \\ 
$[\ion{O}{i}]$~6300.3~\AA\ & 8.3e-4 (1.3e-3) & ---- (1.7e-7) & 2.9e-3 (2.7e-3) & 8.7e-5 (9.5e-3) & 3.8e-3 (1.4e-2) \\ 
$[\ion{S}{iii}]$~6312.06~\AA\ & 1.6e-2 (2.5e-2) & 4.2e-7 (1.7e-6) & 1.5e-2 (1.8e-3) & 2.6e-6 (2.9e-4) & 3.1e-2 (2.7e-2) \\ 
$[\ion{N}{ii}]$~6548.05~\AA\ & 1.5e-1 (2.4e-1) & 2.0e-5 (5.1e-5) & 3.e-1 (4.e-1) & 8.9e-4 (9.8e-2) & 4.5e-1 (7.4e-1) \\ 
$\ion{H}{i}$~6562.81~\AA\ & 1.3e+0 (2.1e+0) & 1.8e-1 (6.6e-2) & 1.4e+0 (4.8e-1) & 1.6e-3 (1.8e-1) & 2.9e+0 (2.8e+0) \\ 
$\ion{C}{ii}$~6580.0~\AA\ & 1.1e-4 (1.8e-4) & 7.6e-4 (2.2e-3) & 1.3e-4 (2.7e-5) & 7.5e-8 (8.2e-6) & 1.0e-3 (2.4e-3) \\ 
$[\ion{N}{ii}]$~6583.45~\AA\ & 4.5e-1 (7.1e-1) & 6.e-5 (1.5e-4) & 8.8e-1 (1.2e+0) & 2.6e-3 (2.9e-1) & 1.3e+0 (2.2e+0) \\ 
$\ion{He}{i}$~6678.15~\AA\ & 1.8e-2 (2.8e-2) & 3.e-3 (2.5e-3) & 2.0e-2 (6.e-3) & 2.4e-5 (2.6e-3) & 4.1e-2 (3.9e-2) \\ 
$[\ion{S}{ii}]$~6716.44~\AA\ & 3.3e-2 (5.3e-2) & 2.5e-6 (9.9e-6) & 4.4e-2 (2.6e-2) & 2.5e-4 (2.7e-2) & 7.7e-2 (1.1e-1) \\ 
$[\ion{S}{ii}]$~6730.82~\AA\ & 5.3e-2 (8.5e-2) & 4.9e-6 (2.1e-5) & 7.1e-2 (4.1e-2) & 3.7e-4 (4.0e-2) & 1.2e-1 (1.7e-1) \\ 
$\ion{He}{i}$~7065.22~\AA\ & 1.4e-2 (2.3e-2) & 1.3e-3 (1.2e-3) & 1.6e-2 (3.8e-3) & 1.4e-5 (1.5e-3) & 3.2e-2 (2.9e-2) \\ 
$[\ion{Ar}{iii}]$~7135.79~\AA\ & 8.6e-2 (1.4e-1) & ---- (----) & 9.8e-2 (1.8e-2) & 5.7e-5 (6.2e-3) & 1.8e-1 (1.6e-1) \\ 
$\ion{C}{ii}$~7231.0~\AA\** & 1.8e-5 (2.8e-5) & 2.8e-4 (8.7e-4) & 2.1e-5 (5.0e-6) & 1.4e-8 (1.6e-6) & 3.2e-4 (9.0e-4) \\ 
$[\ion{O}{ii}]$~7332.0~\AA\ & 6.9e-3 (1.1e-2) & 2.5e-3 (2.3e-3) & 1.2e-2 (4.8e-3) & 1.0e-5 (1.1e-3) & 2.2e-2 (1.9e-2) \\ 
$[\ion{S}{iii}]$~9068.62~\AA\ & 2.9e-1 (4.6e-1) & 1.3e-6 (4.0e-6) & 3.2e-1 (7.1e-2) & 1.2e-4 (1.3e-2) & 6.1e-1 (5.5e-1) \\ 
$[\ion{S}{iii}]$~9530.62~\AA\ & 7.3e-1 (1.2e+0) & 3.3e-6 (1.0e-5) & 8.1e-1 (1.8e-1) & 3.0e-4 (3.4e-2) & 1.5e+0 (1.4e+0) \\ 
$[\ion{C}{i}]$~9850.26~\AA\ & 3.5e-4 (5.5e-4) & ---- (2.5e-8) & 5.5e-4 (4.3e-4) & 1.5e-5 (1.7e-3) & 9.1e-4 (2.7e-3) \\ 
$\ion{He}{i}$~10830.3~\AA\ & 3.6e-1 (5.7e-1) & 3.1e-2 (3.e-2) & 3.9e-1 (9.0e-2) & 3.2e-4 (3.5e-2) & 7.8e-1 (7.2e-1) \\ 
$[\ion{Ar}{iii}]$~9.0~$\mu$m & 8.5e-2 (1.4e-1) & 4.3e-2 (1.0e-1) & 1.1e-1 (3.1e-2) & 1.1e-4 (1.2e-2) & 2.4e-1 (2.8e-1) \\ 
$[\ion{S}{iv}]$~10.5~$\mu$m & 1.0e+0 (1.6e+0) & 7.9e-1 (3.1e-1) & 5.8e-1 (1.2e-6) & 2.6e-5 (2.8e-3) & 2.4e+0 (1.9e+0) \\ 
$[\ion{N}{ii}]$~12.2~$\mu$m & 1.9e-3 (3.1e-3) & 7.5e-4 (1.5e-2) & 4.1e-3 (6.6e-3) & 2.5e-5 (2.8e-3) & 6.9e-3 (2.8e-2) \\ 
$[\ion{Ne}{ii}]$~12.8~$\mu$m & 2.1e-2 (3.3e-2) & 4.4e-2 (7.2e-1) & 5.8e-2 (1.8e-1) & 5.8e-4 (6.4e-2) & 1.2e-1 (1.0e+0) \\ 
$[\ion{Ne}{iii}]$~15.6~$\mu$m & 8.3e-1 (1.3e+0) & 1.1e+0 (8.1e-1) & 8.3e-1 (----) & 1.4e-4 (1.5e-2) & 2.8e+0 (2.1e+0) \\ 
$[\ion{S}{iii}]$~18.7~$\mu$m & 3.6e-1 (5.7e-1) & 2.9e-1 (5.5e-1) & 4.3e-1 (1.3e-1) & 2.8e-4 (3.0e-2) & 1.1e+0 (1.3e+0) \\ 
$[\ion{Ar}{iii}]$~21.8~$\mu$m & 5.4e-3 (8.7e-3) & 8.4e-4 (1.5e-3) & 6.8e-3 (1.9e-3) & 6.8e-6 (7.5e-4) & 1.3e-2 (1.3e-2) \\ 
$[\ion{O}{iv}]$~25.9~$\mu$m & 4.9e-1 (7.8e-1) & 7.5e-1 (2.6e-1) & 2.7e-4 (----) & ---- (8.7e-7) & 1.2e+0 (1.0e+0) \\ 
$[\ion{S}{iii}]$~33.5~$\mu$m & 1.4e-1 (2.3e-1) & 1.4e-1 (2.5e-1) & 1.7e-1 (4.2e-2) & 1.1e-4 (1.2e-2) & 4.5e-1 (5.4e-1) \\ 
$[\ion{Ne}{iii}]$~36.0~$\mu$m & 7.3e-2 (1.2e-1) & 5.1e-2 (3.6e-2) & 7.4e-2 (----) & 1.2e-5 (1.3e-3) & 2.e-1 (1.5e-1) \\ 
$[\ion{O}{iii}]$~51.8~$\mu$m & 7.9e-1 (1.3e+0) & 5.9e-1 (3.1e-1) & 6.1e-1 (----) & 3.4e-4 (3.7e-2) & 2.e+0 (1.6e+0) \\ 
$[\ion{N}{iii}]$~57.3~$\mu$m & 3.5e-1 (5.5e-1) & 2.9e-1 (2.5e-1) & 3.2e-1 (3.7e-4) & 1.3e-4 (1.4e-2) & 9.6e-1 (8.1e-1) \\ 
$[\ion{O}{iii}]$~88.3~$\mu$m & 1.7e-1 (2.7e-1) & 1.2e-1 (6.e-2) & 1.2e-1 (----) & 6.9e-5 (7.6e-3) & 4.1e-1 (3.4e-1) \\ 
\hline

%% file: table2_s3.tex
<T$_e$> [K] & 9653  & 2826    & 6913 & 6539 \\ 
<n$_e$> [cm$^{{-3}}$] & 2767  & 15570    & 2412  & 2670  \\ 
12+log(O/H) & 8.75  & 10.85 & 8.75  & 8.75 \\ 
$\Omega$/4$\pi$* & 0.50 & 0.50 & 0.50  & 0.50 \\ 
H$^+$/H & 0.99  & 1.00 & 0.96  & 0.97  \\ 
O$^+$/O & 0.08  & 0.36  & 0.96 & 0.66 \\ 
O$^{{++}}$/O & 0.85  & 0.64  & 0.00  & 0.31  \\ 
O$^{{+3}}$/O & 0.07  & 0.01 & 0.00& 0.00  \\ 
O mass (10$^{{-4}}$) [M$_\odot$] & 6.6  & 7.5  & 5.8 & 1.2  \\ 
Volume [10$^{{50}}$cm$^{{-3}}$] & 345  & 3.2  & 306.3  & 64.2 \\ 
\% mass & 44.2  & 8.3   & 39.3 & 8.2 \\ 
\hline

%% file: paper.bbl
\begin{thebibliography}{}
\makeatletter
\relax
\def\mn@urlcharsother{\let\do\@makeother \do\$\do\&\do\#\do\^\do\_\do\%\do\~}
\def\mn@doi{\begingroup\mn@urlcharsother \@ifnextchar [ {\mn@doi@}
  {\mn@doi@[]}}
\def\mn@doi@[#1]#2{\def\@tempa{#1}\ifx\@tempa\@empty \href
  {http://dx.doi.org/#2} {doi:#2}\else \href {http://dx.doi.org/#2} {#1}\fi
  \endgroup}
\def\mn@eprint#1#2{\mn@eprint@#1:#2::\@nil}
\def\mn@eprint@arXiv#1{\href {http://arxiv.org/abs/#1} {{\tt arXiv:#1}}}
\def\mn@eprint@dblp#1{\href {http://dblp.uni-trier.de/rec/bibtex/#1.xml}
  {dblp:#1}}
\def\mn@eprint@#1:#2:#3:#4\@nil{\def\@tempa {#1}\def\@tempb {#2}\def\@tempc
  {#3}\ifx \@tempc \@empty \let \@tempc \@tempb \let \@tempb \@tempa \fi \ifx
  \@tempb \@empty \def\@tempb {arXiv}\fi \@ifundefined
  {mn@eprint@\@tempb}{\@tempb:\@tempc}{\expandafter \expandafter \csname
  mn@eprint@\@tempb\endcsname \expandafter{\@tempc}}}
  
\bibitem[\protect\citeauthoryear{{Bohigas}}{{Bohigas}}{2009}]{2009Bohigas_rmxa45}
{Bohigas} J.,  2009, \rmxaa, 45, 107

\bibitem[\protect\citeauthoryear{{Borkowski}, {Harrington}, {Blair}  \&
  {Bregman}}{{Borkowski} et~al.}{1994}]{1994Borkowski_apj435}
{Borkowski} K.~J.,  {Harrington} J.~P.,  {Blair} W.~P.,   {Bregman} J.~D.,
  1994, \mn@doi [\apj] {10.1086/174849}, 435, 722

\bibitem[\protect\citeauthoryear{{Butler} \& {Zeippen}}{{Butler} \&
  {Zeippen}}{1989}]{1989Butler_aap208}
{Butler} K.,  {Zeippen} C.~J.,  1989, \aap, 208, 337

\bibitem[\protect\citeauthoryear{{Delgado-Inglada}, {Morisset}  \&
  {Stasi{\'n}ska}}{{Delgado-Inglada}
  et~al.}{2014}]{2014Delgado-Inglada_mnra440}
{Delgado-Inglada} G.,  {Morisset} C.,   {Stasi{\'n}ska} G.,  2014, \mnras, 440,
  536

\bibitem[\protect\citeauthoryear{{Ercolano}, {Barlow}, {Storey}, {Liu}, {Rauch}
   \& {Werner}}{{Ercolano} et~al.}{2003}]{2003Ercolano_mnra344}
{Ercolano} B.,  {Barlow} M.~J.,  {Storey} P.~J.,  {Liu} X.,  {Rauch} T.,
  {Werner} K.,  2003, \mnras, 344, 1145

\bibitem[\protect\citeauthoryear{{Ferland} et~al.,}{{Ferland}
  et~al.}{2017}]{2017Ferland_rmxaa53}
{Ferland} G.~J.,  et~al., 2017, \rmxaa, \href
  {http://adsabs.harvard.edu/abs/2017RMxAA..53..385F} {53, 385}

\bibitem[\protect\citeauthoryear{{Froese Fischer} \& {Tachiev}}{{Froese
  Fischer} \& {Tachiev}}{2004}]{2004Froese-Fischer_Atom87}
{Froese Fischer} C.,  {Tachiev} G.,  2004, Atomic Data and Nuclear Data Tables,
  87, 1

\bibitem[\protect\citeauthoryear{{Gaia Collaboration}}{{Gaia
  Collaboration}}{2018}]{2018Gaia_vizier1345}
{Gaia Collaboration} 2018, VizieR Online Data Catalog, \href
  {https://ui.adsabs.harvard.edu/abs/2018yCat.1345....0G} {p. I/345}

\bibitem[\protect\citeauthoryear{{Garc{\'{\i}}a-Rojas}, {Corradi}, {Monteiro},
  {Jones}, {Rodr{\'{\i}}guez-Gil}  \& {Cabrera-Lavers}}{{Garc{\'{\i}}a-Rojas}
  et~al.}{2016}]{2016Garcia-Rojas_apjl824}
{Garc{\'{\i}}a-Rojas} J.,  {Corradi} R.~L.~M.,  {Monteiro} H.,  {Jones} D.,
  {Rodr{\'{\i}}guez-Gil} P.,   {Cabrera-Lavers} A.,  2016, \apjl, 824, L27

\bibitem[\protect\citeauthoryear{{Gesicki}, {Zijlstra}  \&
  {Morisset}}{{Gesicki} et~al.}{2016}]{2016Gesicki_aap585}
{Gesicki} K.,  {Zijlstra} A.~A.,   {Morisset} C.,  2016, \aap, 585, A69

\bibitem[\protect\citeauthoryear{Gomez-Llanos, Morisset, Garcia-Rojas, Jones,
  Wesson, Corradi  \& Boffin}{Gomez-Llanos
  et~al.}{2020}]{2020Gomez-Llanos_arxiv}
Gomez-Llanos V.,  Morisset C.,  Garcia-Rojas J.,  Jones D.,  Wesson R.,
  Corradi R. L.~M.,   Boffin H. M.~J.,  2020, The impact of strong
  recombination on temperature determination in planetary nebulae (\mn@eprint
  {arXiv} {2007.05488})

\bibitem[\protect\citeauthoryear{{Helou} \& {Walker}}{{Helou} \&
  {Walker}}{1988}]{1988iras7H}
{Helou} G.,  {Walker} D.~W.,  1988, in Infrared astronomical satellite (IRAS)
  catalogs and atlases. Volume 7.

\bibitem[\protect\citeauthoryear{{Howarth}}{{Howarth}}{1983}]{1983Howarth_mnra203}
{Howarth} I.~D.,  1983, \mnras, 203, 301

\bibitem[\protect\citeauthoryear{{Jones}, {Wesson}, {Garc{\'\i}a-Rojas},
  {Corradi}  \& {Boffin}}{{Jones} et~al.}{2016}]{2016Jones_mnra455}
{Jones} D.,  {Wesson} R.,  {Garc{\'\i}a-Rojas} J.,  {Corradi} R.~L.~M.,
  {Boffin} H.~M.~J.,  2016, \mn@doi [\mnras] {10.1093/mnras/stv2519}, \href
  {https://ui.adsabs.harvard.edu/abs/2016MNRAS.455.3263J} {455, 3263}

\bibitem[\protect\citeauthoryear{{Kaufman} \& {Sugar}}{{Kaufman} \&
  {Sugar}}{1986}]{1986Kaufman_jpcrd15}
{Kaufman} V.,  {Sugar} J.,  1986, \mn@doi [Journal of Physical and Chemical
  Reference Data] {10.1063/1.555775}, \href
  {http://adsabs.harvard.edu/abs/1986JPCRD..15..321K} {15, 321}

\bibitem[\protect\citeauthoryear{{Kisielius}, {Storey}, {Ferland}  \&
  {Keenan}}{{Kisielius} et~al.}{2009}]{2009Kisielius_mnra397}
{Kisielius} R.,  {Storey} P.~J.,  {Ferland} G.~J.,   {Keenan} F.~P.,  2009,
  \mnras, 397, 903

\bibitem[\protect\citeauthoryear{{Liu}, {Storey}, {Barlow}, {Danziger}, {Cohen}
   \& {Bryce}}{{Liu} et~al.}{2000}]{2000Liu_mnra312}
{Liu} X.-W.,  {Storey} P.~J.,  {Barlow} M.~J.,  {Danziger} I.~J.,  {Cohen} M.,
   {Bryce} M.,  2000, \mnras, 312, 585

\bibitem[\protect\citeauthoryear{{Liu} et~al.,}{{Liu}
  et~al.}{2001a}]{2001Liu_mnra323}
{Liu} X.,  et~al., 2001a, \mnras, 323, 343

\bibitem[\protect\citeauthoryear{{Liu}, {Luo}, {Barlow}, {Danziger}  \&
  {Storey}}{{Liu} et~al.}{2001b}]{2001Liu_mnras327}
{Liu} X.,  {Luo} S.,  {Barlow} M.~J.,  {Danziger} I.~J.,   {Storey} P.~J.,
  2001b, \mnras, 327, 141

\bibitem[\protect\citeauthoryear{{Luridiana}, {Morisset}  \&
  {Shaw}}{{Luridiana} et~al.}{2012}]{2012Luridiana_283}
{Luridiana} V.,  {Morisset} C.,   {Shaw} R.~A.,  2012, in IAU Symposium. pp
  422--423

\bibitem[\protect\citeauthoryear{{Luridiana}, {Morisset}  \&
  {Shaw}}{{Luridiana} et~al.}{2013}]{2013Luridiana_ascl}
{Luridiana} V.,  {Morisset} C.,   {Shaw} R.~A.,  2013, {PyNeb: Analysis of
  emission lines}, Astrophysics Source Code Library (\mn@eprint {ascl}
  {1304.021})

\bibitem[\protect\citeauthoryear{{Mendoza}}{{Mendoza}}{1983}]{1983Mendoza_103}
{Mendoza} C.,  1983, in {Flower} D.~R.,  ed.,  IAU Symposium Vol. 103,
  Planetary Nebulae. pp 143--172

\bibitem[\protect\citeauthoryear{{Mendoza} \& {Zeippen}}{{Mendoza} \&
  {Zeippen}}{1982}]{1982Mendoza_mnra198}
{Mendoza} C.,  {Zeippen} C.~J.,  1982, \mnras, 198, 127

\bibitem[\protect\citeauthoryear{{Morisset}}{{Morisset}}{2013}]{2013Morisset_}
{Morisset} C.,  2013, {pyCloudy: Tools to manage astronomical Cloudy
  photoionization code}

\bibitem[\protect\citeauthoryear{{Morisset} \& {Pequignot}}{{Morisset} \&
  {Pequignot}}{1996}]{1996Morisset_aap312}
{Morisset} C.,  {Pequignot} D.,  1996, \aap, 312, 135

\bibitem[\protect\citeauthoryear{{Nicholls}, {Dopita}  \&
  {Sutherland}}{{Nicholls} et~al.}{2012}]{2012Nicholls_apj752}
{Nicholls} D.~C.,  {Dopita} M.~A.,   {Sutherland} R.~S.,  2012, \apj, 752, 148

\bibitem[\protect\citeauthoryear{{Nussbaumer} \& {Storey}}{{Nussbaumer} \&
  {Storey}}{1984}]{1984Nussbaumer_Astr56}
{Nussbaumer} H.,  {Storey} P.~J.,  1984, Astronomy and Astrophysics Supplement
  Series, \href {https://ui.adsabs.harvard.edu/abs/1984A&AS...56..293N} {56,
  293}

\bibitem[\protect\citeauthoryear{{O'Dell}, {Henney}  \& {Ferland}}{{O'Dell}
  et~al.}{2005}]{2005AJ....130..172O}
{O'Dell} C.~R.,  {Henney} W.~J.,   {Ferland} G.~J.,  2005, \mn@doi [\aj]
  {10.1086/430803}, \href
  {https://ui.adsabs.harvard.edu/abs/2005AJ....130..172O} {130, 172}

\bibitem[\protect\citeauthoryear{{Pe{\~n}a}, {Ruiz-Escobedo},
  {Rechy-Garc{\'{\i}}a}  \& {Garc{\'{\i}}a-Rojas}}{{Pe{\~n}a}
  et~al.}{2017}]{2017Pena_mnras472}
{Pe{\~n}a} M.,  {Ruiz-Escobedo} F.,  {Rechy-Garc{\'{\i}}a} J.~S.,
  {Garc{\'{\i}}a-Rojas} J.,  2017, \mn@doi [\mnras] {10.1093/mnras/stx1991},
  \href {http://adsabs.harvard.edu/abs/2017MNRAS.472.1182P} {472, 1182}
    
    
\bibitem[\protect\citeauthoryear{{Peimbert}}{{Peimbert}}{1967}]{1967Peimbert_apj150}
{Peimbert} M.,  1967, \apj, 150, 825

\bibitem[\protect\citeauthoryear{{Peimbert}}{{Peimbert}}{1971}]{1971Peimbert_Bole6}
{Peimbert} M.,  1971, Boletin de los Observatorios Tonantzintla y Tacubaya, 6,  29
  
\bibitem[\protect\citeauthoryear{{Peimbert}, {Peimbert}  \&
  {Delgado-Inglada}}{{Peimbert} et~al.}{2017}]{2017Peimbert_pasp129}
{Peimbert} M.,  {Peimbert} A.,   {Delgado-Inglada} G.,  2017, \mn@doi [\pasp]
  {10.1088/1538-3873/aa72c3}, \href
  {http://adsabs.harvard.edu/abs/2017PASP..129h2001P} {129, 082001}
  
\bibitem[\protect\citeauthoryear{{Pequignot}, {Petitjean}  \&
  {Boisson}}{{Pequignot} et~al.}{1991}]{1991Pequignot_aap251}
{Pequignot} D.,  {Petitjean} P.,   {Boisson} C.,  1991, \aap, 251, 680

\bibitem[\protect\citeauthoryear{{P{\'e}quignot}, {Amara}, {Liu}, {Barlow},
  {Storey}, {Morisset}, {Torres-Peimbert}  \& {Peimbert}}{{P{\'e}quignot}
  et~al.}{2002}]{2002Pequignot_12}
{P{\'e}quignot} D.,  {Amara} M.,  {Liu} X.,  {Barlow} M.~J.,  {Storey} P.~J.,
  {Morisset} C.,  {Torres-Peimbert} S.,   {Peimbert} M.,  2002, in
  {W.~J.~Henney, J.~Franco, \& M.~Martos} ed.,  Revista Mexicana de Astronomia
  y Astrofisica Conference Series Vol. 12, Revista Mexicana de Astronomia y
  Astrofisica Conference Series. pp 142--143

\bibitem[\protect\citeauthoryear{{P{\'e}quignot}, {Liu}, {Barlow}, {Storey}  \&
  {Morisset}}{{P{\'e}quignot} et~al.}{2003}]{2003Pequignot_209}
{P{\'e}quignot} D.,  {Liu} X.,  {Barlow} M.~J.,  {Storey} P.~J.,   {Morisset}
  C.,  2003, in {S.~Kwok, M.~Dopita, \& R.~Sutherland} ed.,  IAU Symposium Vol.
  209, Planetary Nebulae: Their Evolution and Role in the Universe. p.~347

\bibitem[\protect\citeauthoryear{{Podobedova}, {Kelleher}  \&
  {Wiese}}{{Podobedova} et~al.}{2009}]{2009Podobedova_Jour38}
{Podobedova} L.~I.,  {Kelleher} D.~E.,   {Wiese} W.~L.,  2009, Journal of
  Physical and Chemical Reference Data, 38, 171

\bibitem[\protect\citeauthoryear{{Pottasch} et~al.,}{{Pottasch}
  et~al.}{1984}]{1984Pottasch_apjl278}
{Pottasch} S.~R.,  et~al., 1984, \apjl, 278, L33

\bibitem[\protect\citeauthoryear{{Ramsbottom} \& {Bell}}{{Ramsbottom} \&
  {Bell}}{1997}]{1997Ramsbottom_ADNDT66}
{Ramsbottom} C.~A.,  {Bell} K.~L.,  1997, \mn@doi [Atomic Data and Nuclear Data
  Tables] {10.1006/adnd.1997.0741}, \href
  {http://adsabs.harvard.edu/abs/1997ADNDT..66...65R} {66, 65}
  
\bibitem[\protect\citeauthoryear{{Skrutskie} et~al.,}{{Skrutskie}
  et~al.}{2006}]{2006Skrutskie_aj131}
{Skrutskie} M.~F.,  et~al., 2006, \mn@doi [\aj] {10.1086/498708}, \href
  {https://ui.adsabs.harvard.edu/abs/2006AJ....131.1163S} {131, 1163}

\bibitem[\protect\citeauthoryear{{Smits}}{{Smits}}{1996}]{1996Smits_mnra278}
{Smits} D.~P.,  1996, \mn@doi [\mnras] {10.1093/mnras/278.3.683}, \href
  {http://adsabs.harvard.edu/abs/1996MNRAS.278..683S} {278, 683}

\bibitem[\protect\citeauthoryear{{Storey} \& {Hummer}}{{Storey} \&
  {Hummer}}{1995}]{1995Storey_mnra272}
{Storey} P.~J.,  {Hummer} D.~G.,  1995, \mnras, 272, 41

\bibitem[\protect\citeauthoryear{{Storey} \& {Zeippen}}{{Storey} \&
  {Zeippen}}{2000}]{2000Storey_mnra312}
{Storey} P.~J.,  {Zeippen} C.~J.,  2000, \mnras, 312, 813

\bibitem[\protect\citeauthoryear{{Storey}, {Sochi}  \& {Badnell}}{{Storey}
  et~al.}{2014}]{2014Storey_mnras441}
{Storey} P.~J.,  {Sochi} T.,   {Badnell} N.~R.,  2014, \mn@doi [\mnras]
  {10.1093/mnras/stu777}, \href
  {http://adsabs.harvard.edu/abs/2014MNRAS.441.3028S} {441, 3028}

\bibitem[\protect\citeauthoryear{{Storey}, {Sochi}  \& {Bastin}}{{Storey}
  et~al.}{2017}]{2017Storey_mnra470}
{Storey} P.~J.,  {Sochi} T.,   {Bastin} R.,  2017, \mnras, 470, 379

\bibitem[\protect\citeauthoryear{{Tayal}}{{Tayal}}{2011}]{2011Tayal_apjs195}
{Tayal} S.~S.,  2011, \apjs, 195, 12

\bibitem[\protect\citeauthoryear{{Tayal} \& {Zatsarinny}}{{Tayal} \&
  {Zatsarinny}}{2010}]{2010Tayal_apjs188}
{Tayal} S.~S.,  {Zatsarinny} O.,  2010, \apjs, 188, 32

\bibitem[\protect\citeauthoryear{{Torres-Peimbert}, {Peimbert}  \&
  {Pena}}{{Torres-Peimbert} et~al.}{1990}]{1990Torres-Peimbert_aap233}
{Torres-Peimbert} S.,  {Peimbert} M.,   {Pena} M.,  1990, \aap, 233, 540

\bibitem[\protect\citeauthoryear{{Tsamis} \& {P{\'e}quignot}}{{Tsamis} \&
  {P{\'e}quignot}}{2005}]{2005Tsamis_mnra364}
{Tsamis} Y.~G.,  {P{\'e}quignot} D.,  2005, \mnras, 364, 687

\bibitem[\protect\citeauthoryear{{Tsamis} \& {P{\'e}quignot}}{{Tsamis} \&
  {P{\'e}quignot}}{2006}]{2006pnbm.conf..192T}
{Tsamis} Y.~G.,  {P{\'e}quignot} D.,  2006, in {Stanghellini} L.,  {Walsh}
  J.~R.,   {Douglas} N.~G.,  eds, Planetary Nebulae Beyond the Milky Way.
  p.~192, \mn@doi{10.1007/3-540-34270-2_25}

\bibitem[\protect\citeauthoryear{{Tsamis}, {Barlow}, {Liu}, {Storey}  \&
  {Danziger}}{{Tsamis} et~al.}{2004}]{2004Tsamis_mnra353}
{Tsamis} Y.~G.,  {Barlow} M.~J.,  {Liu} X.,  {Storey} P.~J.,   {Danziger}
  I.~J.,  2004, \mnras, 353, 953

\bibitem[\protect\citeauthoryear{{Tylenda}}{{Tylenda}}{2003}]{2003IAUS..209..389T}
{Tylenda} R.,  2003, in {Kwok} S.,  {Dopita} M.,   {Sutherland} R.,  eds,  IAU
  Symposium Vol. 209, Planetary Nebulae: Their Evolution and Role in the
  Universe. p.~389
  
\bibitem[\protect\citeauthoryear{{Viegas} \& {Clegg}}{{Viegas} \&
  {Clegg}}{1994}]{1994Viegas_mnra271}
{Viegas} S.~M.,  {Clegg} R.~E.~S.,  1994, \mnras, 271, 993

\bibitem[\protect\citeauthoryear{{Wesson}, {Jones}, {Garc{\'{\i}}a-Rojas},
  {Boffin}  \& {Corradi}}{{Wesson} et~al.}{2018}]{2018Wesson_mnra480}
{Wesson} R.,  {Jones} D.,  {Garc{\'{\i}}a-Rojas} J.,  {Boffin} H.~M.~J.,
  {Corradi} R.~L.~M.,  2018, \mn@doi [\mnras] {10.1093/mnras/sty1871}, \href
  {http://adsabs.harvard.edu/abs/2018MNRAS.480.4589W} {480, 4589}

\bibitem[\protect\citeauthoryear{{Yuan}, {Liu}, {P{\'e}quignot}, {Rubin},
  {Ercolano}  \& {Zhang}}{{Yuan} et~al.}{2011}]{2011Yuan_mnra411}
{Yuan} H.,  {Liu} X.,  {P{\'e}quignot} D.,  {Rubin} R.~H.,  {Ercolano} B.,
  {Zhang} Y.,  2011, \mnras, 411, 1035

  \makeatother
\end{thebibliography}
